\documentclass[preprint,12pt]{elsarticle}

\usepackage{lineno,hyperref}
\usepackage{xcolor}
\usepackage{amssymb}
\usepackage{amsmath}
\usepackage{siunitx}
\usepackage{subfigure}
\usepackage{color, colortbl}
\usepackage{multirow}
\usepackage{float}
\usepackage{graphicx}
\usepackage{epsfig}
\usepackage{epstopdf}
\usepackage{caption}
\usepackage{booktabs,threeparttable}
\usepackage{bm}
\usepackage{lscape}

\usepackage{textcomp}

\usepackage{pifont}
\usepackage[normalem]{ulem}

\newcommand{\xmark}{\ding{55}}

\graphicspath{{./Figures/}}
\definecolor{Gray}{gray}{0.9}

\journal{Nuclear Physics B}

\begin{document}

\begin{frontmatter}



\title{Mechanically reconfigurable layered GRIN lenses for mid-infrared and mm-wave bands}

\author[FirstAuthorsAffiliation,SecondAuthorsAffiliation]{Keivan Kaboutari}
\author[FirstAuthorsAffiliation]{Xiu Liu}
\author[FirstAuthorsAffiliation]{Sheng Shen\corref{mycorrespondingauthor1}}
\cortext[mycorrespondingauthor1]{Corresponding author}
\ead{shengshe@andrew.cmu.edu}
\author[SecondAuthorsAffiliation]{Pedro Pinho}
\author[SecondAuthorsAffiliation]{Stanislav Maslovski\corref{mycorrespondingauthor2}}
\cortext[mycorrespondingauthor2]{Corresponding author}
\ead{stanislav.maslovski@ua.pt}
\address[FirstAuthorsAffiliation]{Department of Mechanical Engineering, Carnegie Mellon University, Pittsburgh, PA 15213, United States of America}
\address[SecondAuthorsAffiliation]{Instituto de Telecomunica\c{c}\~{o}es and Department of Electronics, Telecommunications and Informatics, University of Aveiro, 3810-193 Aveiro, Portugal}




\begin{abstract}
    This paper presents a concept of mechanically adjustable planar Gradient-index (GRIN) lens that can be used to realize reconfigurable focusing at mid-infrared (mid-IR) or millimeter-wave (mm-wave) frequencies. The concept is based on a multilayer design rooted in effective medium theory, where spatial refractive index profiles are synthesized using first-kind Chebyshev polynomials. Symmetric lateral displacement of the layers enables continuous modulation of the effective refractive index distribution, allowing controlled transitions between divergent, afocal, and convergent wavefronts. Numerical simulations demonstrate accurate focal tuning at mid-IR and mm-waves. In the mid-IR regime, focal lengths of about 56~$\mu\text{m}$ and 24~$\mu\text{m}$ are achieved with relative errors of around 4.4\% and 10\%, respectively. In the mm-wave regime, the lens produces focal lengths of about 158~mm and 128~mm with improved accuracies of 2.3\% and 6.0\%, respectively. The reconstructed refractive index profiles exhibit high similarity to ideal convex lens distributions, confirming the validity of the proposed synthesis approach. The proposed GRIN lens concept offers a scalable and physically interpretable approach to reconfigurable focusing, with potential applications in multimodal imaging, adaptive sensing, and telecommunication systems.
\end{abstract}

\begin{graphicalabstract}
    \centering
    \includegraphics[width=1\linewidth,trim={0.0cm 0.0cm 0.0cm 0.0cm},clip=true]{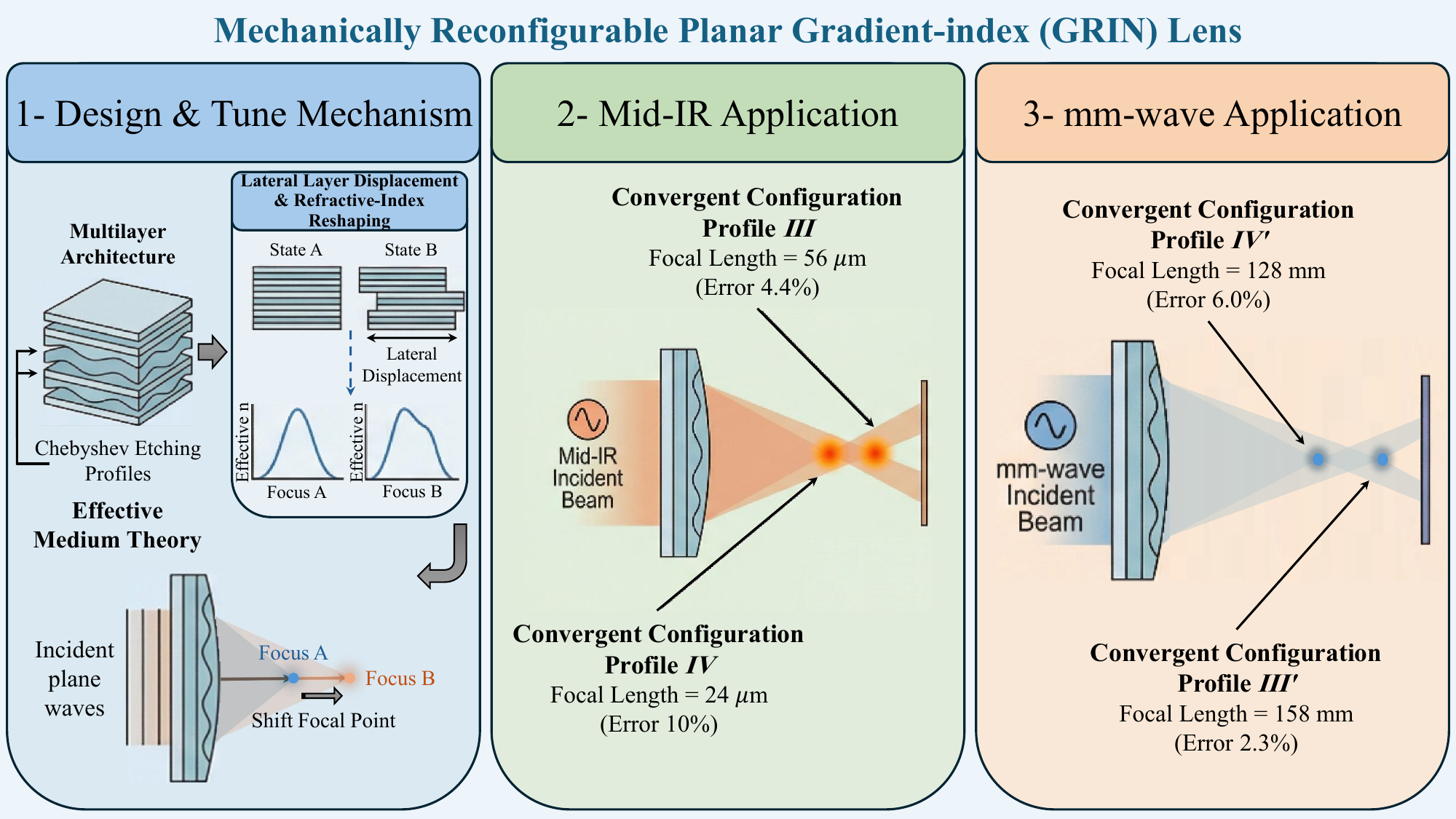}
\end{graphicalabstract}

\begin{highlights}
    \item Planar multilayer GRIN lens enables dynamic tuning for mid-IR and mm-wave bands.
    \item Chebyshev polynomials provide optimal basis to approximate smooth index gradients.
    \item Focal distance and location are continuously controlled via lateral layer shifts.
    \item Translates the mechanical Alvarez lens tuning principle into the GRIN domain.
    \item Decouples base focusing power from the reconfigurable tuning mechanism.
\end{highlights}


\end{frontmatter}



\section{Introduction}
    Lenses are fundamental optical components that manipulate electromagnetic waves and light through refraction. They enable applications in the microwave~\cite{Tang_2021, Kaboutari_2025} and optical~\cite{Hanninen_2019} spectrums. Convex and concave lenses are the two primary types. Each type has distinct refractive index properties and applications. Understanding their behavior is essential in imaging~\cite{Datta_2022} and sensing~\cite{Bagheri_2025}, photography~\cite{Kasztelanic_2019}, ophthalmology~\cite{Shen_2015, Koch_2004}, microscopy~\cite{Zhao_2025}, vision correction~\cite{Williams_2025}, and laser~\cite{Shabairou_2024} systems. Beamforming and focusing systems provide precise control over electromagnetic waves and light propagation. They improve performance in radio~\cite{Chen_2011, Meng_2018} and optical systems~\cite{Wei_2025}, non-destructive testing~\cite{Gopalsami_1994}, radar systems~\cite{Tang_2021, Raman_1998}, biomedical~\cite{Paul_2023}, photonics~\cite{Halir_2020}, Light Detection and Ranging (LiDAR)~\cite{Li_2022}, Optical Coherence Tomography (OCT)~\cite{Kang_2020}, and optical engineering~\cite{Fröch_2023} applications. 
    
    Gradient-index (GRIN) lenses are important in these applications because they manipulate electromagnetic waves and light through a spatially varying refractive index. This removes the need for complex curved surfaces. Compared with conventional lenses, GRIN lenses offer compact, lightweight, and aberration-corrected solutions, making them suitable for miniaturized and integrated systems~\cite{Kaboutari_2025}. Beamforming with GRIN lenses enables dynamic steering and shaping of beams, which is critical in engineering and industrial applications. Their strong focusing capability also improves resolution in endoscopy and microscopy, supporting advances in biomedical imaging and laser surgery~\cite{Koch_2004, Guo_2020}. In addition, GRIN lenses are useful in industrial laser processing and telecommunications, where precise beam control improves efficiency and signal integrity.
    
    Fröch et al.~\cite{Fröch_2023} present a Meta-optical Fiber Endoscope (MOFIE) for real-time full-color imaging. The design uses inverse-designed meta-optics to reduce chromatic aberrations. It optimizes a multi-chromatic Modulation Transfer Function (MTF) to ensure broadband performance across visible wavelengths. The device is fabricated with silicon nitride nanopillars and integrated with a coherent fiber bundle. Compared with traditional GRIN lenses, it achieves a 33\% reduction in rigid tip length. Whiting et al.~\cite{Whiting_2023} optimize GRIN lenses to improve reflectarray feed performance by reducing size and increasing aperture efficiency. The lenses have spatially varying permittivities and are designed using numerical optimization methods, including adjoint sensitivity and Covariance Matrix Adaptation-Evolution Strategy (CMA-ES) algorithms. Additive manufacturing is used to validate the designs.
    
    Blackwell et al.~\cite{Blackwell_2023} present a Quantum Algorithm Emulator (QAE) for the Deutsch-Josza algorithm using an inverse-designed terahertz GRIN lens. The method combines numerical Finite Difference Time-domain (FDTD) simulations with Machine Learning (ML) optimization. This approach improves lens performance. The GRIN lens modulates THz waves together with a silicon oracle subblock to emulate quantum states. ML inverse design is used to optimize hole radii and lens thickness. Moharrami et al.~\cite{Moharrami_2020} propose a methodology for designing tunable GRIN devices based on graphene-dielectric multilayer metamaterials (GDMMs) at terahertz frequencies. The effective dielectric permittivity of GDMMs is spatially modulated by tailoring graphene layers' chemical potential. The Effective Medium Theory (EMT) and FDTD methods are used to analyze planar and cylindrical GDMM structures. A GRIN lens, a focusing reflector, and a cylindrical Generalized Luneburg Lens (GLL) are employed in order to produce tunable terahertz photonic jets.
    
    The study presented in~\cite{Rezaeieh_2021} introduces a reconfigurable GRIN lens designed for biomedical torso scanning applications. It offers a potential solution to existing limitations in radar-based imaging methodologies. This innovative approach improves imaging algorithm precision by converting spherical to plane wavefronts, thereby addressing a critical constraint of conventional body-matched antennas that generate spherical rather than planar wavefronts. The reconfigurable mechanism employs fluidic modulation, where water infiltration into internal cavities dynamically alters the lens's local effective permittivity. However, this design necessitates additional hydraulic infrastructure including reservoirs and pumping systems, presenting practical implementation challenges in clinical environments. Furthermore, potential fluid leakage represents a reliability concern that may compromise system performance.
    
    Mousavi et al.~\cite{Mousavi_2023} presents a tapered GRIN lens antenna that enhances electromagnetic wave penetration for near-field imaging of the torso. The methodology involves a four-layer lens filled with engineered dielectric mixtures, optimized using a genetic algorithm to minimize the generalized reflection coefficient. A lens tapered according to total internal reflection theory ensures radiation focuses in a planar manner.
    
    A study in~\cite{Gomez_2008}  presents a reconfigurable planar optical interconnect structure based on GRIN semi-rod lenses for crossover and parallel interconnects. The design exploits on-axis and off-axis multiple imaging within the GRIN substrate to route optical signals, while reconfigurability is achieved using movable transverse mirrors.
    Huang et al.~\cite{Huang_2010} demonstrate a tunable, two-dimensional Liquid Gradient Refractive Index (L-GRIN) microlens for out-of-plane light focusing. Fabricated via standard soft lithography, the planar microfluidic chamber achieves 2D point focusing by establishing a hyperbolic secant refractive index profile via lateral molecular diffusion between co-flowing {\tt CaCl$_{\rm 2}$} solution and deionized water. Adjusting the inlet flow rate ratios dynamically tunes the gradient contrast and shifts the focal position along the axis perpendicular to the flow. The device provides a low-consumption, highly repeatable optofluidic mechanism suited for lab-on-a-chip applications. Zeng et al. \cite{Zeng_2014} propose a graphene plasmonic metasurface by patterning the underlying dielectric layer of monolayer graphene. Operating in the terahertz regime, this platform realizes an electronically tunable GRIN lens capable of high-performance wave focusing, collimating, and guiding at a deep subwavelength scale.
    
    In this paper, a reconfigurable GRIN lens employing a multi-layered architecture with spatially varying refractive indices engineered via EMT is presented. Each layer is etched with groove profiles defined by first-kind Chebyshev polynomials, ensuring smooth refractive index gradients for precise reconfigurable focusing. Symmetric lateral displacement of layers with mirrored refractive index profiles enables dynamic focal tuning, which can be achieved with piezoelectric or Micro-electromechanical (MEMS) based positioning for Mid-Infrared (mid-IR) and stepper motor-driven actuation for millimeter-wave (mm-wave) systems. A closely related analogy of this architecture in the domain of non-GRIN lenses is the Alvarez lens~\cite{Babington_2015}.
    
    Traditional Alvarez lenses achieve focal tunability by laterally displacing complementary cubic phase plates, producing a quadratic phase profile whose optical power varies linearly with displacement while the element spacing remains fixed~\cite{Babington_2015}. In contrast, our architecture implements a similar tuning principle in a planar GRIN lens by replacing surface curvature-induced phase modulation with a spatially varying effective refractive index. A stationary second-order Chebyshev layer provides the fixed focusing power, while two laterally displaced third-order Chebyshev layers modulate the quadratic coefficient of the effective index profile. Unlike conventional Alvarez lenses, which rely on freeform surfaces and cancellation of cubic sag terms, the proposed design exploits the orthogonality of the Chebyshev basis and uses effective medium theory to realize the required index distribution in a lithographically compatible planar platform.
    
    The proposed GRIN lens can be fabricated using conventional-dielectric materials. Numerical simulations validate phase coherence and focal accuracy, while Geometric Optics (GO) analysis confirms adaptive focusing capabilities. The design integrates a thin metallic collimator to mitigate edge effects and enhance focal spot resolution. Overall, the proposed methodology represents a hybrid of GRIN optics and the known operating principle of the Alvarez lens, albeit deviating from the standard paraxial cubic approximation, by decoupling the base focusing power from the focal tuning mechanism. As a result, the proposed planar structure preserves the nominal lens quality while enabling focal-length control.    
    
    Furthermore, the joint investigation of the mid-IR and mm-wave regimes that is done in this work confirms frequency scalability of the proposed GRIN-based focal tunability mechanism. Since the lens structure is derived from the EMT-based refractive index synthesis, the proposed concept is inherently wavelength-normalized. As a result, the same design principles apply across vastly different spectral domains, despite differences in absolute wavelength, material platforms, and fabrication constraints.
\section{Structure of Reconfigurable GRIN Lens}
\begin{figure}[ht!]
        \begin{center}
            \subfigure[]{\label{Fig: Fig01.a}
            \includegraphics[width=0.38\linewidth,trim={0.0cm 0.0cm 0.0cm 0.0cm},clip=true]{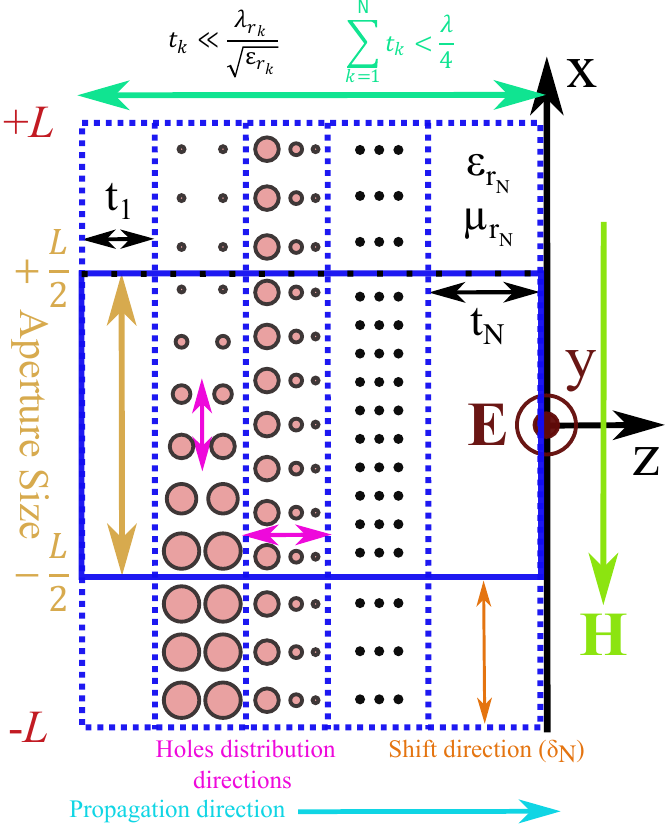}}
            \subfigure[]{\label{Fig: Fig01.b}
            \includegraphics[width=0.65\linewidth,trim={0.0cm 0.0cm 0.0cm 0.0cm},clip=true]{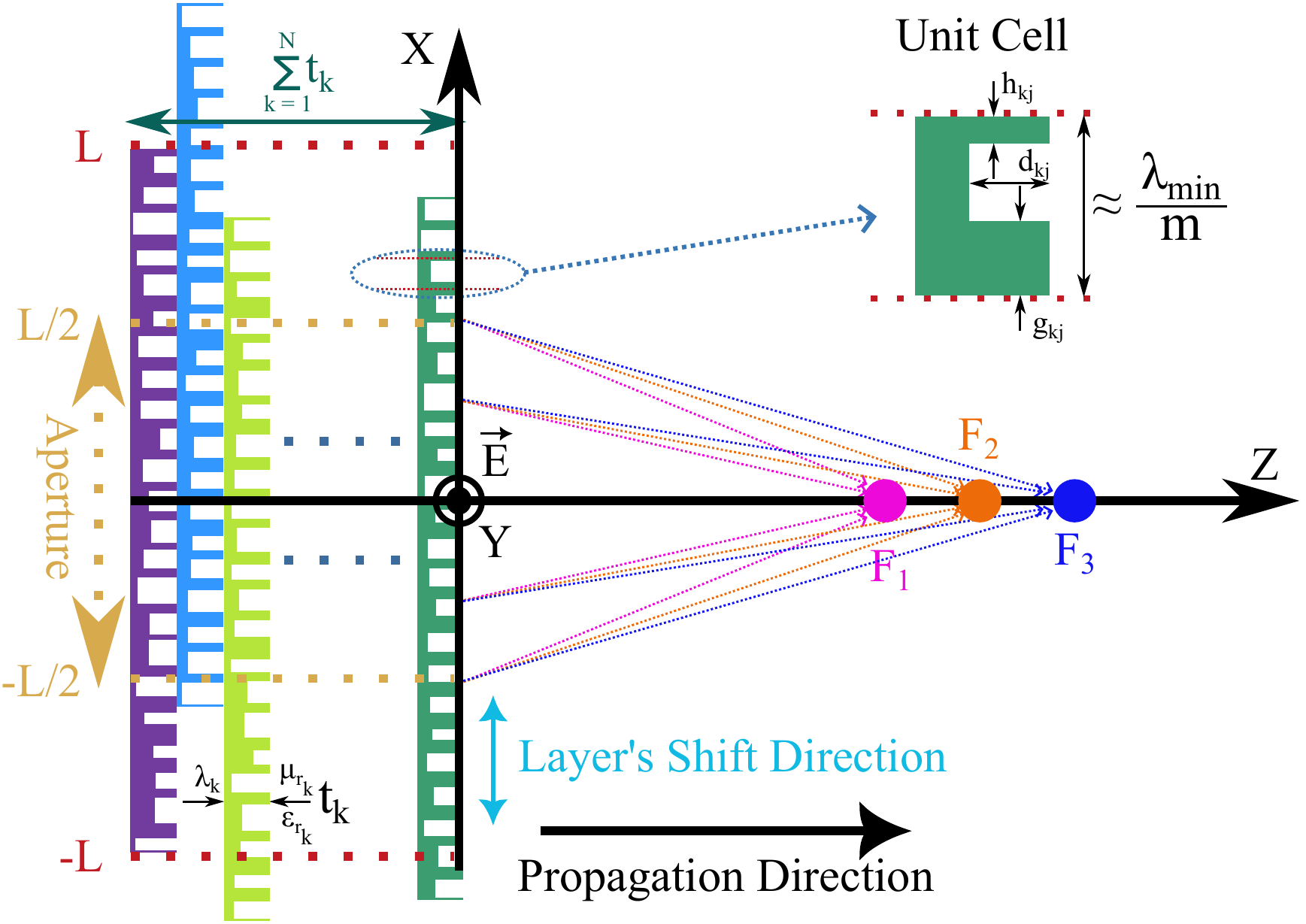}}
            
            \subfigure[]{\label{Fig: Fig01.c}
            \includegraphics[width=0.65\linewidth,trim={0.0cm 0.0cm 0.0cm 0.0cm},clip=true]{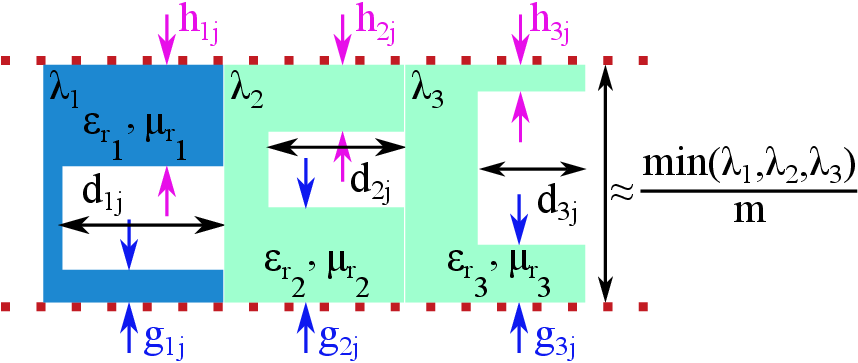}}
        \end{center}
        \vspace{-4.0mm}
        \caption{Geometry of the proposed GRIN lens configuration: (a) Structure of the proposed laminar GRIN lens; (b) Structural schematics and the geometrical optics ray picture; (c) Supercell of the GRIN lens. Here, $\varepsilon_{\rm r_{\it k}}$ and $\mu_{\rm r_{\it k}}$ are the relative permittivity and permeability of material $k$. Additionally, $N$, $k$, and $j$ represent the number of layers in GRIN lens and layer and groove indices, respectively.}
        \label{Fig: Fig01}
    \end{figure}

    In this study, a focal-length tuning GRIN lens concept  with a repositionable focal point in the radiative near field zone is being developed for the mid-IR and mm-wave bands. In order to achieve such reconfigurable focusing, the lens is divided into several layers, each layer consisting of materials with different refractive indices, as Figs.~\ref{Fig: Fig01.a} and~\ref{Fig: Fig01.b}. To manipulate the effective refractive index of the GRIN lens, the materials of its layers are etched so as to control the local effective refractive index by employing EMT~\cite{Slovick_2014, Alu_2011}.
    
    Specifically, EMT is adopted to model the macroscopic properties of the lens layers. As shown in Fig.~\ref{Fig: Fig01}, the corrugations are oriented parallel to the $y$-axis aligned with the direction of the electric field of the incident wave. To achieve a targeted refractive index distribution, the grooves may be filled with either air or alternative materials possessing distinct dielectric properties. However, the etched layers should be capable of being mechanically moved, resulting in a reconfigurable focal point.
    Figure~\ref{Fig: Fig01.b} also illustrates a representative unit cell of the layered structure. The unit cell’s $x$-dimension (height $l$) is supposed to be a fraction of the wavelength, to ensure applicability of EMT. In what follows, $\lambda_{\rm 0}$ denotes the free space wavelength at the operating frequency. The wavelength in a dielectric with the refractive index $n_{\rm i} > 1$ is shorter: $\lambda_{\rm i} = \lambda_0/n_{\rm i}$. In this study, the parameter $m$ in Fig.~\ref{Fig: Fig01.b} was fixed at $m=2$ ($l\leq\lambda_{\rm min}/2$) for mid-IR and microwave applications at 150~THz and 40~GHz frequencies to ensure sufficiently smooth spatial variations in the layer refractive indices along the $x$-axis, adequately smaller than $\lambda_{\rm 0}$ (where $\lambda_{\rm min}$ is the min$(\lambda_{1},\lambda_{2},\dots,\lambda_{N})$). This ensures that the sub-wavelength unit cell condition is satisfied for the highest-index layer without grooves, representing the worst-case scenario. Introducing grooves reduces the effective refractive index, increasing the wavelength in the material and making the unit cell even smaller relative to the wavelength. The specific dimensions $h_{kj}$, $d_{kj}$, and $g_{kj}$ are analytically derived to achieve the target local refractive index at each position of the unit cell, determined by the requirements of the graded index profile.
    
    Through comprehensive simulation and optimization, a reconfigurable GRIN lens based on a three-layer architecture has been developed. The design incorporates: ($i$) a fixed high refractive index layer that primarily governs the transmitted field's phase profile, and ($ii$) two independently movable layers composed of identical low refractive index materials with laterally mirrored groove structures. As a result, it enables axial focal adjustment by symmetrically displacing the second and third layers in opposite lateral directions. With an aperture size $L$, each layer extends $2L$ along the lateral dimension, permitting maximum opposing displacements of $\pm L/2$ for the adjustable layers while maintaining structural integrity and optical functionality. This configuration provides an accurate focal control.
    \subsection{Effective Medium Theory-Guided Material Selection}
\begin{figure}[ht!]
            \begin{center}
                \subfigure[]{\label{Fig: Fig02.a}
                \includegraphics[height=0.25\linewidth,trim={0.0cm 0.0cm 0.0cm 0.0cm},clip=true]{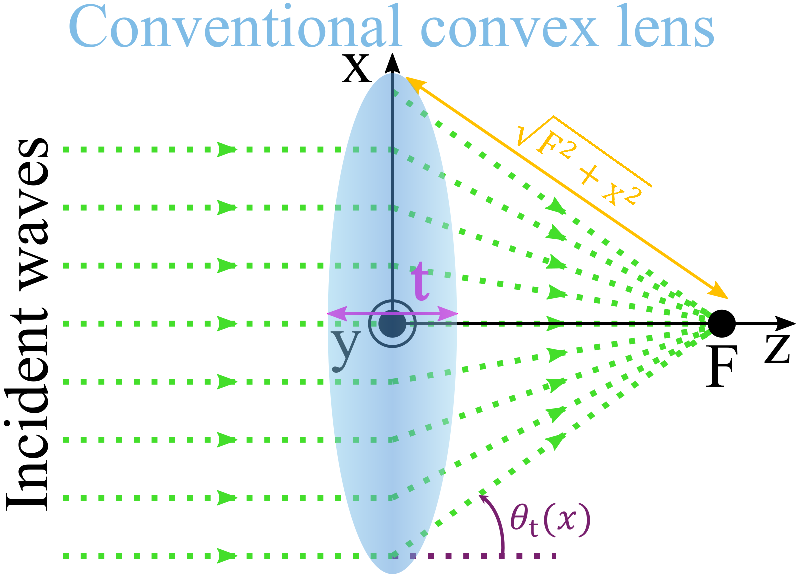}}
                \subfigure[]{\label{Fig: Fig02.b}
                \includegraphics[height=0.25\linewidth,trim={0.0cm 0.0cm 0.0cm 0.0cm},clip=true]{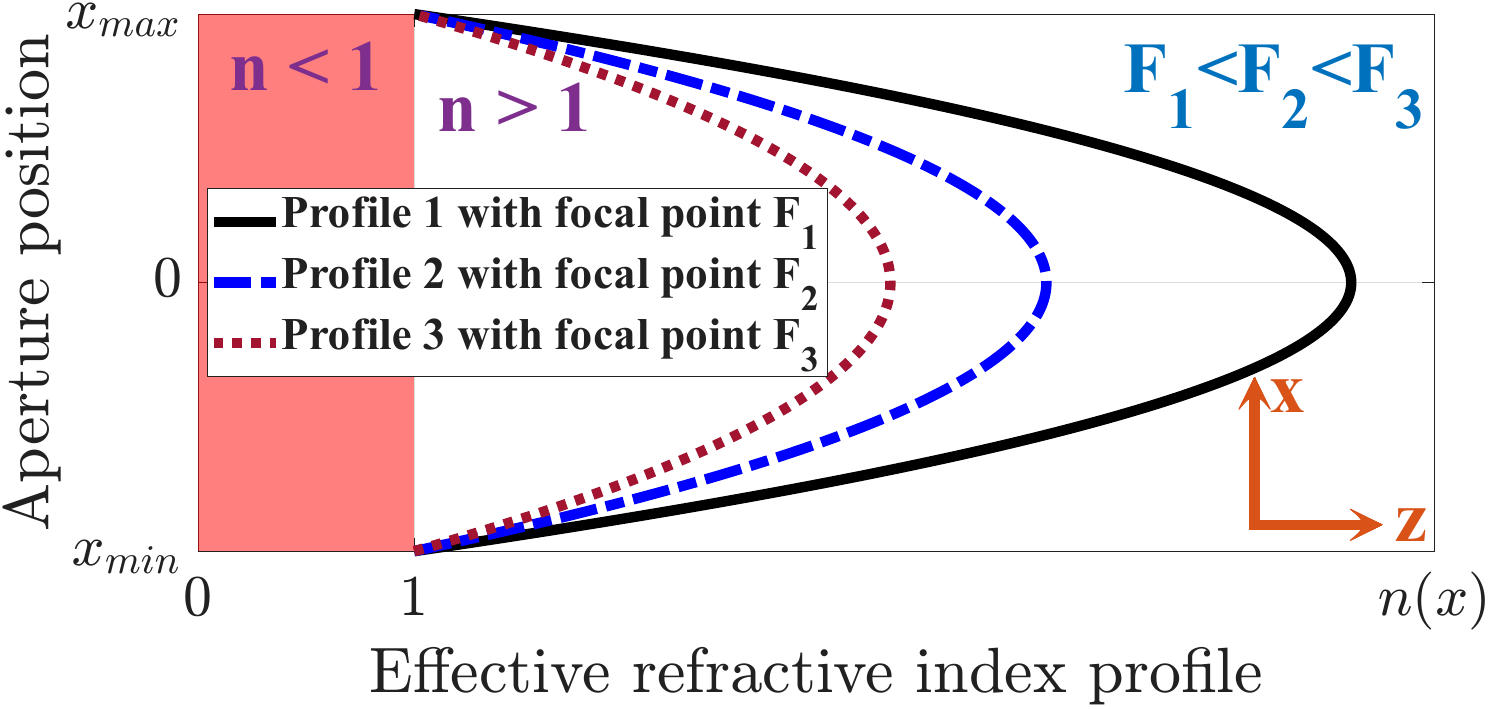}}
                
                \subfigure[]{\label{Fig: Fig02.c}
                \includegraphics[height=0.25\linewidth,trim={0.0cm 0.0cm 0.0cm 0.0cm},clip=true]{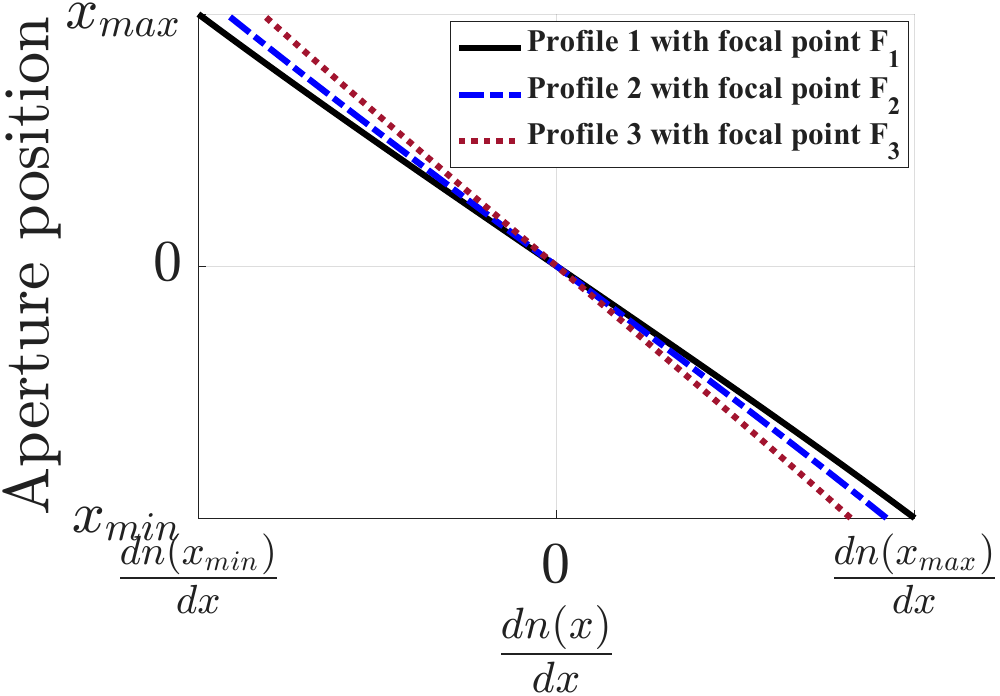}}
                \subfigure[]{\label{Fig: Fig02.d}
                \includegraphics[height=0.25\linewidth,trim={0.0cm 0.0cm 0.0cm 0.0cm},clip=true]{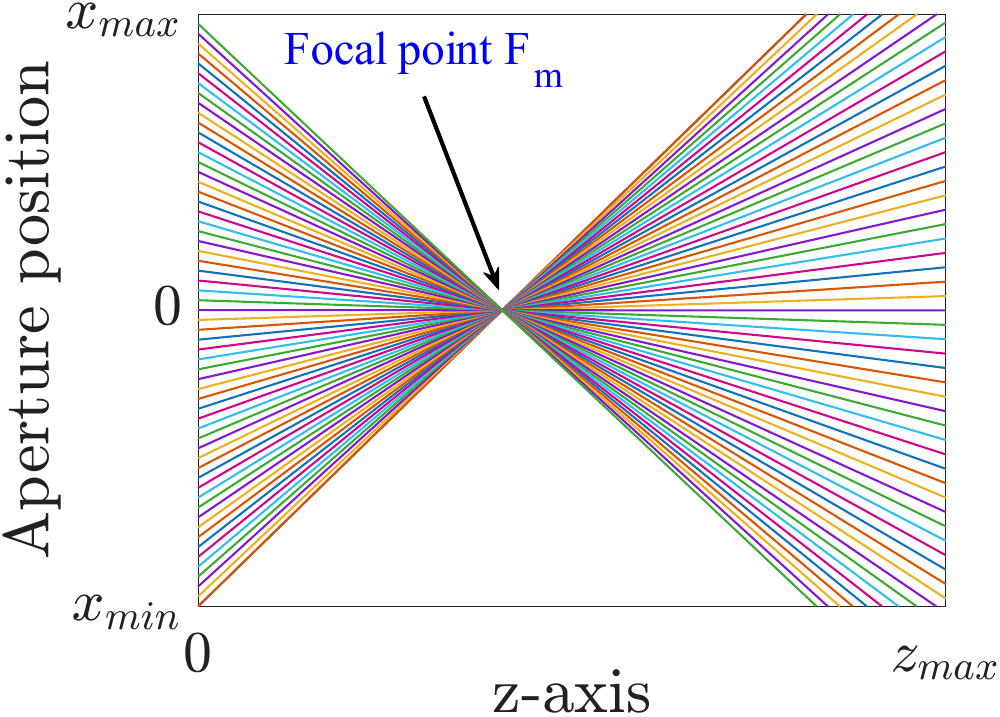}}
            \end{center}
            \vspace{-4.0mm}
            \caption{Refractive properties of a conventional convex lens: (a) Ray diagram illustrating light focusing to an arbitrary focal point via surface curvature-induced refraction, (b) Quadratic effective refractive index profile along the lens's optical axis ($x$-direction), demonstrating the spatial variation required for beam convergence, (c) Aperture-dependent gradient of the refractive index, and (d) GO ray picture of an ideal convex lens.}
            \label{Fig: Fig02}
        \end{figure}

        Conventional convex lenses focus light to a focal point due to refraction at their curved surfaces, causing all incident rays to converge, Fig.~\ref{Fig: Fig02.a}. The curvature of the surfaces ensures spatially dependent refraction angles, directing rays toward the focal point. If one uses a planar lens instead, a similar behavior can be achieved with a GRIN lens having quadratic effective refractive index profile along the $x$-axis.
        
        By emulating this effective refractive index profile, it becomes possible to engineer a structure capable of beam focusing at distinct positions, Fig.~\ref{Fig: Fig02.b}.
        The refractive index distribution required to focus beams at an arbitrary axial point (Eq.~\ref{Eq: Eq01}), $F$, is given by (See Supplementary Material):
        \begin{equation}
            n(x)=\frac{\sqrt{F^{\rm 2}+x_{\rm max}^{\rm 2}}-\sqrt{F^{\rm 2}+x^{\rm 2}}}{t(x)}+1+\delta.
            \label{Eq: Eq01}
        \end{equation}
        In cases where the focal point does not reside on the $z$-axis, the structure can be displaced upward or downward along the $x$-axis to attain the intended focal location. The effective refractive index profiles shown in the figure have different focusing points along the $z$-axis. In other words, such profiles represent ideal cases for displacing beams toward the desired focal points along the $z$-axis. 
        
        The achievable focal point repositioning range is governed by the extremal values of the effective refractive index in the $x$-direction. Consequently, the fabrication of a GRIN lens with tunable focal points necessitates careful selection of materials whose refractive indices can faithfully reproduce this target index profile. Thus, material choice is critical to realizing the desired graded-index distribution for reconfigurable focusing.
        
        Figure~\ref{Fig: Fig02.c} depicts the refractive index gradient across the aperture for an effective GRIN lens modeled off a conventional convex lens. As can be seen, the variation rate of the refractive index is constant. This analysis reveals that a quasi-linear variation of the effective index of refraction correlates strongly with the ability of a GRIN lens to converge the incident rays to a single focal spot. A faster variation in the refraction index results in a lens with a greater focusing power and shorter focal distance.
        
        A GO ray picture for a conventional convex lens (after passing the lens) is shown in Fig.~\ref{Fig: Fig02.d}. As can be seen, the lens produces a sharp focal point. In this case, the standard GO formulation is rooted in the Snell's Law of refraction, which can be reformulated for a GRIN lens as follows:
        \begin{equation}
            \theta_{\rm t}(x)=\sin^{\rm -1}{\left(\frac{dn(x)}{dx}t(x) + \frac{dt(x)}{dx}n(x)\right)},
            \label{Eq: Eq02}
        \end{equation}
        where $\theta_{t}(x)$ is the transmitted ray angle. Note that the lens thickness $t(x) = {\rm const}$ in a flat GRIN lens and the second term disappears. The above result assumes that the lens is illuminated by a bunch of rays parallel to its optical axis.
        
        On the other hand, the effective refractive index distribution for the considered GRIN lens, which is required to focus beams at any desired axial position (Figs.~\ref{Fig: Fig03.g} to~\ref{Fig: Fig03.l}) can be determined using EMT as follows:
        \begin{equation}
            n^{\rm eff.}(x)=\sum_{k=1}^{N}\mathbf{T}_{k}n_{\rm eff.}^{k}(x),
            \label{Eq: Eq03}
        \end{equation}
        where $n_{\rm eff.}^{k}(x)$ (Figs.~\ref{Fig: Fig03.a} to~\ref{Fig: Fig03.f}) and $\mathbf{T}_{k}$ are the effective refractive index and relative thickness of the $k^{\rm th}$ layer, respectively. The relative thickness of the $k^{\rm th}$ layer is defined as:
        \begin{equation*}
            \mathbf{T}_{k}=\displaystyle\frac{\displaystyle t_{k}}{\displaystyle \sum_{i=1}^{N}t_{i}}.
        \end{equation*}
    \subsection{Etching Profile Optimization for Refractive Index Engineering}
        To attain the target effective refractive index profile along the $x$-axis, suitable basis functions for the refractive index distribution in each layer must be selected. These distributions are realized by etching grooves of varying depths and heights in each layer. By shifting the layers along the $x$-axis, distinct focal points can be generated along the propagation direction (the $z$-axis).
        
\begin{figure}[ht!]
            \begin{center}
                \subfigure[]{\label{Fig: Fig03.a}
                \includegraphics[width=0.3\linewidth,trim={0.0cm 0.0cm 0.0cm 0.0cm},clip=true]{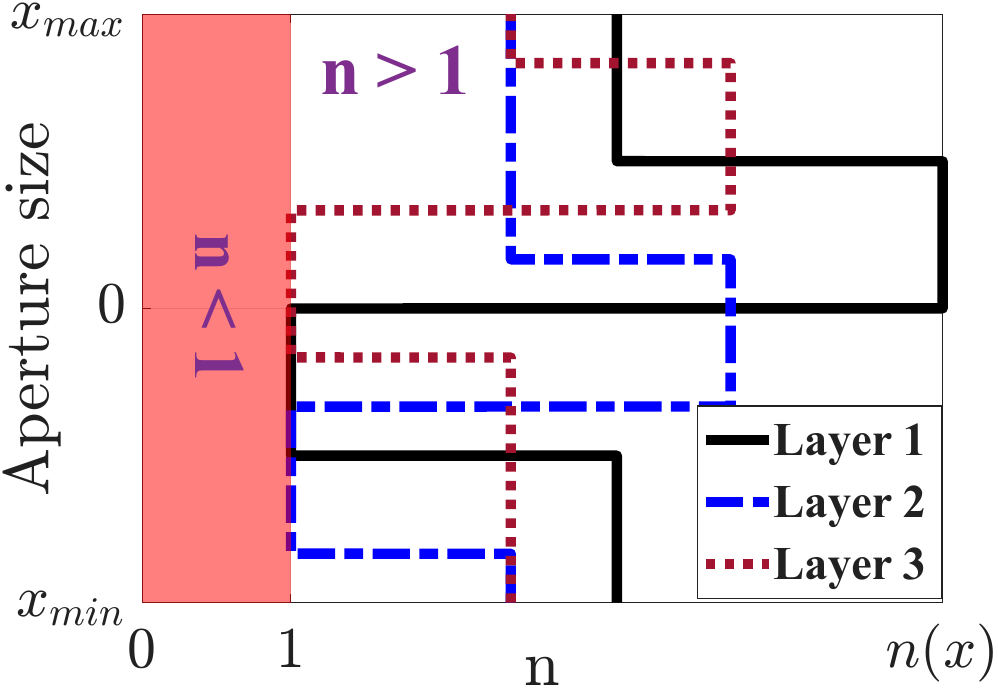}}
                \subfigure[]{\label{Fig: Fig03.b}
                \includegraphics[width=0.3\linewidth,trim={0.0cm 0.0cm 0.0cm 0.0cm},clip=true]{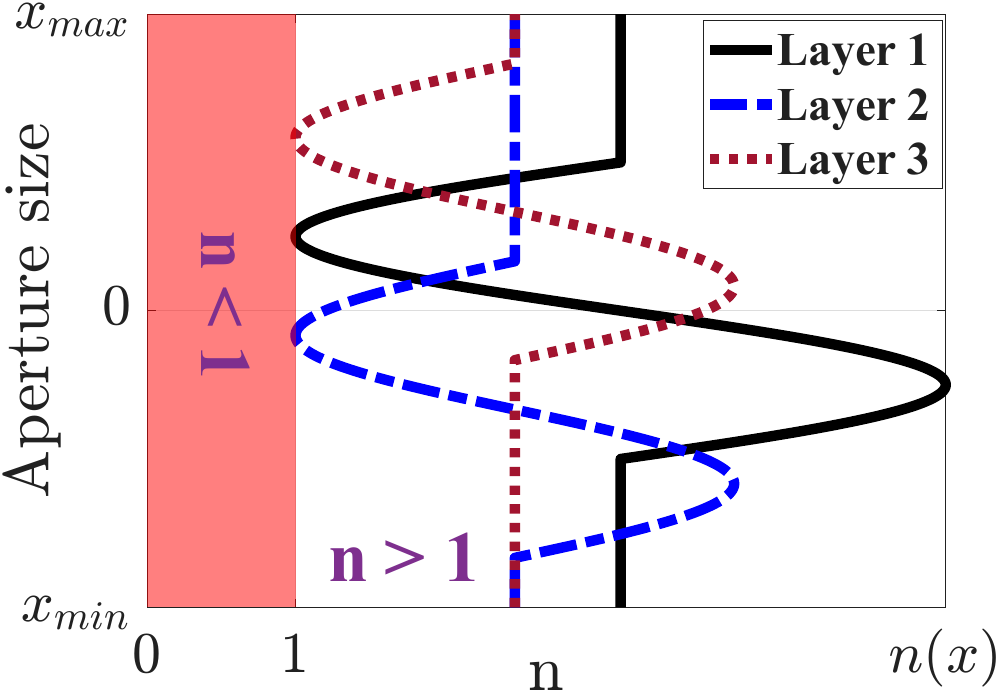}}
                \subfigure[]{\label{Fig: Fig03.c}
                \includegraphics[width=0.3\linewidth,trim={0.0cm 0.0cm 0.0cm 0.0cm},clip=true]{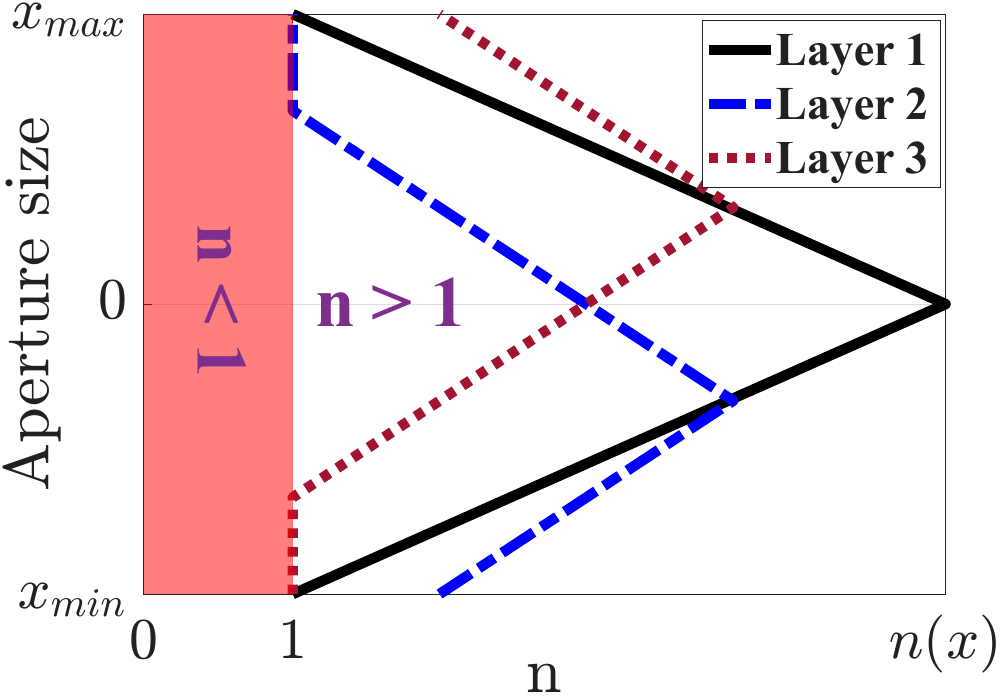}}
                
                \subfigure[]{\label{Fig: Fig03.d}
                \includegraphics[width=0.3\linewidth,trim={0.0cm 0.0cm 0.0cm 0.0cm},clip=true]{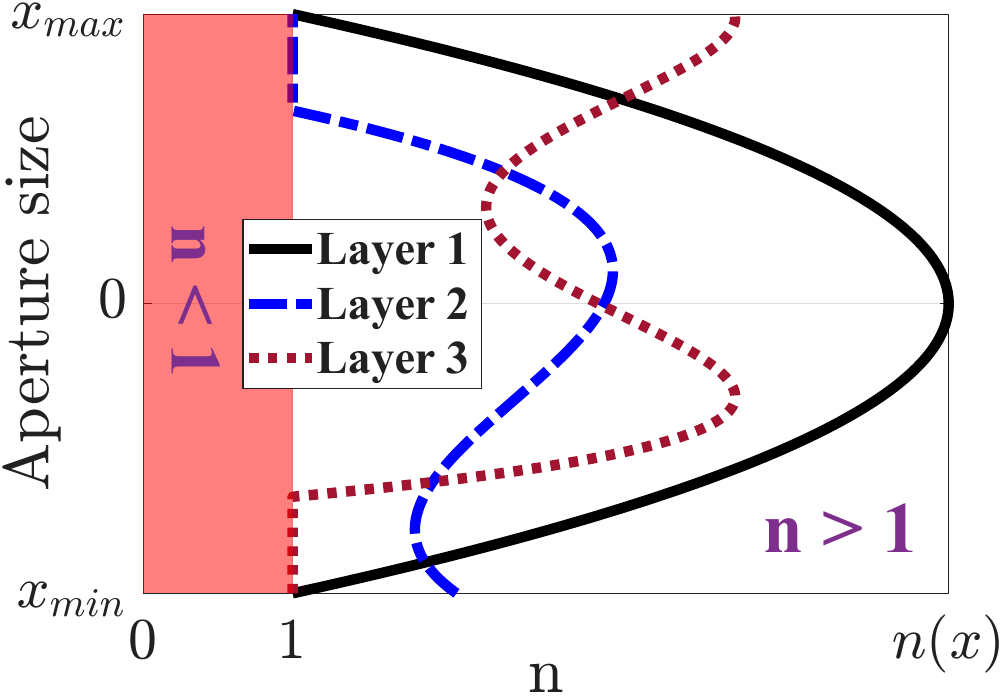}}
                \subfigure[]{\label{Fig: Fig03.e}
                \includegraphics[width=0.3\linewidth,trim={0.0cm 0.0cm 0.0cm 0.0cm},clip=true]{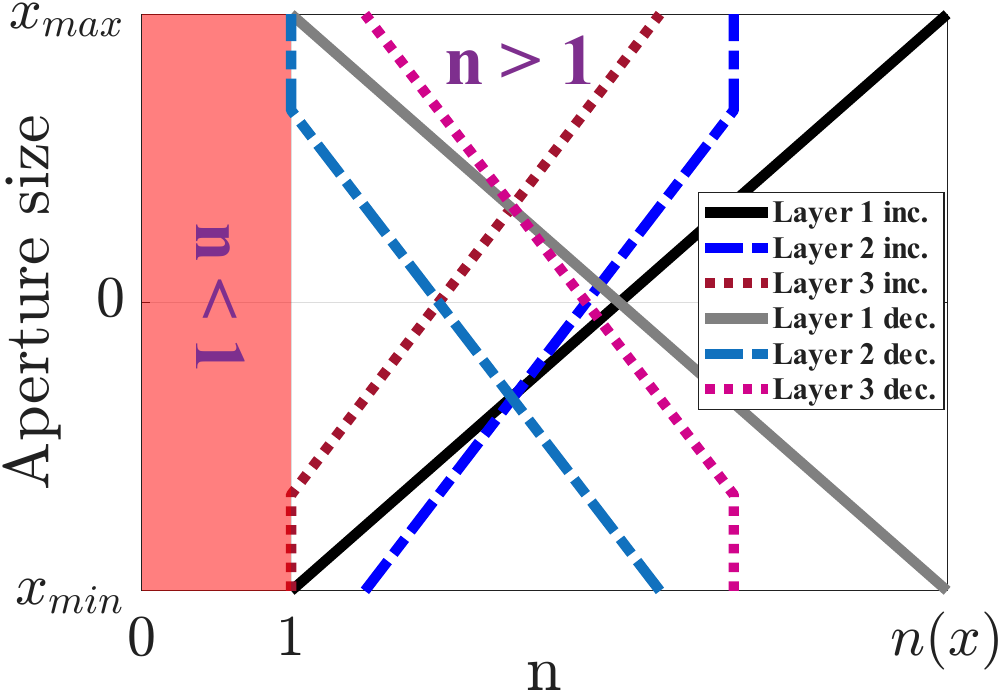}}
                \subfigure[]{\label{Fig: Fig03.f}
                \includegraphics[width=0.3\linewidth,trim={0.0cm 0.0cm 0.0cm 0.0cm},clip=true]{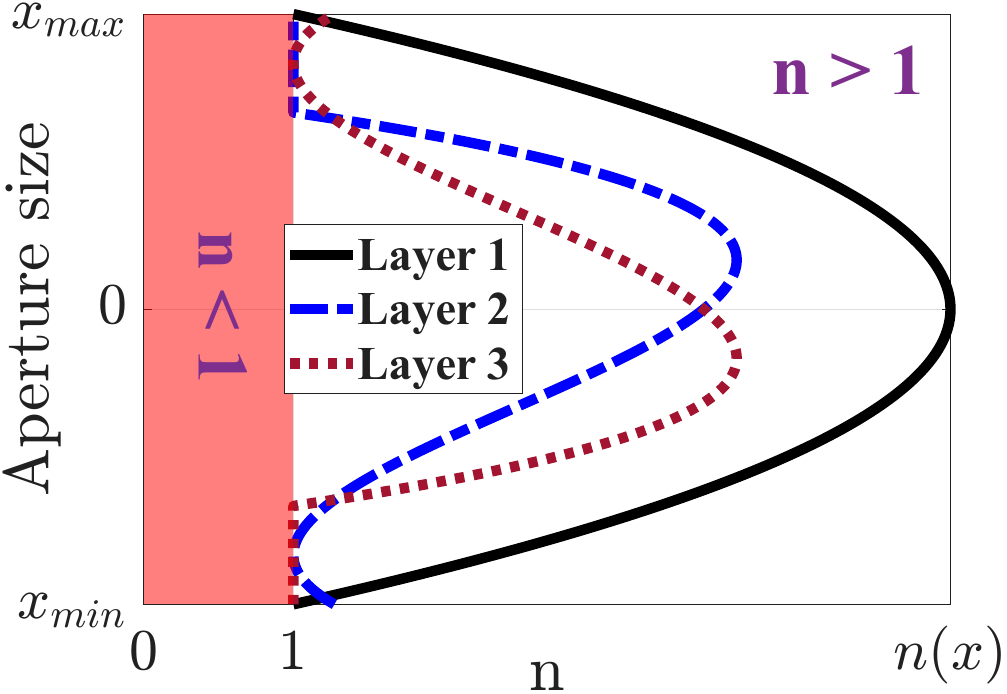}}
                
                \subfigure[]{\label{Fig: Fig03.g}
                \includegraphics[width=0.3\linewidth,trim={0.0cm 0.0cm 0.0cm 0.0cm},clip=true]{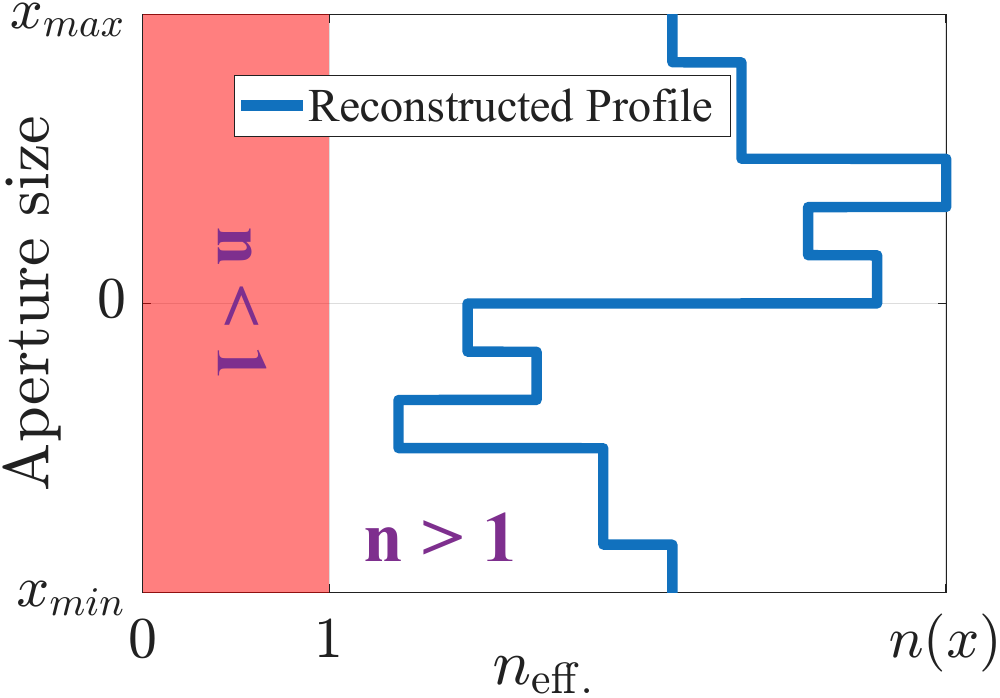}}
                \subfigure[]{\label{Fig: Fig03.h}
                \includegraphics[width=0.3\linewidth,trim={0.0cm 0.0cm 0.0cm 0.0cm},clip=true]{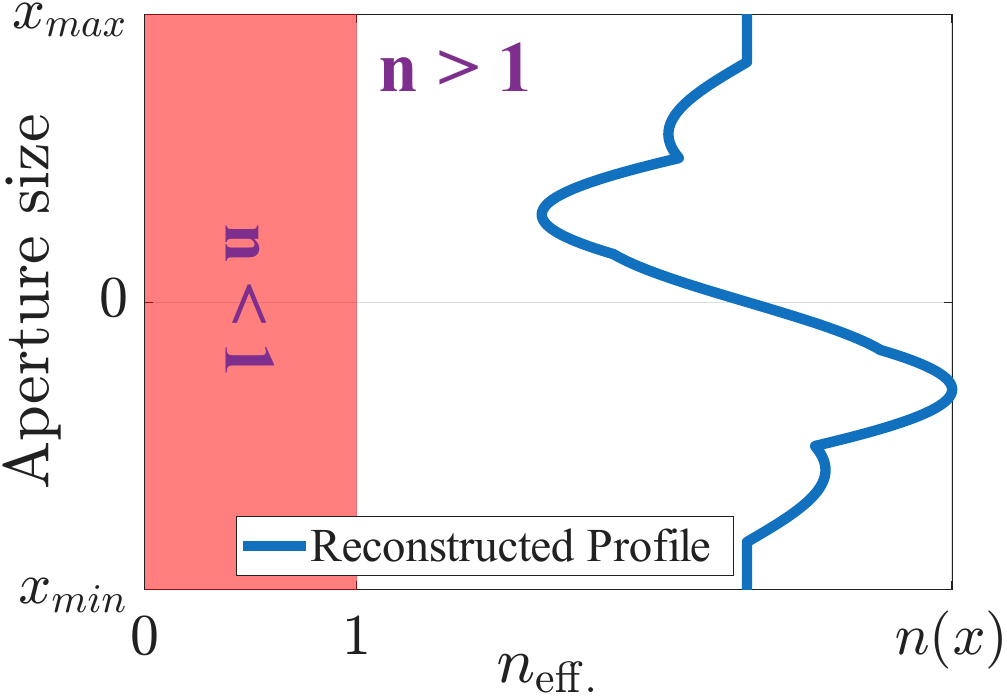}}
                \subfigure[]{\label{Fig: Fig03.i}
                \includegraphics[width=0.3\linewidth,trim={0.0cm 0.0cm 0.0cm 0.0cm},clip=true]{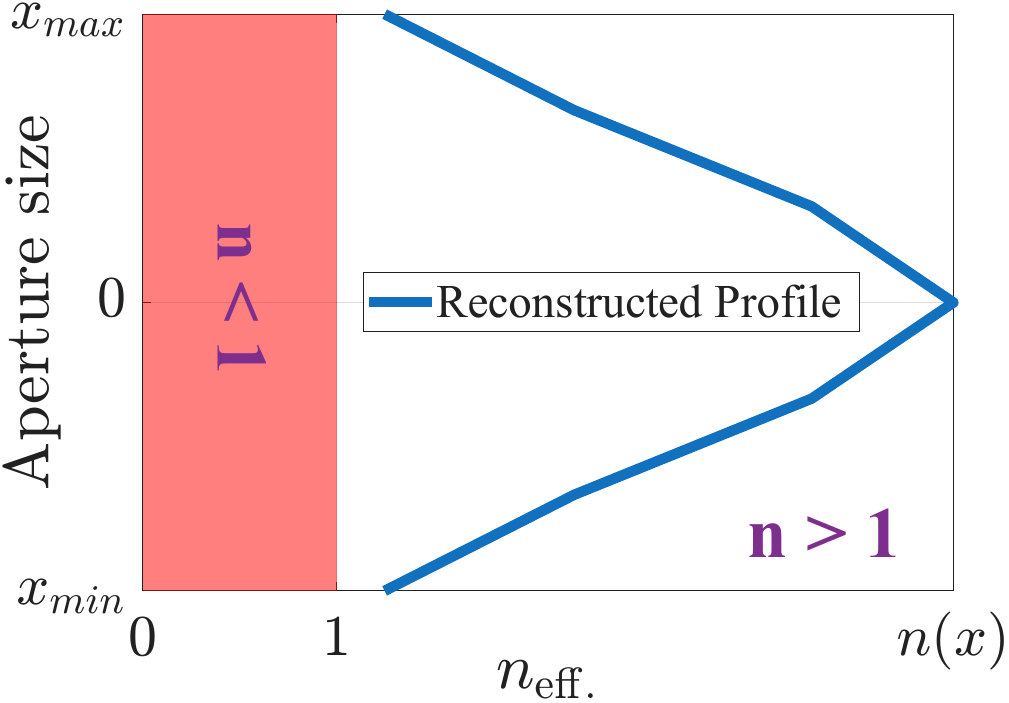}}
                
                \subfigure[]{\label{Fig: Fig03.j}
                \includegraphics[width=0.3\linewidth,trim={0.0cm 0.0cm 0.0cm 0.0cm},clip=true]{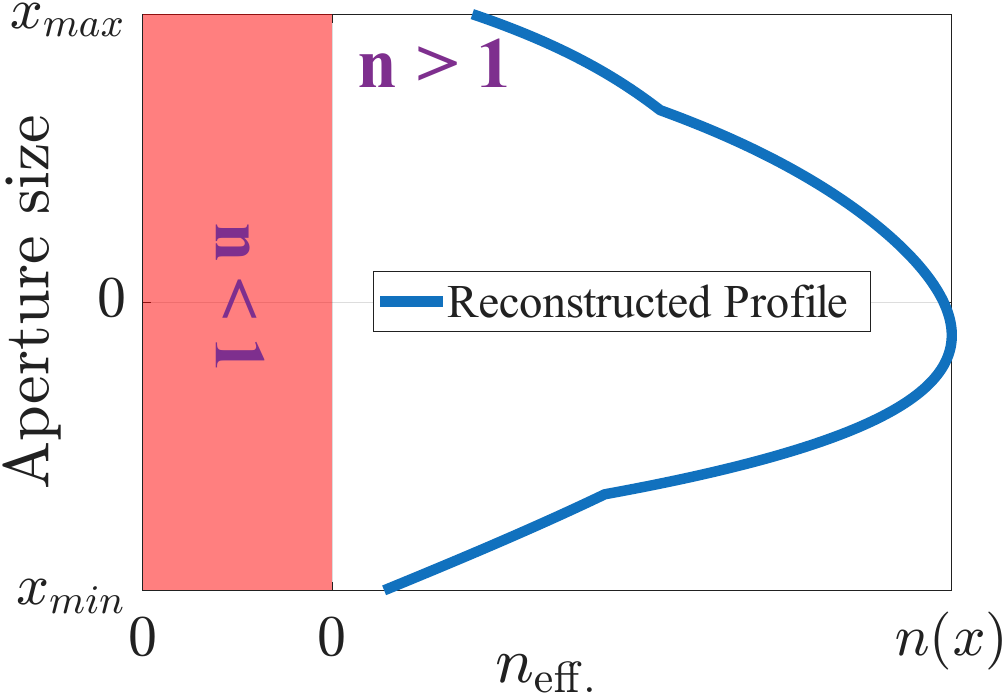}}
                \subfigure[]{\label{Fig: Fig03.k}
                \includegraphics[width=0.3\linewidth,trim={0.0cm 0.0cm 0.0cm 0.0cm},clip=true]{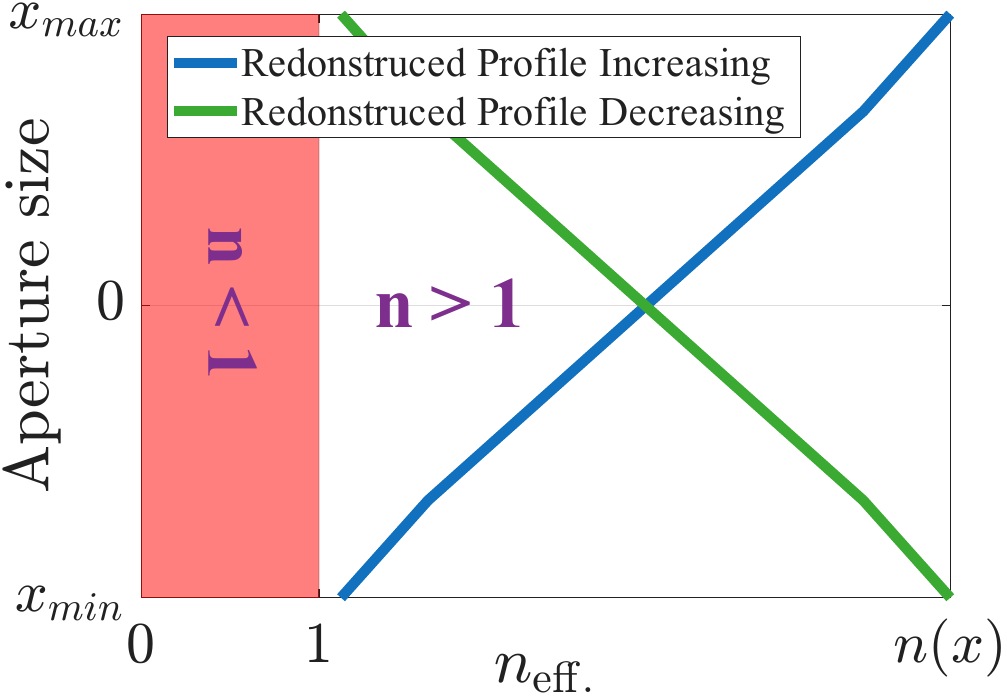}}
                \subfigure[]{\label{Fig: Fig03.l}
                \includegraphics[width=0.3\linewidth,trim={0.0cm 0.0cm 0.0cm 0.0cm},clip=true]{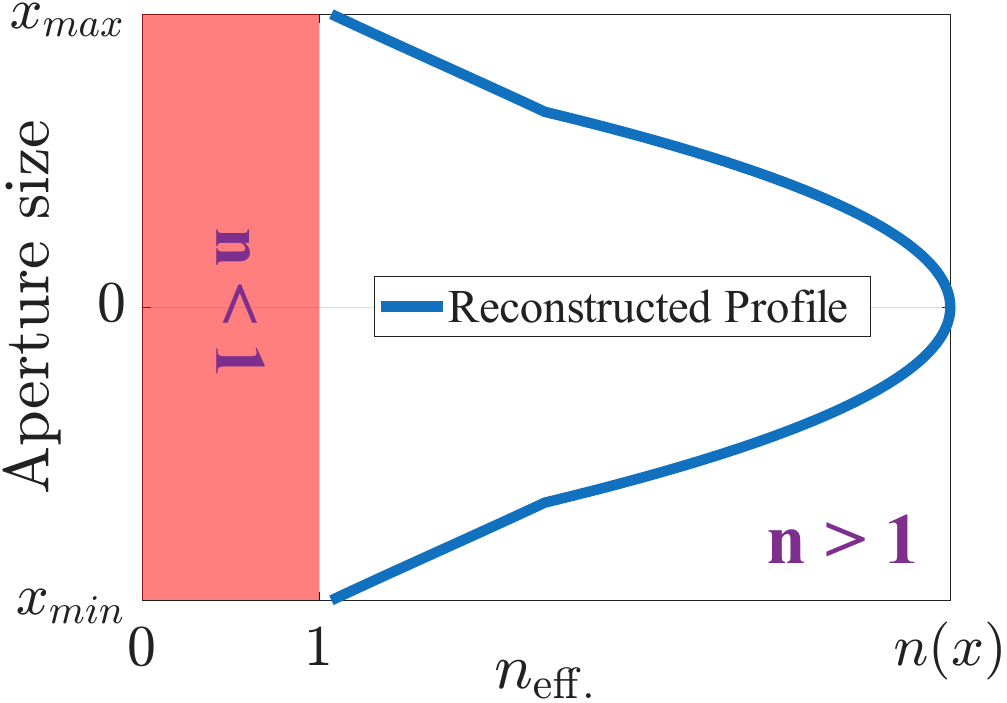}}
            \end{center}
            \vspace{-4.0mm}
            \caption{The refractive index distributions across the GRIN lens layers are presented for: (a) Haar wavelets, (b) sinusoidal, (c) sawtooth, (d) Legendre polynomials, (e) linear, and (f) first-kind Chebyshev polynomial profiles in a three-layer configuration with mutual interlayer shifts. Panels (g)-(l) correspondingly display the effective refractive index profiles reconstructed by implementing each of these basis functions in the proposed GRIN lens design.}
            \label{Fig: Fig03}
        \end{figure}

        The selection of basis functions and layer displacements is crucial and should be optimized to accurately produce the intended refractive index profile. This section examines various basis functions, including Haar wavelets, sinusoidal, sawtooth, Legendre polynomial, linear, and first-kind Chebyshev polynomial, assessing their ability to approximate and induce smooth refractive index variations. Figure~\ref{Fig: Fig03} illustrates the refractive index profiles of these basis functions for a three-layered lens, along with the resulting effective refractive index profile. Figures~\ref{Fig: Fig03.a} to~\ref{Fig: Fig03.f} illustrate the selected basis functions as groove-etching profiles applied to the GRIN lens layers. Correspondingly, Figs.~\ref{Fig: Fig03.g} to~\ref{Fig: Fig03.l} present the resulting effective refractive index profiles achieved through this approach. The findings reveal that, except for the first-kind Chebyshev polynomial, none of the tested basis functions yielded a satisfactory profile at least in a three-layered GRIN lens configuration (see Section~\ref{Lab: labMetric}).
        
        Given the effective refractive index profile achieved with the proposed GRIN lens (Figs.~\ref{Fig: Fig03.f} and~\ref{Fig: Fig03.l}), the first-kind Chebyshev polynomials emerge as a promising functional basis, offering a pathway for practical implementations. In the considered GRIN lens with three layers, the basis functions are implemented separately in each of the layers. The necessary overall profile is then obtained with controlled lateral displacements of $0$, $X$, and $-X$ units along the $x$-axis, corresponding to the successive layers in the multilayer GRIN lens. Specifically, the first layer remains stationary to serve as the primary contributor to the effective refractive index profile, while the second and third layers are symmetrically displaced along the lateral direction, i.e., with equal magnitude but opposite signs, to enable tunability of the effective refractive index.
        
        Such symmetric displacement can be implemented via piezoelectric or Micro-electromechanical (MEMS) based nanopositioning systems for mid-IR spectrum or stepper motor-driven systems for the mm-wave regime, providing submicron- or micron-scale actuation resolutions, respectively. In THz and infrared applications, piezoelectric or MEMS-based nanopositioning systems offer superior displacement control, achieving 1–5~nm lateral shifts via voltage modulation. Precision microstepping in the mm-wave band with digital pulse-controlled angular rotation converted into linear translation via lead screws may produce 1–5~$\mu\text{m}$ incremental steps. 
        
\begin{figure}[ht!]
            \begin{center}
                \subfigure[]{\label{Fig: Fig04.a}
                \includegraphics[height=0.4\linewidth,trim={0.0cm 0.0cm 0.0cm 0.0cm},clip=true]{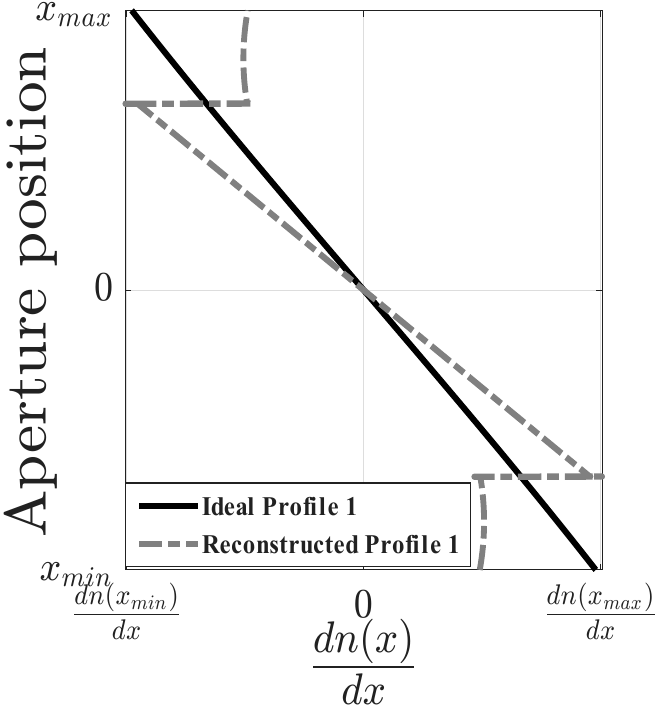}}
                \subfigure[]{\label{Fig: Fig04.b}
                \includegraphics[height=0.4\linewidth,trim={0.0cm 0.0cm 0.0cm 0.0cm},clip=true]{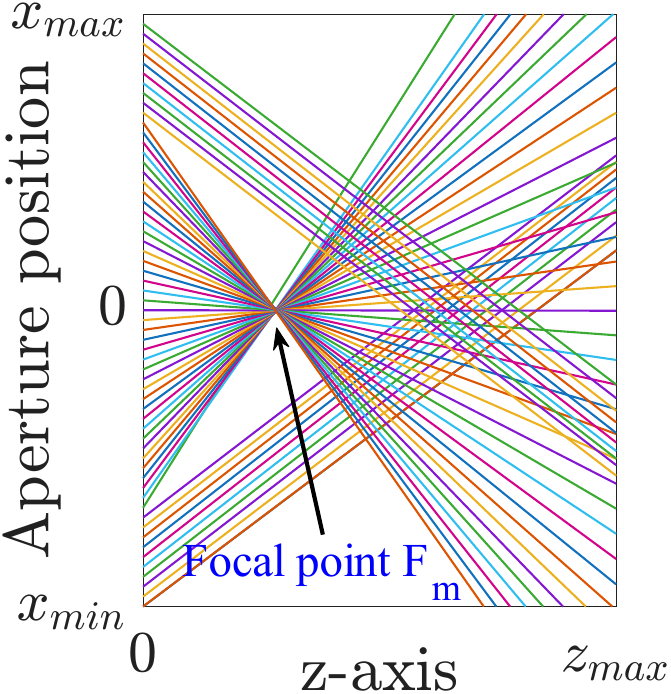}}
            \end{center}
            \vspace{-4.0mm}
            \caption{Refractive index gradient and geometric optics profiles wrt aperture: (a) Refractive index gradient profile across the GRIN lens aperture, and (b) Geometric optics of the proposed GRIN lens for an arbitrary focal point.}
            \label{Fig: Fig04}
        \end{figure}

        Figure~\ref{Fig: Fig04.a} presents the refractive index gradient profile across the GRIN lens aperture for the refractive index distribution illustrated in Fig.~\ref{Fig: Fig03.l}, with the conventional convex lens gradient profiles (ideal cases) provided as a reference. It seems that, despite the presence of irregularities (infinite gradients) at certain points in the lens, their effects on overall optical performance are negligible due to their marginal spatial extent. Analysis reveals that when the reconstructed GRIN lens gradient profile closely matches the ideal convex lens gradient, the resulting focal spot exhibits significantly improved sharpness and focal length accuracy. A discrepancy is observed between the gradient index slope in the ideal and reconstructed case, Fig.~\ref{Fig: Fig04.a}. As a result, due to mismatches in the gradient index slope between the ideal and reconstructed profiles, as well as the mentioned irregularities, the GRIN lens exhibits longitudinal aberration, which may lead to multifocal characteristics (i.e., introducing aberrations compared to the uniform focusing behavior of traditional lenses).
        
        This deviation arises because the reconstructed effective refractive index does not precisely align with the ideal profile, as illustrated in Figs.~\ref{Fig: Fig05.e} and~\ref{Fig: Fig07.e} (Profiles $III-IV$ and $III'-IV'$). The divergence between the ideal and intended case becomes more significant as the deviation increases. Furthermore, certain reconstructed effective refractive index profiles may deviate from the required refractive index limit at their boundaries. As a result, these instances may contribute to discrepancies between the ideal and intended focusing profiles (Profiles $III$ and $III'$ in Figs.~\ref{Fig: Fig05.e} and~\ref{Fig: Fig07.e}). The geometric optics focal point for the given refractive index in Fig.~\ref{Fig: Fig03.l} with specific value of $X$ is depicted in Fig.~\ref{Fig: Fig04.b}. It reveals a discernible discrepancy between the ideal focal point produced by a conventional lens (Fig.~\ref{Fig: Fig02.d}) and the semi-blurred focal point achieved by the proposed GRIN lens.
    \subsection{Effective Medium Theory-Based Etching Depth Determination}
        It is worth reminding that in the considered GRIN lens concept, the grooves within the first mechanically fixed layer must be engineered such that their effective refractive index profile adheres to a second-order Chebyshev polynomial. Conversely, the groove profiles for the second and third layers have to be derived using symmetrically mirrored third-order Chebyshev polynomials. Consequently, the lens's overall effective refractive index is obtained through the superposition (i.e., mixing rule of Eq.~\ref{Eq: Eq03}) of the refractive indices contributed by each individual layer.
        
        On the other hand, EMT evaluation of the effective refractive indices of separate GRIN lens layers employs its own mixing rule:
        \begin{equation}
            n_{\rm eff.}^{k}(x)=n_{\rm g}+(n_{k}-n_{\rm g})f_{\rm p_{\it k}}(x),
            \label{Eq: Eq04}
        \end{equation}
        where $n_{\rm g}$, $n_{k}$, and $f_{\rm p_{\it k}}(x)$ represent the refractive indices of the grooves and layers material, and the unit cell packing factor at the $k^{\rm th}$ layer, respectively. The initial groove height for all grooves is set to $l/2$, i.e., $h_{kj}+g_{kj}=l/2$. 
        
        Note that the groove depth must be determined based on both packing factors and the required refractive index for each unit cell. In certain design configurations, the optimal groove depth approaches the corresponding layer thickness to achieve the target refractive index. However, to preserve the structural integrity of the patterned layer, the groove depth is constrained below this theoretical maximum. To compensate, the unit cell arm thicknesses are progressively reduced until the required refractive index is attained, i.e., decrease in $h_{kj}+g_{kj}$ thickness and increase in the height of the grooves. This thickness reduction approach may exhibit a key limitation when the minimal arm thickness cannot satisfy both the refractive index target and mechanical stability requirements in specific design scenarios. As a result, local discrepancies emerge between the achieved and ideal refractive index profiles along the $x$-axis in these specific design scenarios.
        
        The packing factor, also referred to as the filling ratio, for the constituent layers is established by employing basis functions. Accordingly, the specific definition for the packing factor of the unit cells, as illustrated in Fig.~\ref{Fig: Fig01.b}, is provided as:
        \begin{multline}
            f_{\rm p_{\it kj}}(x)=\frac{lt_{k}-(l-(h_{kj}+g_{kj}))d_{kj}}{lt_{k}}=\\
                         1-\frac{(l-(h_{kj}+g_{kj}))d_{kj}}{lt_{k}}=\\
                         1-\displaystyle\frac{\left(\frac{\lambda_{\rm min}}{m}-(h_{kj}+g_{kj})\right)d_{kj}}{\frac{\lambda_{\rm min}}{m}t_{k}}
            \label{Eq: Eq05}
        \end{multline}
        $t_{k}$, $h_{kj}$, $g_{kj}$, and $d_{kj}$ represent the respective unit cell thicknesses along the $z$-axis, upper and lower arm thicknesses along the $x$-axis, and groove depth for associated unit cells, respectively. The indices $k$ and $j$ denote the layer number and unit cell number within each layer, respectively. Meanwhile, the packing factor exhibits a correlation with the refractive indices of the materials comprising the GRIN lens layers and grooves, Eq.~\ref{Eq: Eq04}. Based on Eqs.~\ref{Eq: Eq04} and~\ref{Eq: Eq05}, the groove depth for unit cells can be determined as follows:
        \begin{equation*}
            d_{kj}=\frac{\frac{\lambda_{\rm min}}{m}t_{k}}{\frac{\lambda_{\rm min}}{m}-(h_{kj}+g_{kj})}\frac{n_{k}-n_{\rm eff.}^{k}(x)}{n_{k}-n_{\rm g}}.
        \end{equation*}
\section{Numerical Validation vs. Theoretical Predictions}
    As mentioned previously, two three-layer GRIN lenses with aperture sizes of $L=32$~$\mu\text{m}$ and $L=64$~mm are to be examined under lateral displacement conditions. This is done to evaluate the adaptive focusing performance of the proposed design within the 149 to 151~THz and 39 to 41~GHz frequency bands, wherein each lens layer was shifted by $[0,X,-X]$ units.
    
    For the GRIN lens design targeting the specified spectra, two distinct dielectric substrates are adopted during the optimization process. In the mid-IR spectrum, silicon ({\tt Si}) serves as the substrate for the first layer, whereas zinc selenide ({\tt ZnSe}) is employed as the substrate material for the second and third layers. The refractive index behavior of {\tt Si} and {\tt ZnSe} materials in relation to the operating frequency band at 150~THz has been adopted from the sources of Refs.~\cite{RefractiveIndexInfo,FRANTA2017405, Querry_1987}. The refractive indices of the {\tt Si} and {\tt ZnSe} materials at 150~THz are 3.4545 (with an extinction coefficient of $k\approx 6.8797\times10^{-11}$) and 2.4310 ($k\approx 0$), respectively. However, for the mm-wave range, the first layer utilizes the Rogers RO3010 substrate, which has a refractive index of $\sqrt{10.2}$. In contrast, the second and third layers are constructed using the Rogers RO3006 substrate, characterized by a refractive index of $\sqrt{6.15}$.
    
    The respective layer thicknesses for the proposed mid-IR and mm-wave GRIN lenses are 40, 20, and 20 micro-inches (mid-IR), and 40, 20, and 20 milli-inches (mm-wave), respectively. In the full-structure simulations, the boundary conditions are set to open, with additional space added to minimize deceptive reflections along the $x$- and $z$-directions. However, the boundary conditions along the $y$-axis are set to the electric field to constrain the field components parallel to the domain interfaces properly. This combination ensures accurate modeling of wave propagation and field confinement in the simulated structure.
    \subsection{Basis Function Selection}
        \label{Lab: labMetric}
        Candidate basis functions were evaluated by comparing the reconstructed refractive-index profile with the target lens profile using the Correlation Coefficient, Cosine Similarity, RMSE, MAE, Euclidean Distance, PRDN, and Maximum Comparison (maximum absolute error), and the Closeness $\Psi$, defined in Eq.~\ref{Eq: Eq06}:
        \begin{equation}
            \Psi=\int_{-L/2}^{L/2}|n^{\rm eff.}(x)-n_{\rm Ideal}(x)|dx.
            \label{Eq: Eq06}
        \end{equation}
        These metrics quantify global reconstruction fidelity and local deviations that affect optical performance. Their results were correlated with the focal-length error ($\Delta F$) obtained from full-wave/GO simulations. Based on these mathematical and optical metrics, first-kind Chebyshev polynomials were selected as the basis functions for the proposed tunable GRIN lens (Table~\ref{Tab: Table01}).
        \begin{table*}[ht!]
            \tiny
            \centering
            \caption{Quantitative comparison of metrics for various basis functions.}
            \vspace*{2mm}
            \setlength{\tabcolsep}{3pt}
            \begin{tabular}{c|c|c|c|c|c|c|c}
                \toprule
                \toprule
                 \rowcolor{Gray}
                 \multicolumn{8}{c}{Profiles $III$ and $III'$} \\
                \toprule
                 Similarity &  & Haar &  &  & Legendre &  & Chebyshev \\
                 metric & \multirow{-2}{*}{Spectrum} & wavelets & \multirow{-2}{*}{Sinusoidal} & \multirow{-2}{*}{Sawtooth} & polynomial & \multirow{-2}{*}{Linear} & polynomial \\
                \hline
                \rowcolor{Gray}
                 Euclidean & mid-IR & 57.87 & 43.97 & \textcolor{purple}{18.34} & 28.38 & 39.06 & \textbf{\textcolor{teal}{23.66}} \\
                 \rowcolor{Gray}
                 distance & mm-wave & 64.71 & 54.05 & 16.66 & 17.28 & 58.79 & \textbf{\textcolor{teal}{12.66}} \\
                \hline
                  & mid-IR & 1.02 & 0.78 & \textcolor{purple}{0.32} & 0.5 & 0.69 & \textbf{\textcolor{teal}{0.42}} \\
                 \multirow{-2}{*}{RMSE} & mm-wave & 0.81 & 0.68 & 0.21 & 0.22 & 0.73 & \textbf{\textcolor{teal}{0.16}} \\
                \hline
                 \rowcolor{Gray}
                 Correlation & mid-IR & 5.57$\times$10$^{\rm -15}$\% & 4.23$\times$10$^{\rm -15}$\% & 96.94\% & 95.83\% & 5.81$\times$10$^{\rm -15}$\% & \textbf{\textcolor{teal}{99.99\%}} \\
                 \rowcolor{Gray}
                 coefficient & mm-wave & 1.75$\times$10$^{\rm -14}$\% & 1.24$\times$10$^{\rm -14}$\% & 96.88\% & 94.62\% & 4.33$\times$10$^{\rm -15}$\% & \textbf{\textcolor{teal}{100\%}} \\
                \hline
                 Mean & mid-IR & 0.97 & 0.71 & \textcolor{purple}{0.22} & 0.49 & 0.50 & \textbf{\textcolor{teal}{0.42}} \\
                 absolute error & mm-wave & 0.61 & 0.51 & 0.19 & 0.9 & 0.63 & \textbf{\textcolor{teal}{0.13}} \\
                \hline
                 \rowcolor{Gray}
                 Cosine & mid-IR & 88.60\% & 93.01\% & 99.48\% & 99.84\% & 94.42\% & \textbf{\textcolor{teal}{99.97\%}} \\
                 \rowcolor{Gray}
                 similarity & mm-wave & 92.56\% & 94.88\% & \textcolor{purple}{99.73\%} & 99.56\% & 93.88\% & \textbf{\textcolor{teal}{99.72\%}} \\
                 \hline
                  & mid-IR & 299.40\% & 227.50\% & \textcolor{purple}{94.90\%} & 146.80\% & 202.10\% & \textbf{\textcolor{teal}{122.42\%}} \\
                 \multirow{-2}{*}{PRDN} & mm-wave & 167.28\% & 139.74\% & 43.06\% & 44.66\% & 151.99\% & \textbf{\textcolor{teal}{32.74\%}} \\
                \hline
                 \rowcolor{Gray}
                  & mid-IR & 0.80 & 0.80 & 0.80 & 0.46 & 0.80 & \textbf{\textcolor{teal}{0.44}} \\
                \rowcolor{Gray}
                 \multirow{-2}{*}{|Max. Comparison|} & mm-wave & 0.22 & 0.22 & 0.22 & \textcolor{purple}{0.13} & 0.22 & \textbf{\textcolor{teal}{0.15}} \\
                \hline
                 Focal-length & mid-IR & - & - & \textcolor{teal}{31.45\%} & \textcolor{teal}{30.81\%} & - & \textbf{\textcolor{teal}{4.37\%}} \\
                 error & mm-wave & - & - & \textcolor{teal}{32.76\%} & \textcolor{teal}{39.95\%} & - & \textbf{\textcolor{teal}{2.27\%}} \\
                 \hline
                 \rowcolor{Gray}
                  & mid-IR & 30.94~$\mu\text{m}$ & 22.66~$\mu\text{m}$ & \textcolor{purple}{7.07~$\mu\text{m}$} & 15.65~$\mu\text{m}$ & 15.97~$\mu\text{m}$ & \textbf{\textcolor{teal}{13.36~$\mu\text{m}$}} \\
                 \rowcolor{Gray}
                 \multirow{-2}{*}{Closeness} & mm-wave & 39.14~mm & 32.58~mm & 11.95~mm & 11.97~mm & 40.46~mm & \textbf{\textcolor{teal}{8.46~mm}} \\
                \toprule
                \toprule
                 \rowcolor{Gray}
                 \multicolumn{8}{c}{Profiles $IV$ and $IV'$} \\
                \toprule
                 Similarity &  & Haar &  &  & Legendre &  & Chebyshev \\
                 metric & \multirow{-2}{*}{Spectrum} & wavelets & \multirow{-2}{*}{Sinusoidal} & \multirow{-2}{*}{Sawtooth} & polynomial & \multirow{-2}{*}{Linear} & polynomial \\
                \hline
                 \rowcolor{Gray}
                 Euclidean & mid-IR & 70.22 & 64.24 & 47.22 & 34.95 & 66.15 & \textbf{\textcolor{teal}{33.07}} \\
                 \rowcolor{Gray}
                 distance & mm-wave & 67.60 & 63.66 & 41.95 & 28.03 & 73.59 & \textbf{\textcolor{teal}{27.78}} \\
                \hline
                  & mid-IR & 1.24 & 1.14 & 0.83 & 0.62 & 1.17 & \textbf{\textcolor{teal}{0.58}} \\
                 \multirow{-2}{*}{RMSE} & mm-wave & 0.85 & 0.80 & 0.52 & \textcolor{purple}{0.35} & 0.92 & \textbf{\textcolor{teal}{0.35}} \\
                \hline
                 \rowcolor{Gray}
                 Correlation & mid-IR & 2.82$\times$10$^{\rm -15}$\% & 5.92$\times$10$^{\rm -15}$\% & 98.59\% & 97.58\% & 2.91$\times$10$^{\rm -15}$\% & \textbf{\textcolor{teal}{98.63\%}} \\
                 \rowcolor{Gray}
                 coefficient & mm-wave & 8.02$\times$10$^{\rm -15}$\% & 2.96$\times$10$^{\rm -14}$\% & 97.95\% & 97.29\% & 5.64$\times$10$^{\rm -15}$\% & \textbf{\textcolor{teal}{98.22\%}} \\
                \hline
                 Mean & mid-IR & 1.08 & 0.98 & 0.75 & 0.56 & 1.04 & \textbf{\textcolor{teal}{0.55}} \\
                 absolute error & mm-wave & 0.69 & 0.68 & 0.49 & \textcolor{purple}{0.32} & 0.82 & \textbf{\textcolor{teal}{0.33}} \\
                \hline
                 \rowcolor{Gray}
                 Cosine & mid-IR & 90.89\% & 93.64\% & 99.64\% & 99.73\% & 92.76\% & \textbf{\textcolor{teal}{99.89\%}} \\
                 \rowcolor{Gray}
                 similarity & mm-wave & 95.13\% & 96.16\% & \textcolor{purple}{99.84\%} & 99.78\% & 93.52\% & \textbf{\textcolor{teal}{99.84\%}} \\
                 \hline
                  & mid-IR & 160.30\% & 146.65\% & 107.79\% & 79.77\% & 151.00\% & \textbf{\textcolor{teal}{75.48\%}} \\
                 \multirow{-2}{*}{PRDN} & mm-wave & 136.96\% & 128.98\% & 84.99\% & 56.80\% & 149.11\% & \textbf{\textcolor{teal}{56.28\%}} \\
                \hline
                 \rowcolor{Gray}
                  & mid-IR & 0.80 & 0.97 & 0.98 & 0.73 & 0.71 & \textbf{\textcolor{teal}{0.62}} \\
                \rowcolor{Gray}
                 \multirow{-2}{*}{|Max. Comparison|} & mm-wave & 0.41 & 0.73 & 0.51 & 0.40 & 0.30 & \textbf{\textcolor{teal}{0.26}} \\
                \hline
                 Focal-length & mid-IR & - & - & \textcolor{teal}{38.77\%} & \textcolor{teal}{41.40\%} & - & \textbf{\textcolor{teal}{9.95\%}} \\
                 error & mm-wave & - & - & \textcolor{teal}{45.94\%} & \textcolor{teal}{31.61\%} & - & \textbf{\textcolor{teal}{6.02\%}} \\
                 \hline
                 \rowcolor{Gray}
                  & mid-IR & 30.50~$\mu\text{m}$ & 31.21~$\mu\text{m}$ & 24.14~$\mu\text{m}$ & 17.99~$\mu\text{m}$ & 33.39~$\mu\text{m}$ & \textbf{\textcolor{teal}{17.61~$\mu\text{m}$}} \\
                 \rowcolor{Gray}
                 \multirow{-2}{*}{Closeness} & mm-wave & 44.18~mm & 43.73~mm & 31.13~mm & \textcolor{purple}{20.78~mm} & 52.60~mm & \textbf{\textcolor{teal}{20.94~mm}} \\
                \toprule
                \toprule
            \end{tabular}
            \begin{minipage}{15cm}
            \vspace{0.05cm}
            \end{minipage}
            \label{Tab: Table01}
        \end{table*}
    \subsection{Achieving Targeted Wavefront Convergence: Effective Refractive Index Profile and Focal Point Analysis}
\begin{figure}[ht!]
            \begin{center}
                \subfigure[]{\label{Fig: Fig05.a}
                \includegraphics[width=0.24\linewidth,trim={0.0cm 0.0cm 0.0cm 0.0cm},clip=true]{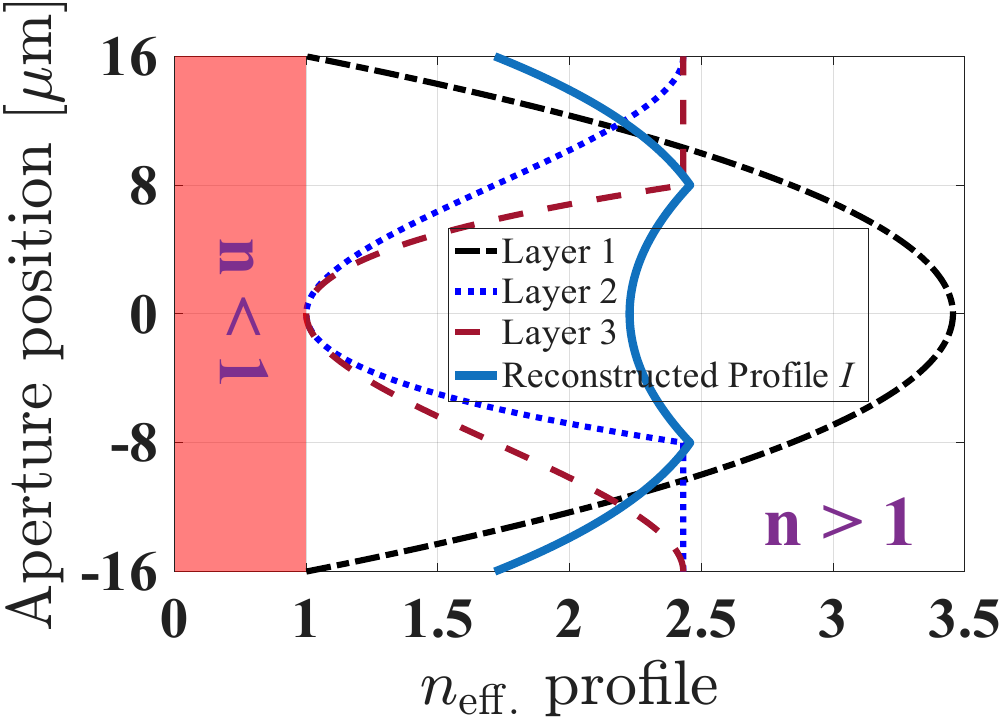}}
                \subfigure[]{\label{Fig: Fig05.b}
                \includegraphics[width=0.24\linewidth,trim={0.0cm 0.0cm 0.0cm 0.0cm},clip=true]{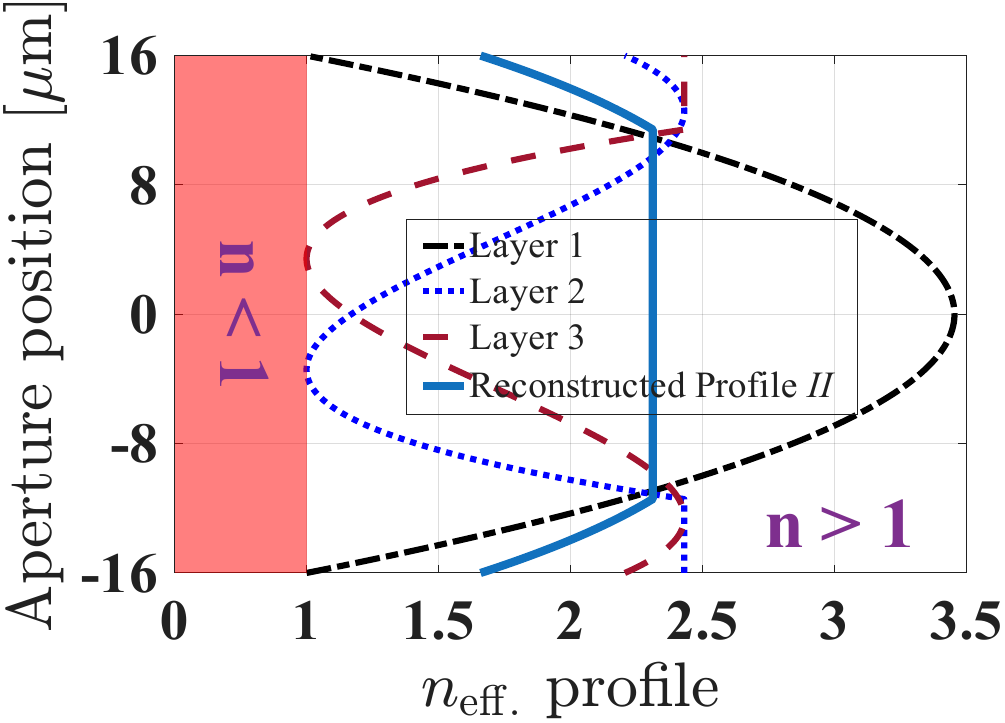}}
                \subfigure[]{\label{Fig: Fig05.c}
                \includegraphics[width=0.24\linewidth,trim={0.0cm 0.0cm 0.0cm 0.0cm},clip=true]{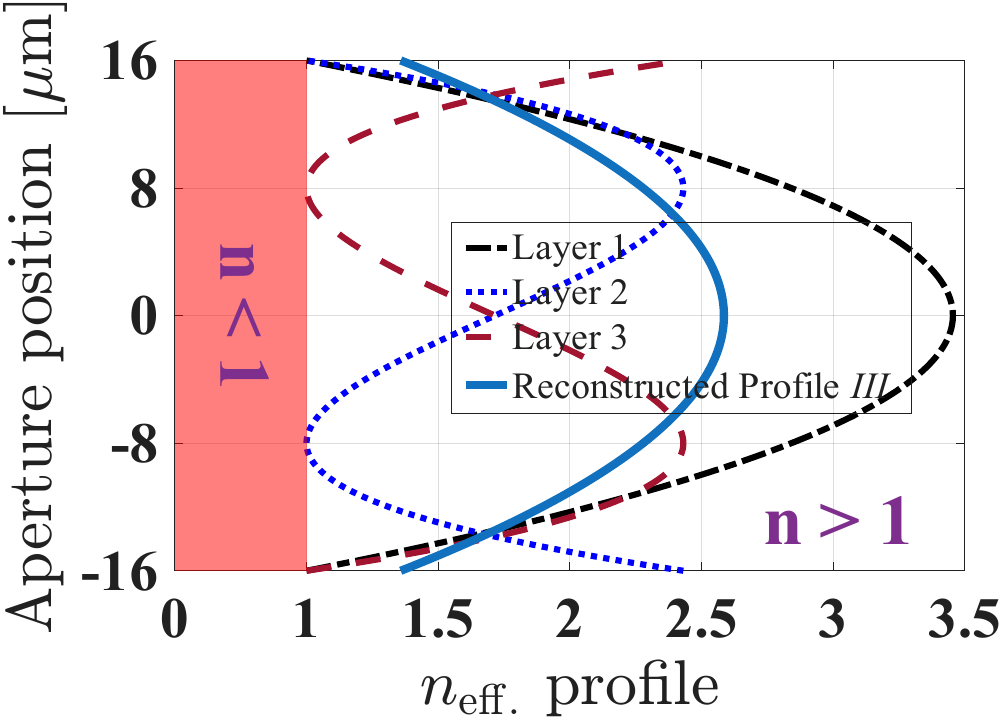}}
                \subfigure[]{\label{Fig: Fig05.d}
                \includegraphics[width=0.24\linewidth,trim={0.0cm 0.0cm 0.0cm 0.0cm},clip=true]{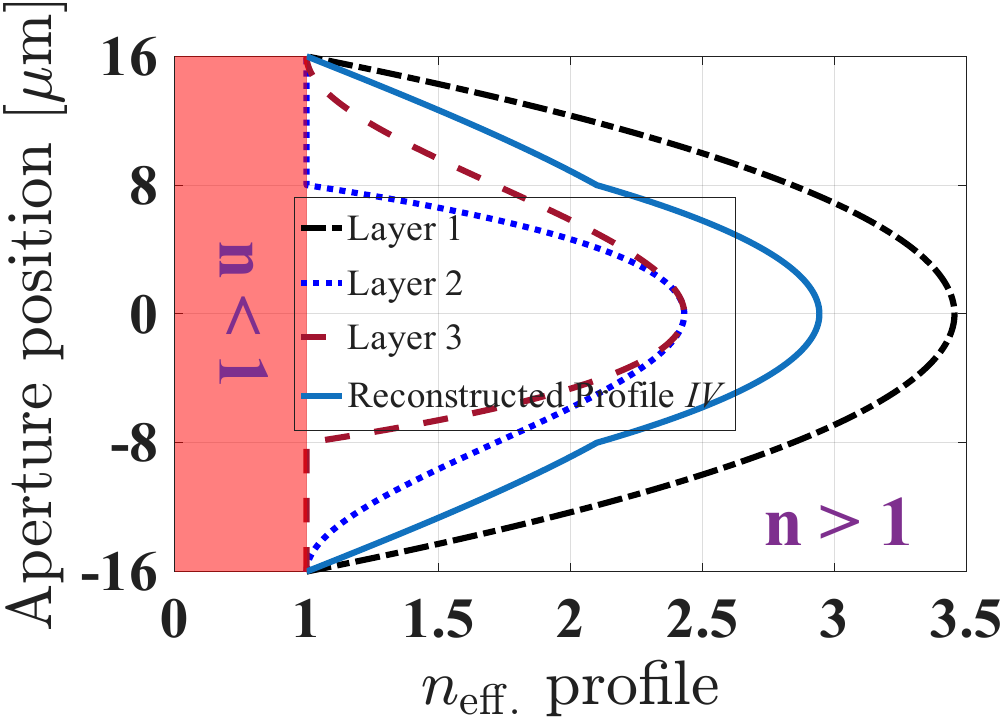}}
                
                \subfigure[]{\label{Fig: Fig05.e}
                \includegraphics[width=0.48\linewidth,trim={0.0cm 0.0cm 0.0cm 0.0cm},clip=true]{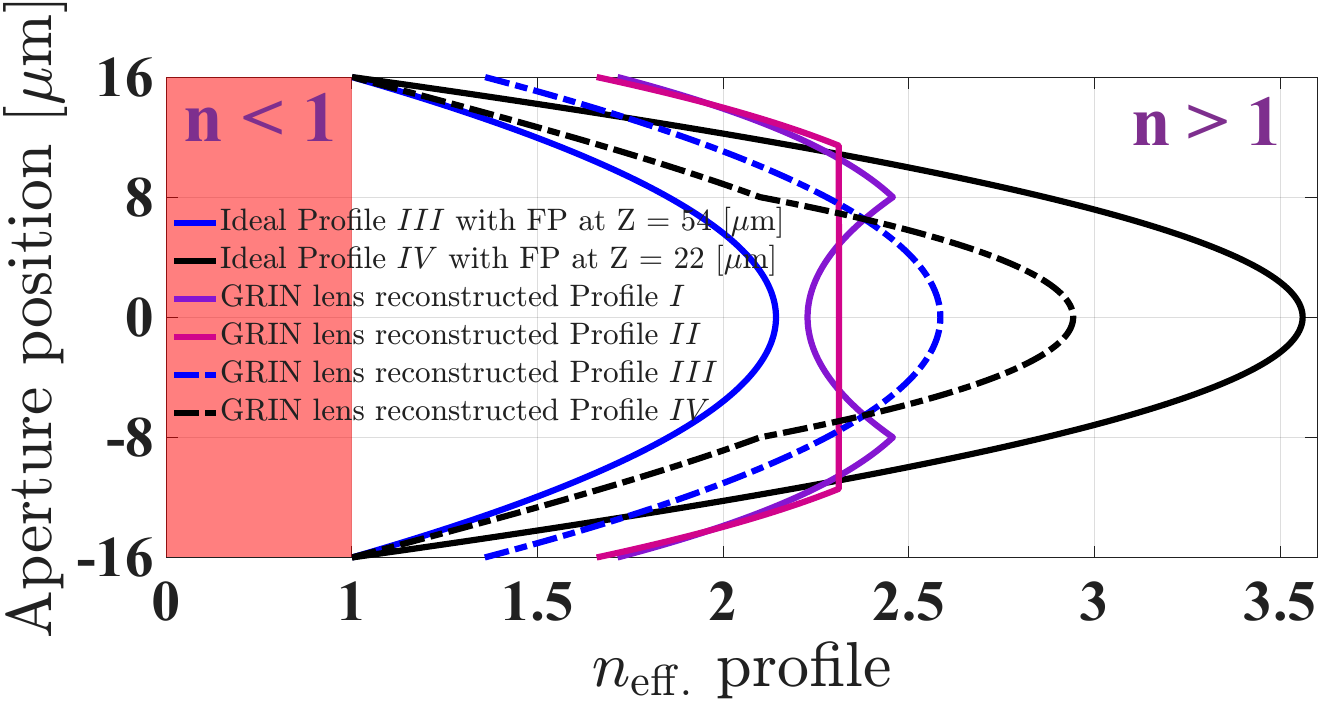}}
                \subfigure[]{\label{Fig: Fig05.f}
                \includegraphics[width=0.48\linewidth,trim={0.0cm 0.0cm 0.0cm 0.0cm},clip=true]{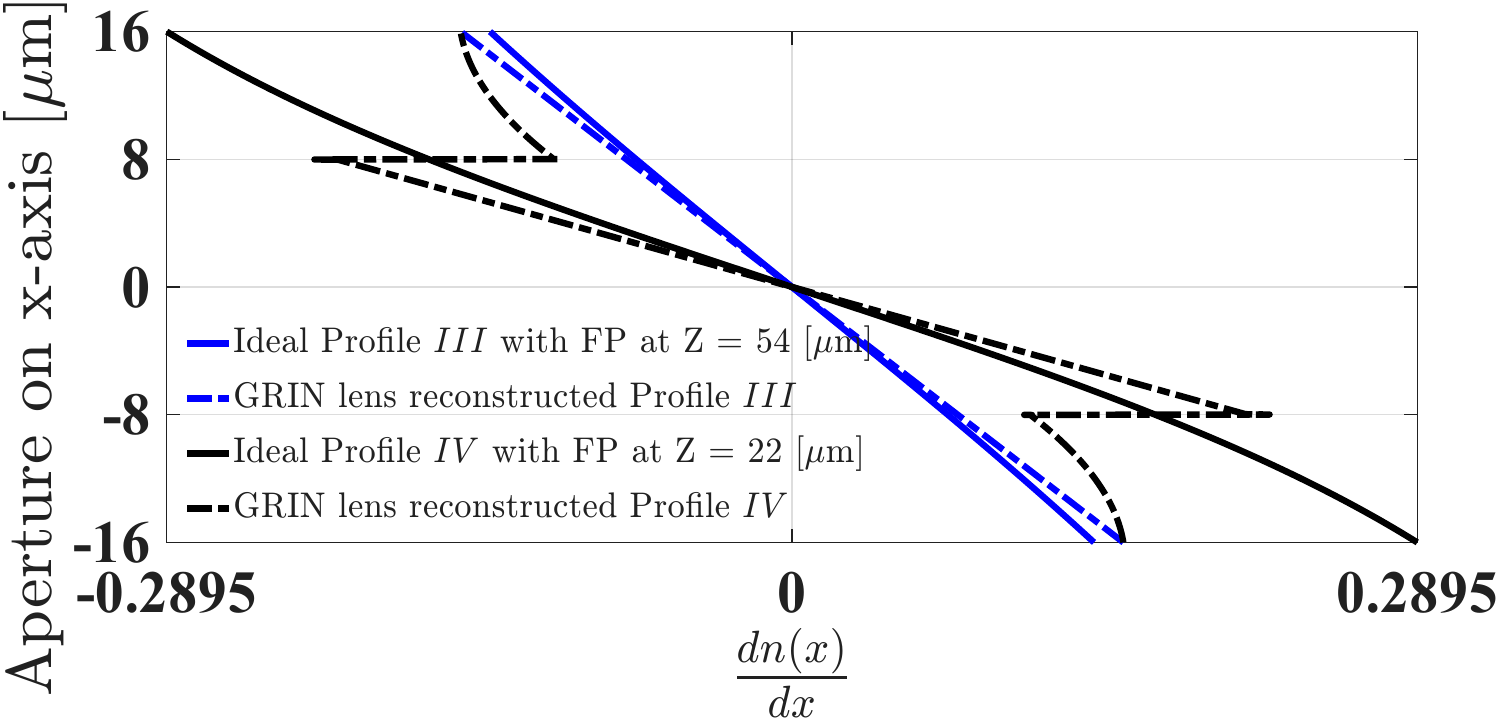}}
                
                \subfigure[]{\label{Fig: Fig05.g}
                \includegraphics[width=0.48\linewidth,trim={0.0cm 0.0cm 0.0cm 0.0cm},clip=true]{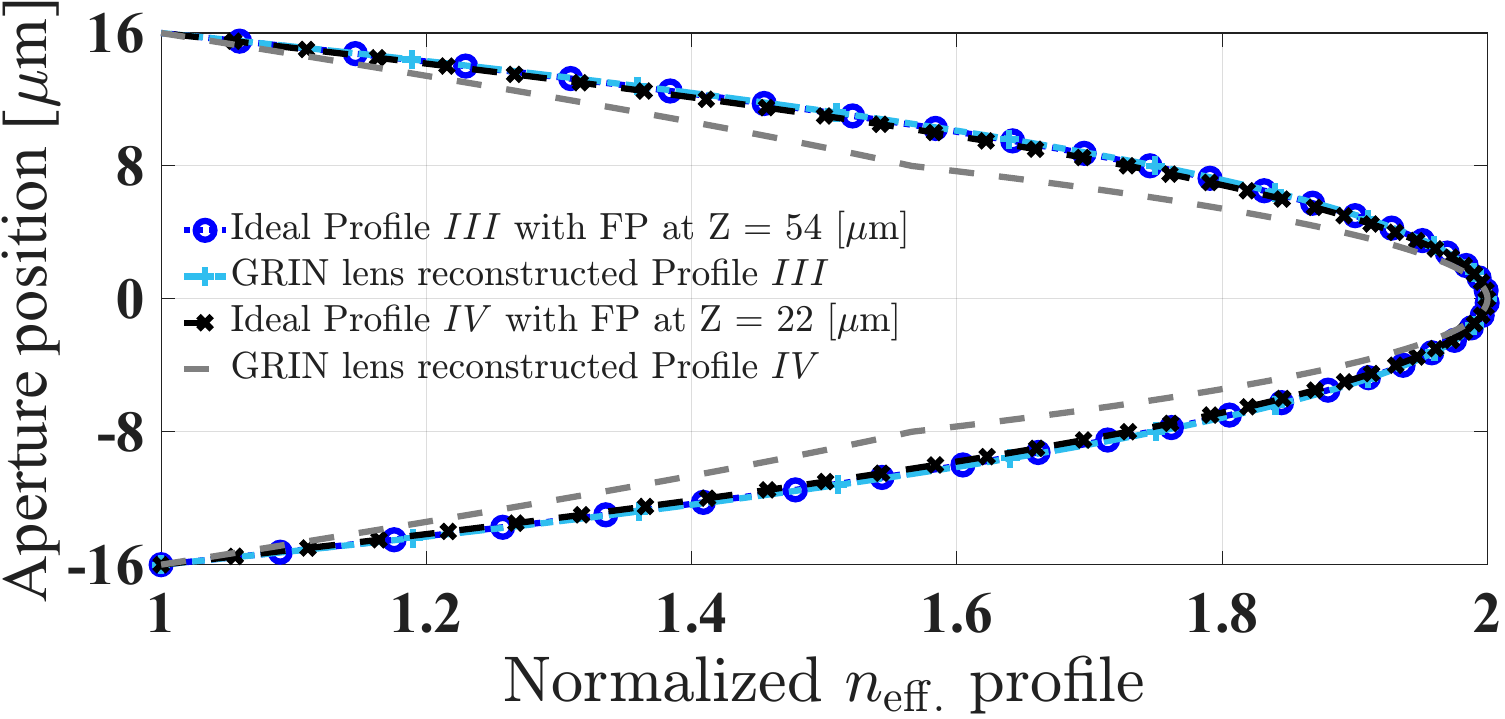}}
                \subfigure[]{\label{Fig: Fig05.h}
                \includegraphics[width=0.48\linewidth,trim={0.0cm 0.0cm 0.0cm 0.0cm},clip=true]{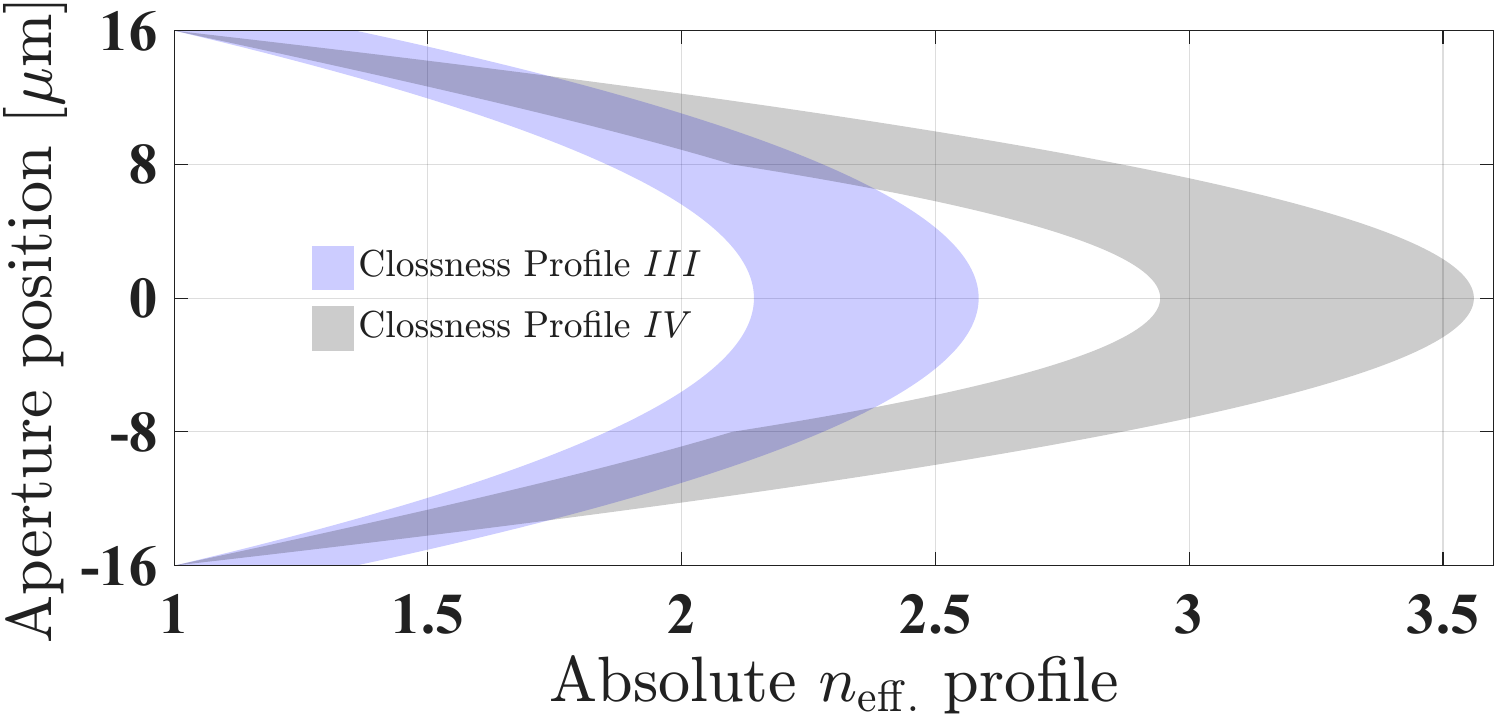}}
            \end{center}
            \vspace{-4.0mm}
            \caption{Comparison of the focusing performance between the ideal lens and the proposed GRIN lens in the mid-IR spectrum. (a)–(d) Illustrate the GRIN lens layers and their corresponding effective refractive index profiles for layer shifts of $[0,8,-8]$~$\mu\text{m}$, $[0,4.57,-4.57]$~$\mu\text{m}$, $[0,0,0]$~$\mu\text{m}$, and $[0,-8,8]$~$\mu\text{m}$ (Profiles $I-IV$), respectively. (e) Presents a comparative analysis of the ideal and attained refractive index profiles, while (f) Depicts the aperture-dependent gradients of the effective refractive indices for both ideal and reconstructed cases. Panels (g) and (h) present the normalized reconstructed effective refractive indices and their closeness, defined as the total spatial difference between them.}
            \label{Fig: Fig05}
        \end{figure}

        A supercell representing the proposed configurations is illustrated in Fig.~\ref{Fig: Fig01.c}. In this design, four distinct configurations (Profiles $I$–$IV$ and $I'$–$IV'$) employing these substrates are realized by applying lateral shifts to the second and third layers. The etching profiles for the GRIN lens layers are generated using first-kind Chebyshev polynomials, Figs.~\ref{Fig: Fig05.a} to~\ref{Fig: Fig05.d} and~\ref{Fig: Fig07.a} to~\ref{Fig: Fig07.d}. Specifically, the grooves in the first (stationary) layer are designed to produce an effective refractive index profile following a second-order Chebyshev polynomial. In contrast, the groove profiles for the second and third layers are determined using third-order Chebyshev polynomials, which are symmetrically mirrored about $x=0$ to achieve alternating maxima and minima in the refractive index distribution. In the unit-cell simulations, the boundary condition at the cell limits in the $x$-axis direction is set to ideal magnetic wall condition. Such a condition enforces zero tangential magnetic field, or equivalently, zero normal electric field derivative at the unit-cell boundary. Open boundaries with additional space are applied at the cell limits along the $z$-axis direction to eliminate reflections.  At the cell limits in the $y$-axis direction, an electric wall condition is used.
        
        To evaluate the adaptive focusing performance of the proposed configuration in the mid-IR spectrum, a three-layer GRIN lens with an aperture size $L=32$~$\mu\text{m}$ was analyzed under lateral displacement conditions of $[0,X,-X]$~$\mu\text{m}$ for each respective layer. Four distinct shift magnitudes were examined for the second and third layers in different scenarios, Profiles $I$-$IV$, spanning $X=[8,4.57,0,-8]$~$\mu\text{m}$. These distinct shift magnitudes enable a systematic assessment of focal tuning capabilities in the mid-IR range. Additionally, to assess the focal-length tuning capabilities within the microwave range, a three-layer GRIN lens featuring an aperture width of $L=64$~mm was investigated under the same lateral misalignments of $[0,X,-X]$~mm. Four different shift values were considered for the second and third layers, Profiles $I'$-$IV'$, covering $X=[12,7.9,0,-12]$~mm, thereby allowing a systematic exploration of focal adjustability in the microwave spectrum. Additionally, an ultrathin metallic collimator with an adjustable aperture was integrated on the illumination side of the GRIN lens to control the incident beam, passing only through the collimator.
        
        Figures~\ref{Fig: Fig05.a}-\ref{Fig: Fig05.d} display the refractive indices of each individual layer, along with their corresponding effective refractive indices for the mid-IR systems under Profiles $I$ through $IV$. Profiles $I$-$IV$ correspond to shift values of $X=[8,4.57,0,-8]$~$\mu\text{m}$ for individual layers. With respect to the reconstructed effective refractive indices of Profiles $I$ and $II$, the transmitted wavefronts are expected to exhibit divergent and afocal behavior, a lens with no net focusing effect (i.e., the transmitted waves should remain parallel). Figure~\ref{Fig: Fig05.e} illustrates the reconstructed effective refractive indices for Profiles $I$-$IV$, along with the ideal refractive indices for conventional convex lenses having focal lengths of 54~$\mu\text{m}$ and 22~$\mu\text{m}$, serving as benchmarks. Figure~\ref{Fig: Fig05.f} presents the aperture-dependent gradients of the effective refractive indices for both the ideal and reconstructed cases. The deviation in the effective refractive index gradient for Profile $III$ is minimal. In contrast, for Profile $IV$, the gradient behavior indicates that the reconstructed effective refractive index will be unable to produce a sharp focal spot. Using cosine similarity as a criterion to compare the shapes of the reconstructed effective refractive indices with ideal profiles for focal lengths of 54~$\mu\text{m}$ and 22~$\mu\text{m}$ yielded similarity values of 99.97\% and 99.89\% for Profiles $III$ and $IV$, respectively. Additionally, when comparing the reconstructed profiles with the proposed ideal cases using normalized effective refractive indices, correlations of 99.99\% and 98.63\% were observed, Fig.~\ref{Fig: Fig05.g}. As a final comparison, the closeness of the profiles was quantified as 13.36~$\mu\text{m}$ and 17.61~$\mu\text{m}$, which represent approximately suitable results for confirming profile similarity, Fig.~\ref{Fig: Fig05.h}.
        
\begin{figure}[ht!]
            \begin{center}
                \subfigure[]{\label{Fig: Fig06.a}
                \includegraphics[width=0.36\linewidth,trim={0.0cm 0.0cm 0.0cm 0.0cm},clip=true]{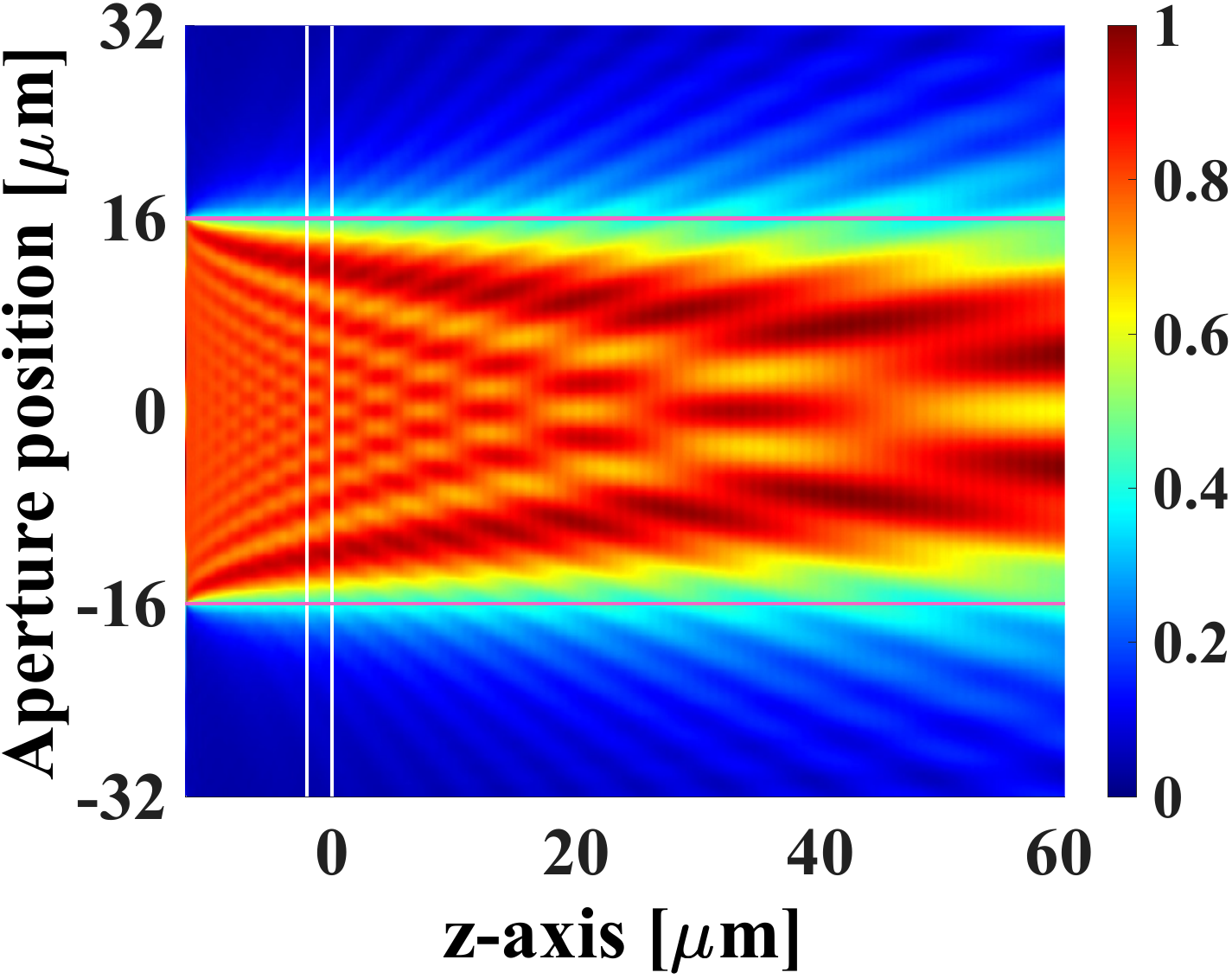}}
                \subfigure[]{\label{Fig: Fig06.b}
                \includegraphics[width=0.36\linewidth,trim={0.0cm 0.0cm 0.0cm 0.0cm},clip=true]{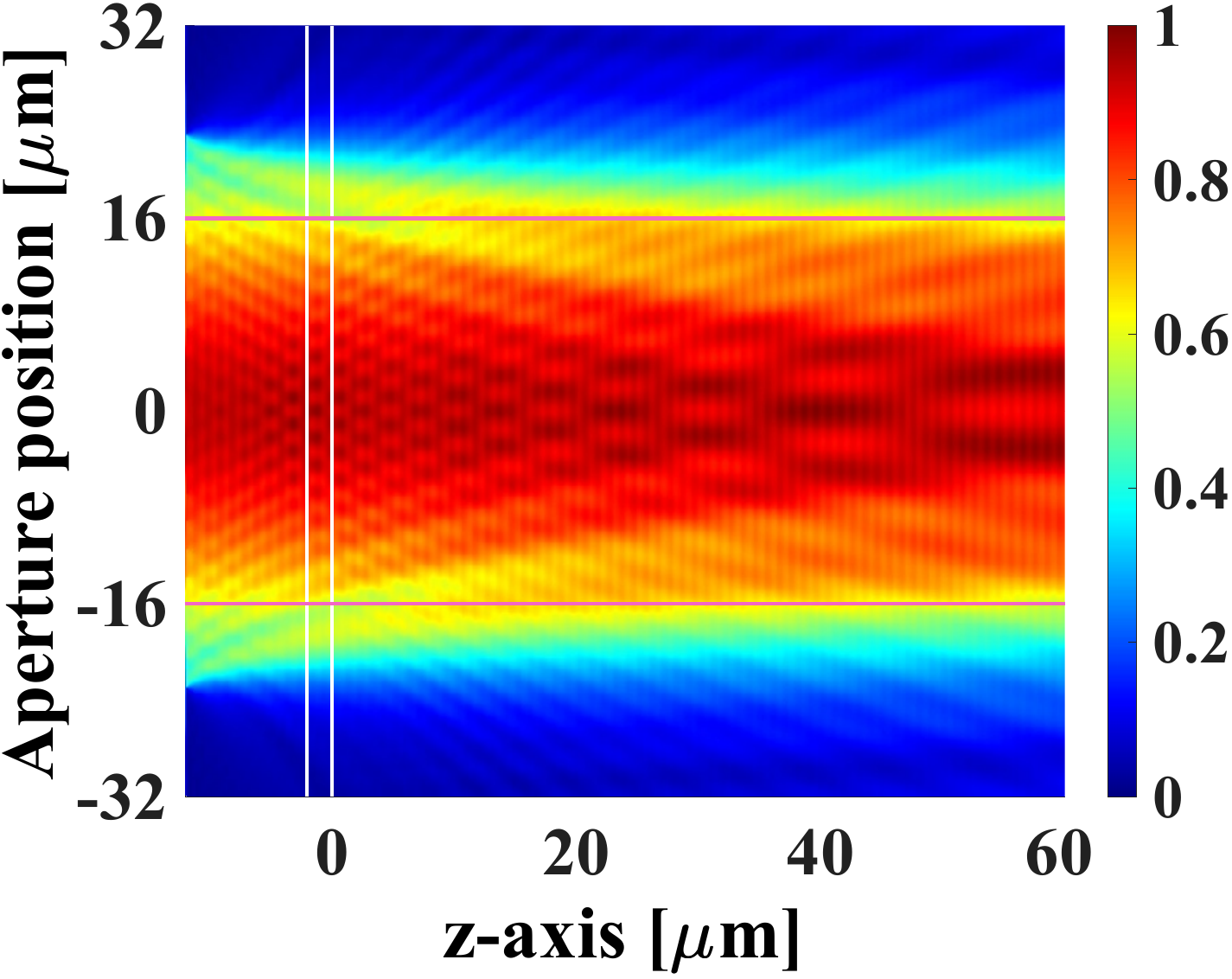}}
                
                \subfigure[]{\label{Fig: Fig06.c}
                \includegraphics[width=0.36\linewidth,trim={1.0cm 0.3cm 1.85cm 1.35cm},clip=true]{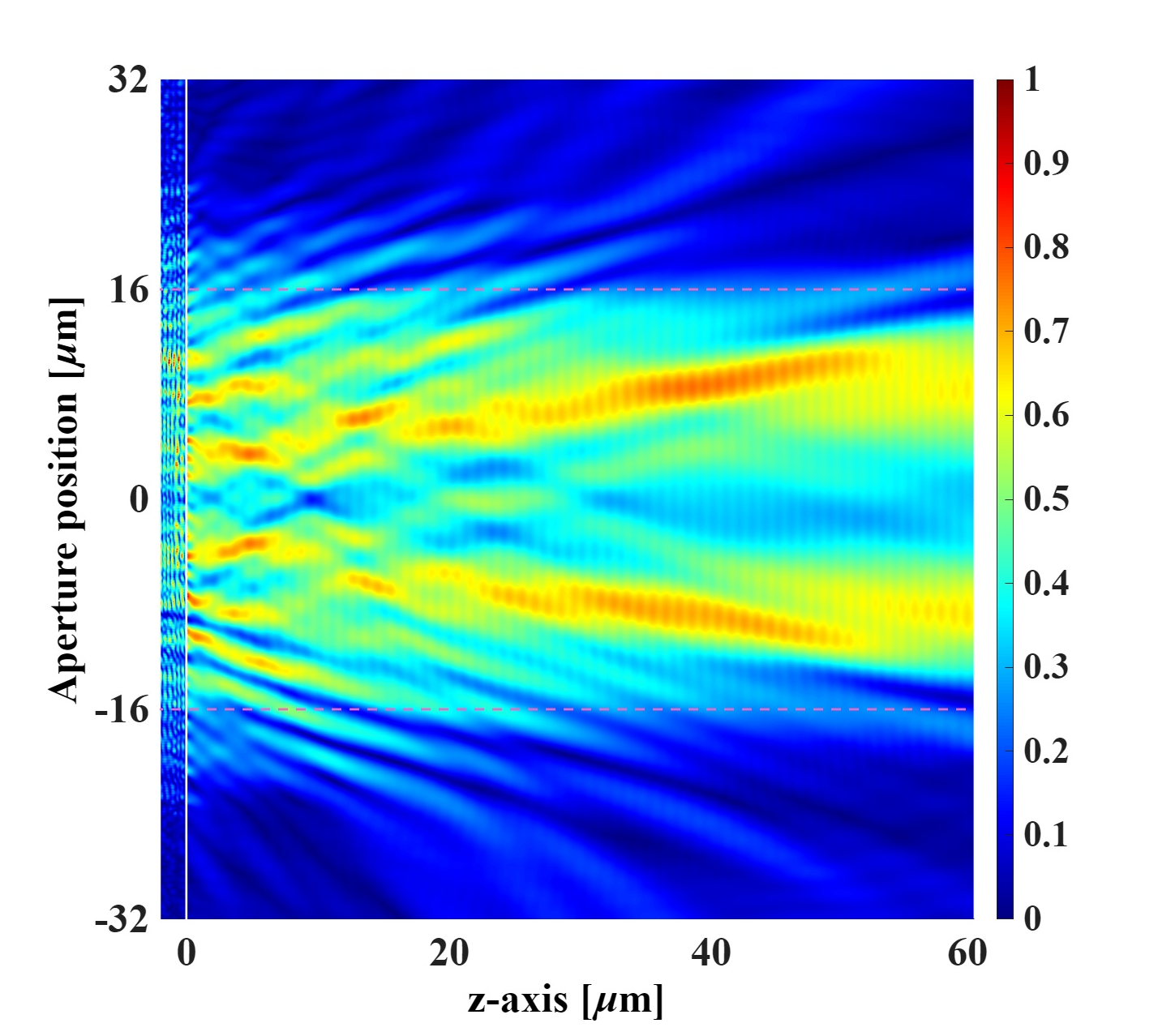}}
                \subfigure[]{\label{Fig: Fig06.d}
                \includegraphics[width=0.36\linewidth,trim={1.0cm 0.3cm 1.85cm 1.35cm},clip=true]{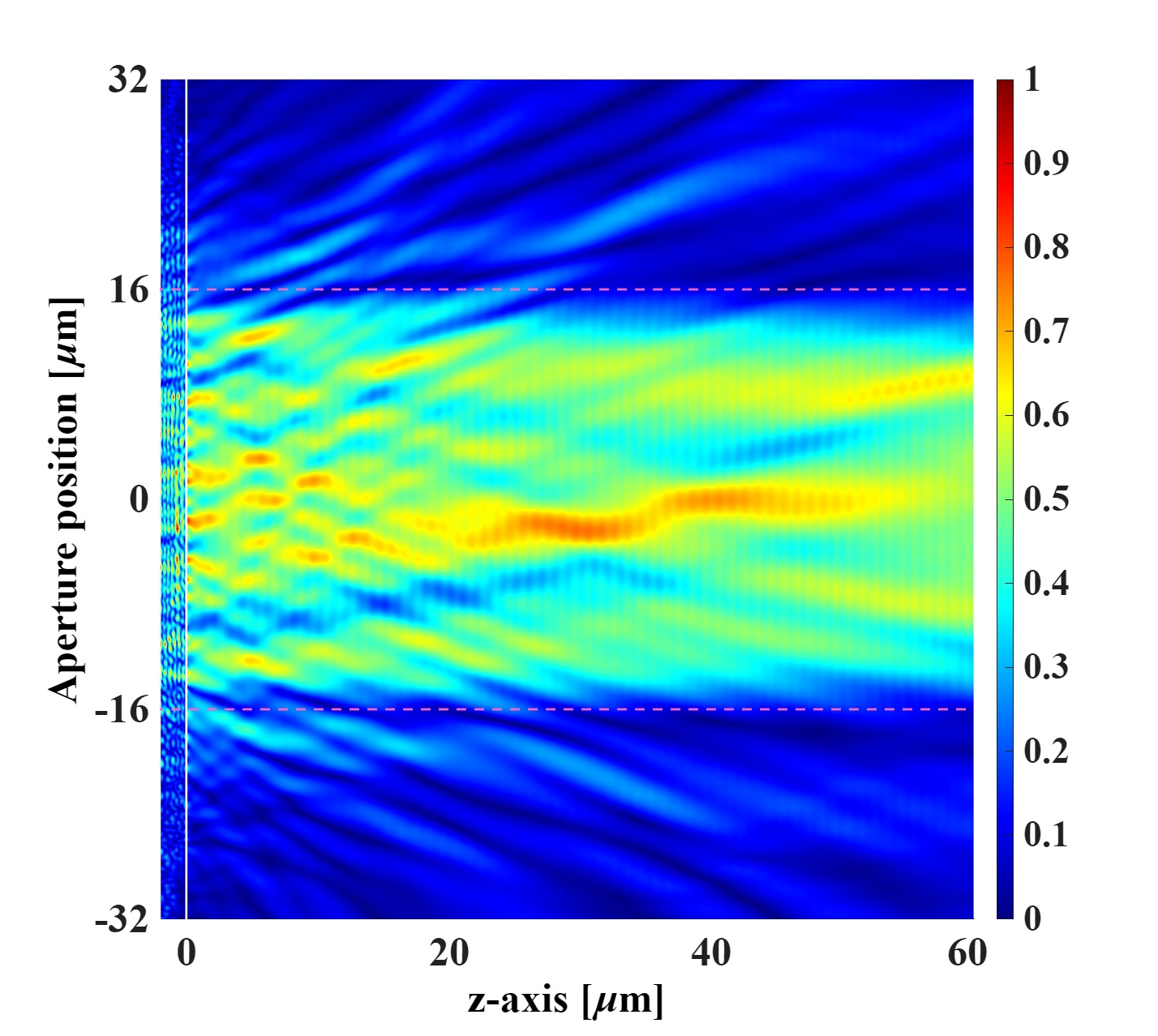}}
                
                \subfigure[]{\label{Fig: Fig06.e}
                \includegraphics[width=0.36\linewidth,trim={1.0cm 0.3cm 1.85cm 1.35cm},clip=true]{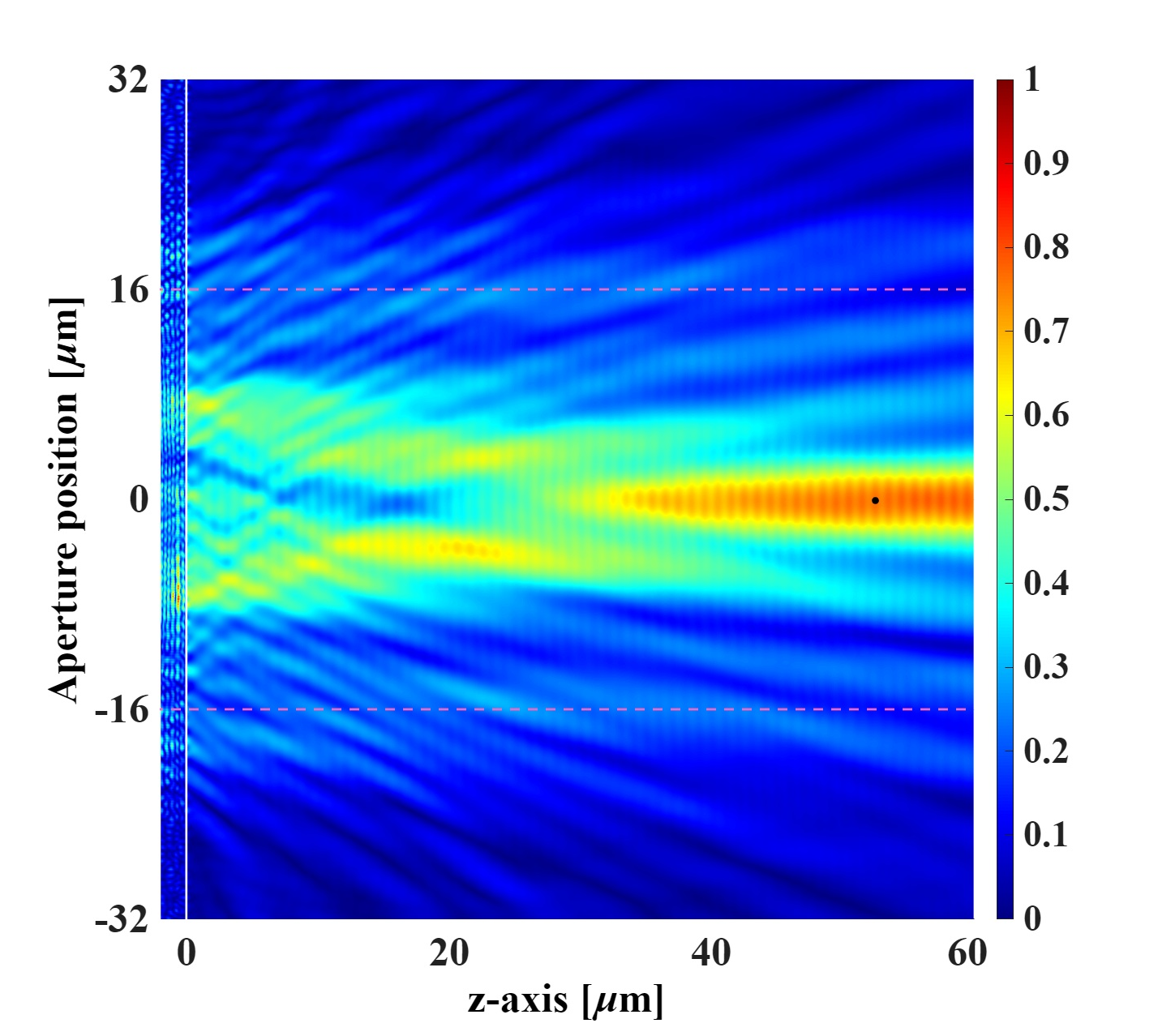}}
                \subfigure[]{\label{Fig: Fig06.f}
                \includegraphics[width=0.36\linewidth,trim={1.0cm 0.3cm 1.85cm 1.35cm},clip=true]{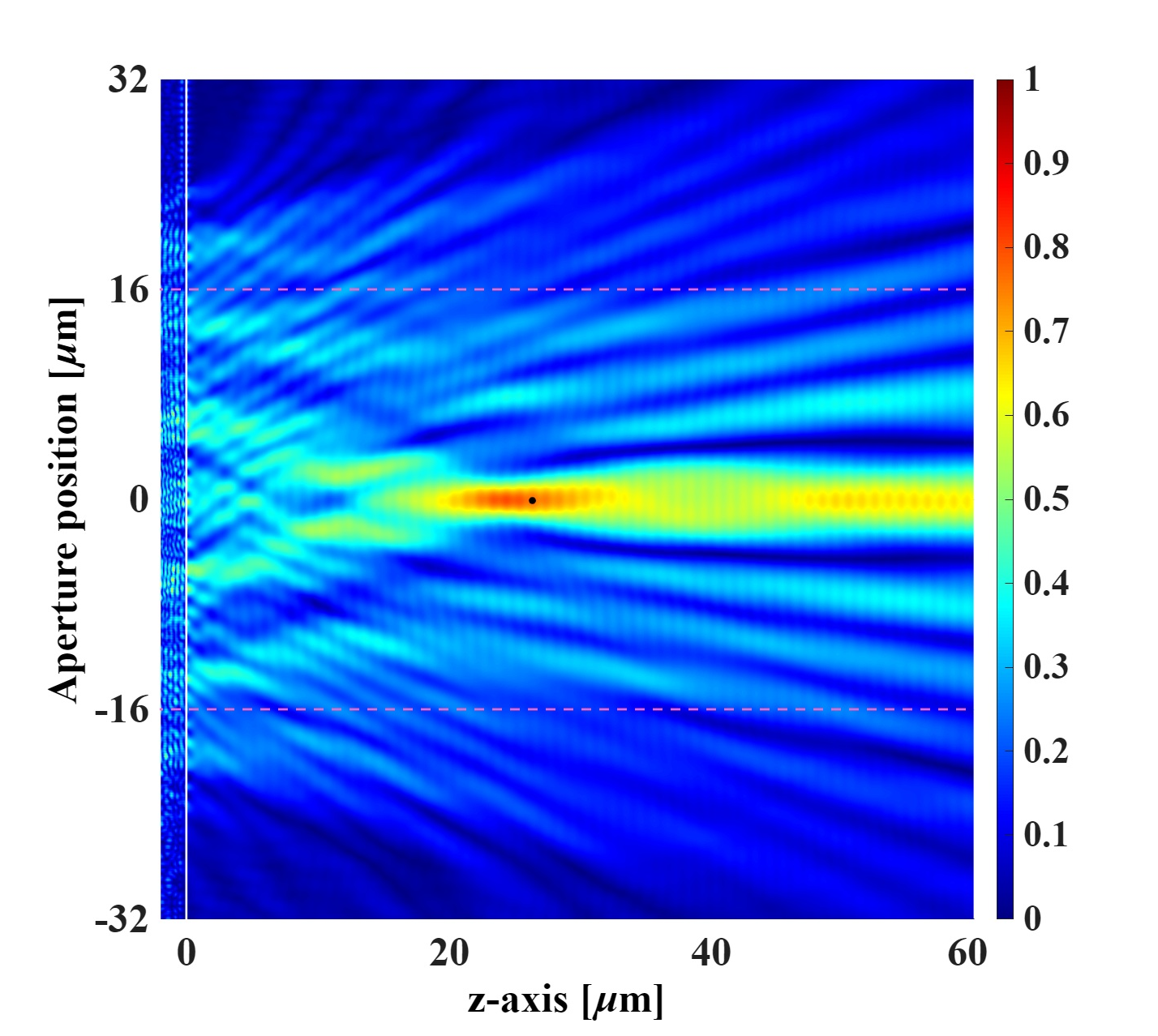}}
            \end{center}
            \vspace{-4.0mm}
            \caption{Normalized electric field ($E$-field) distributions generated by port excitation and the proposed GRIN lens operating in the mid-IR spectrum, 150~THz, without a collimator: (a) Single-port excitation, (b) 20-port excitation with a Gaussian magnitude profile (mean$=$0, standard deviation$=$17), (c) Profile $I$, exhibiting divergent wavefront behavior, (d) Profile $II$, demonstrating parallel wavefront transformation; (e) Profile $III$, yielding a focal length of 56.47~$\mu\text{m}$, and (f) Profile $IV$, producing a focal length of 24.37~$\mu\text{m}$.}
            \label{Fig: Fig06}
        \end{figure}

\begin{figure}[ht!]
            \begin{center}
                \subfigure[]{\label{Fig: Fig07.a}
                \includegraphics[width=0.24\linewidth,trim={0.0cm 0.0cm 0.0cm 0.0cm},clip=true]{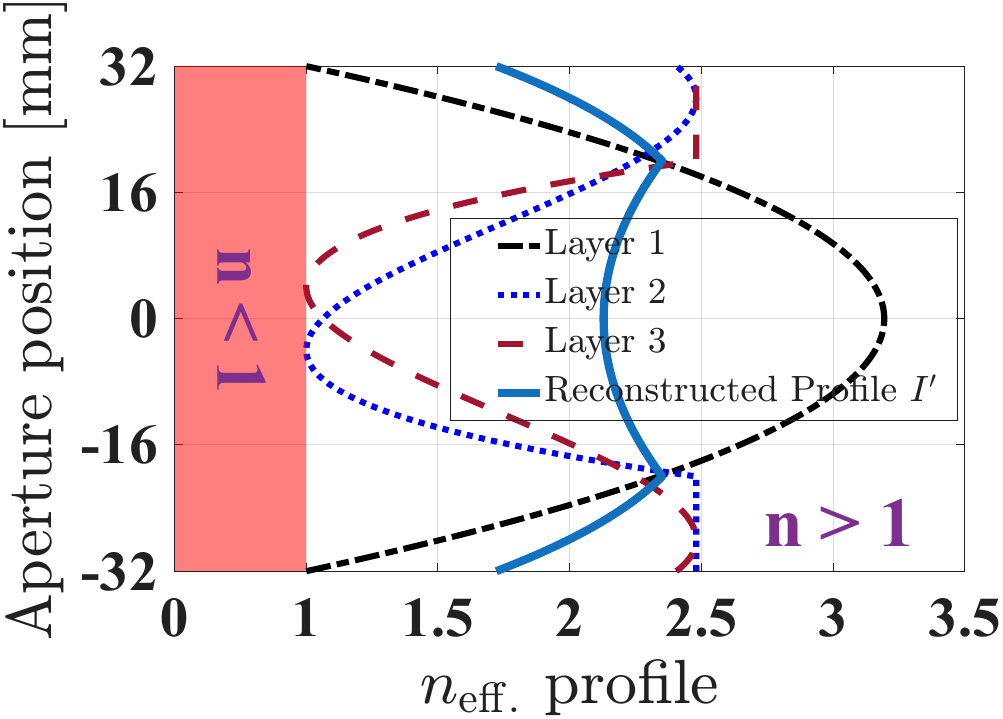}}
                \subfigure[]{\label{Fig: Fig07.b}
                \includegraphics[width=0.24\linewidth,trim={0.0cm 0.0cm 0.0cm 0.0cm},clip=true]{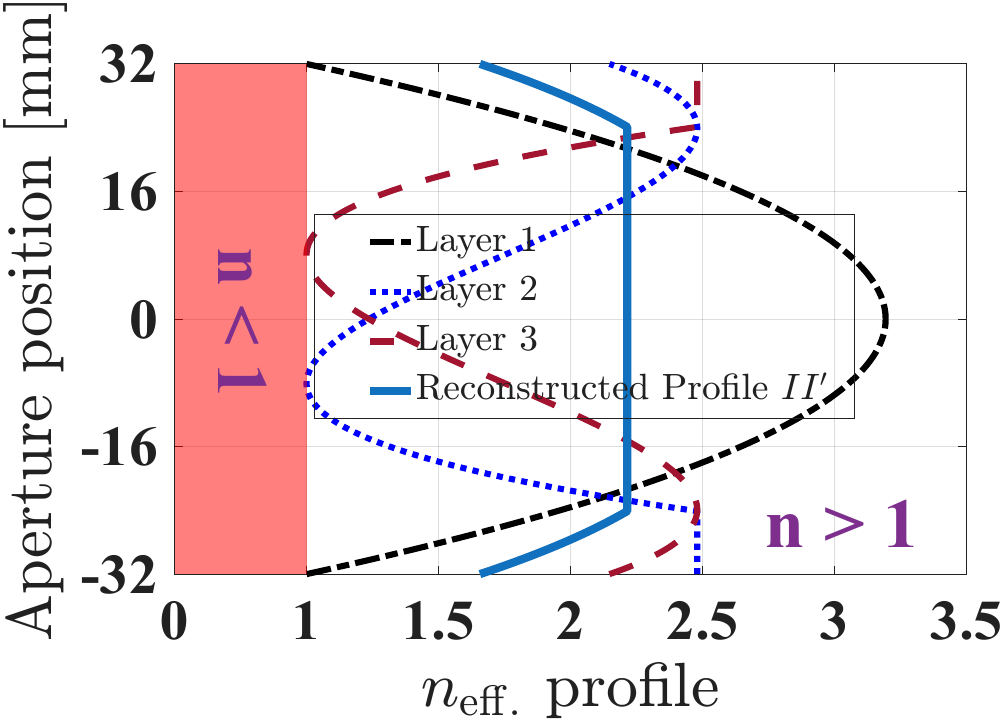}}
                \subfigure[]{\label{Fig: Fig07.c}
                \includegraphics[width=0.24\linewidth,trim={0.0cm 0.0cm 0.0cm 0.0cm},clip=true]{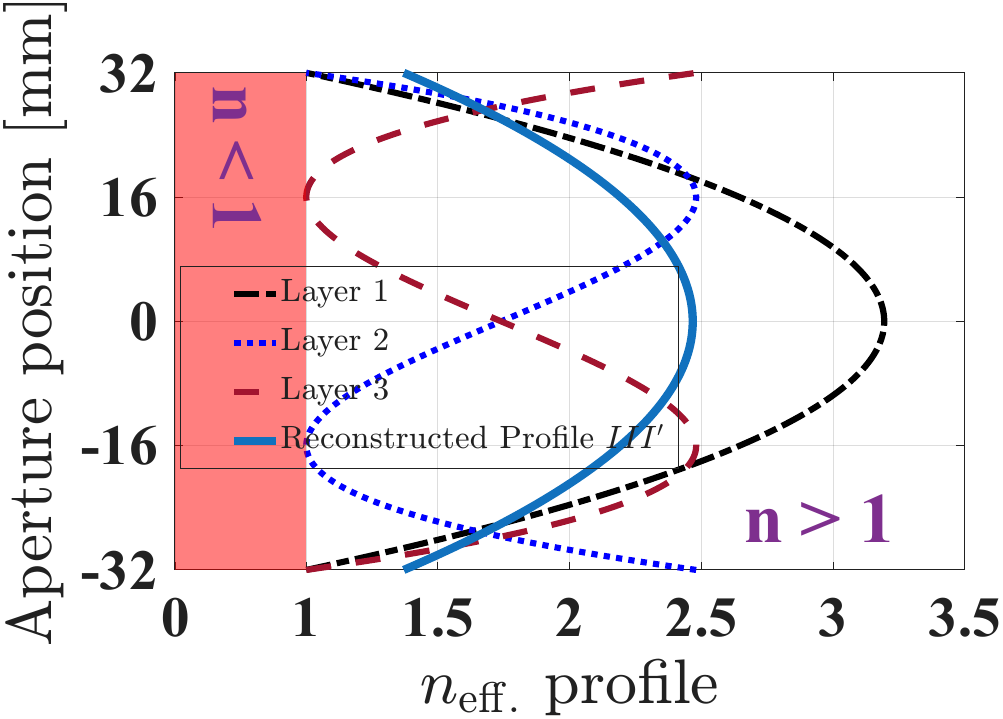}}
                \subfigure[]{\label{Fig: Fig07.d}
                \includegraphics[width=0.24\linewidth,trim={0.0cm 0.0cm 0.0cm 0.0cm},clip=true]{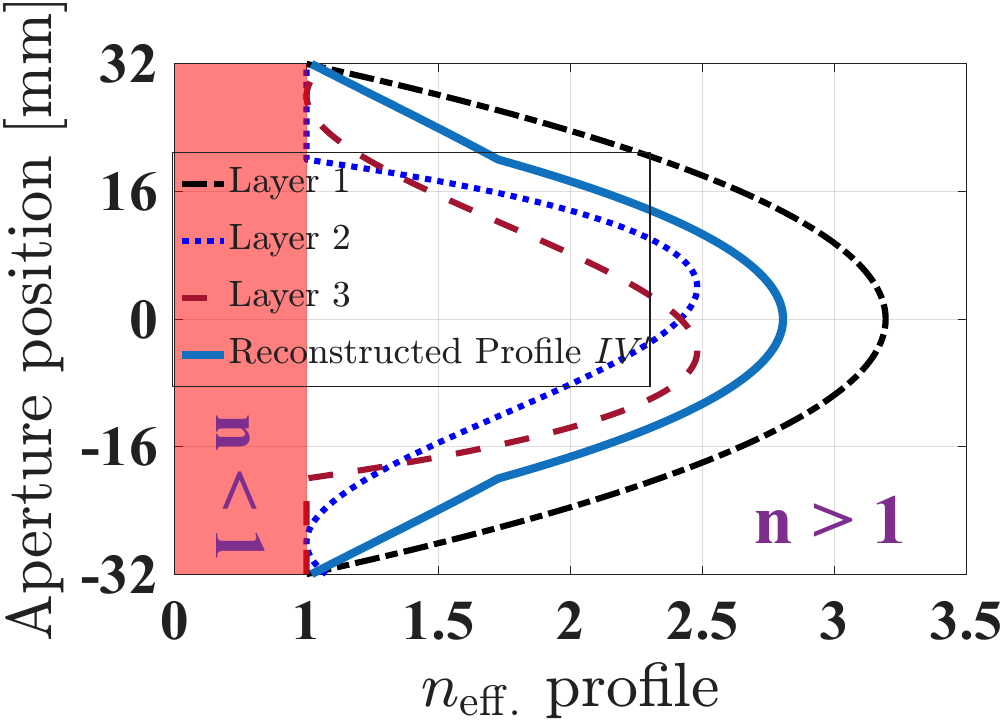}}
                
                \subfigure[]{\label{Fig: Fig07.e}
                \includegraphics[width=0.48\linewidth,trim={0.0cm 0.0cm 0.0cm 0.0cm},clip=true]{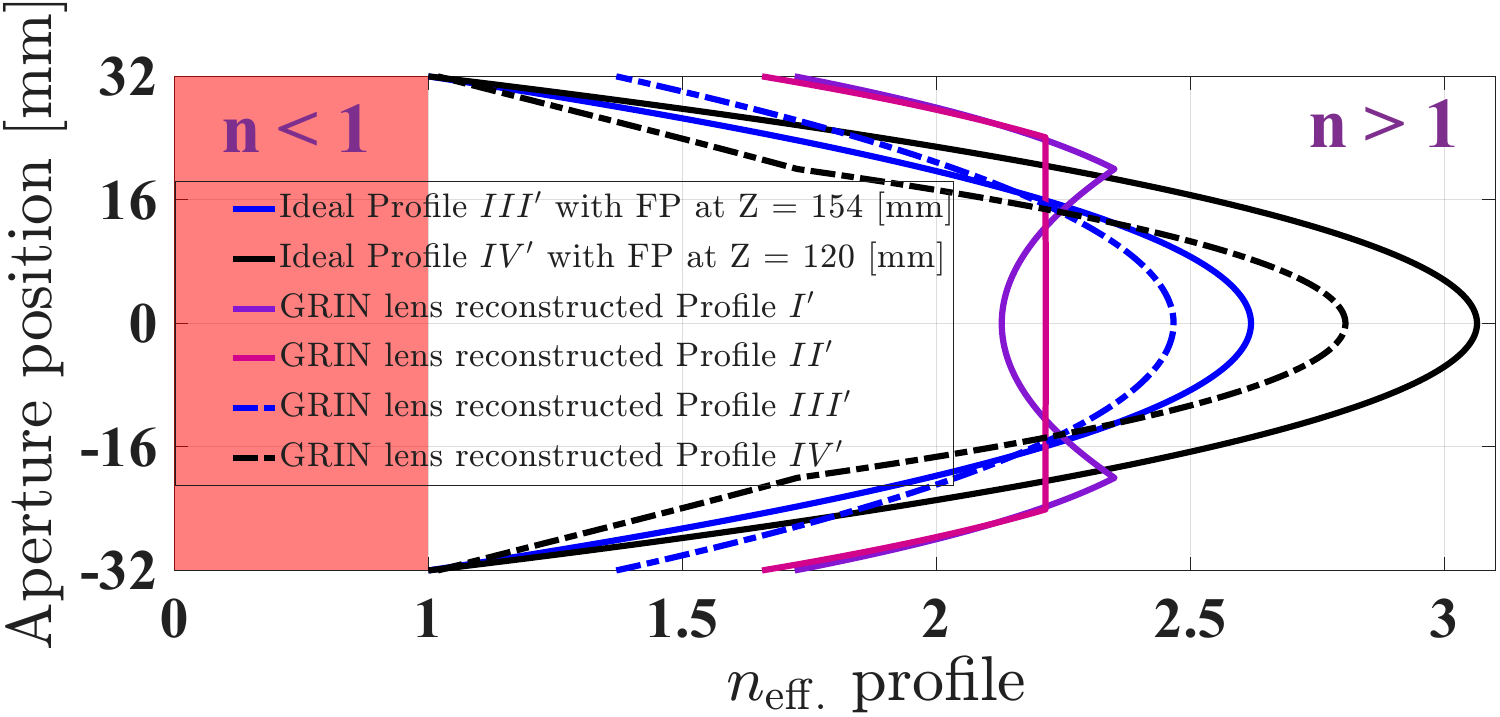}}
                \subfigure[]{\label{Fig: Fig07.f}
                \includegraphics[width=0.48\linewidth,trim={0.0cm 0.0cm 0.0cm 0.0cm},clip=true]{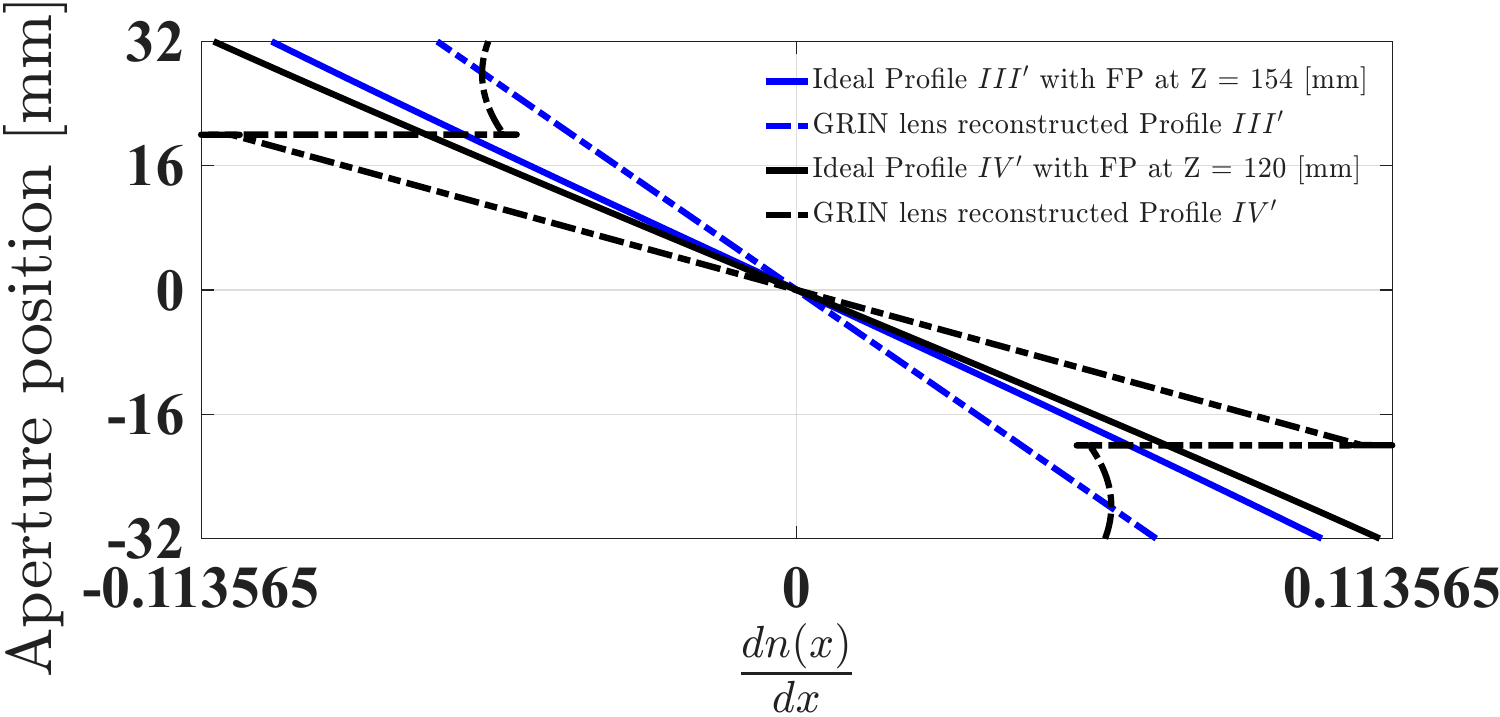}}
                
                \subfigure[]{\label{Fig: Fig07.g}
                \includegraphics[width=0.48\linewidth,trim={0.0cm 0.0cm 0.0cm 0.0cm},clip=true]{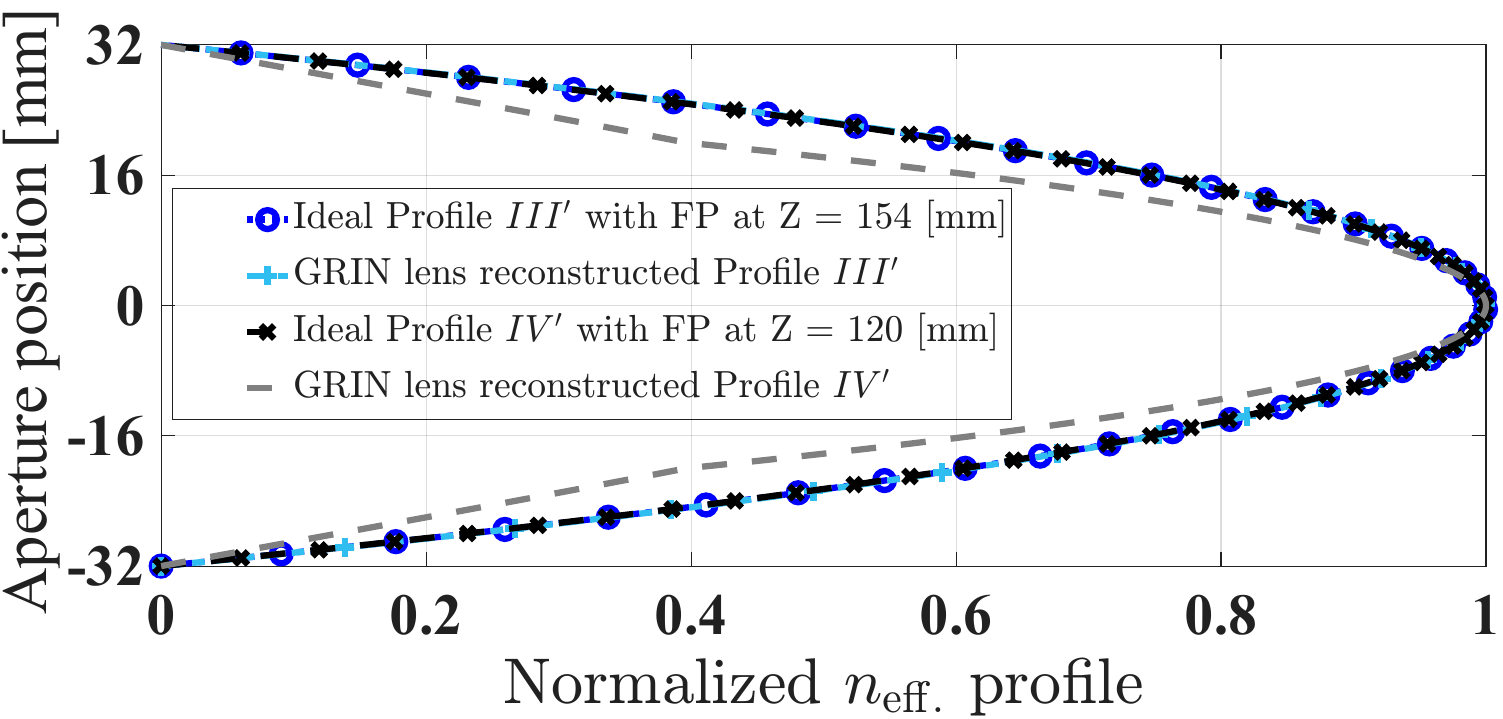}}
                \subfigure[]{\label{Fig: Fig07.h}
                \includegraphics[width=0.48\linewidth,trim={0.0cm 0.0cm 0.0cm 0.0cm},clip=true]{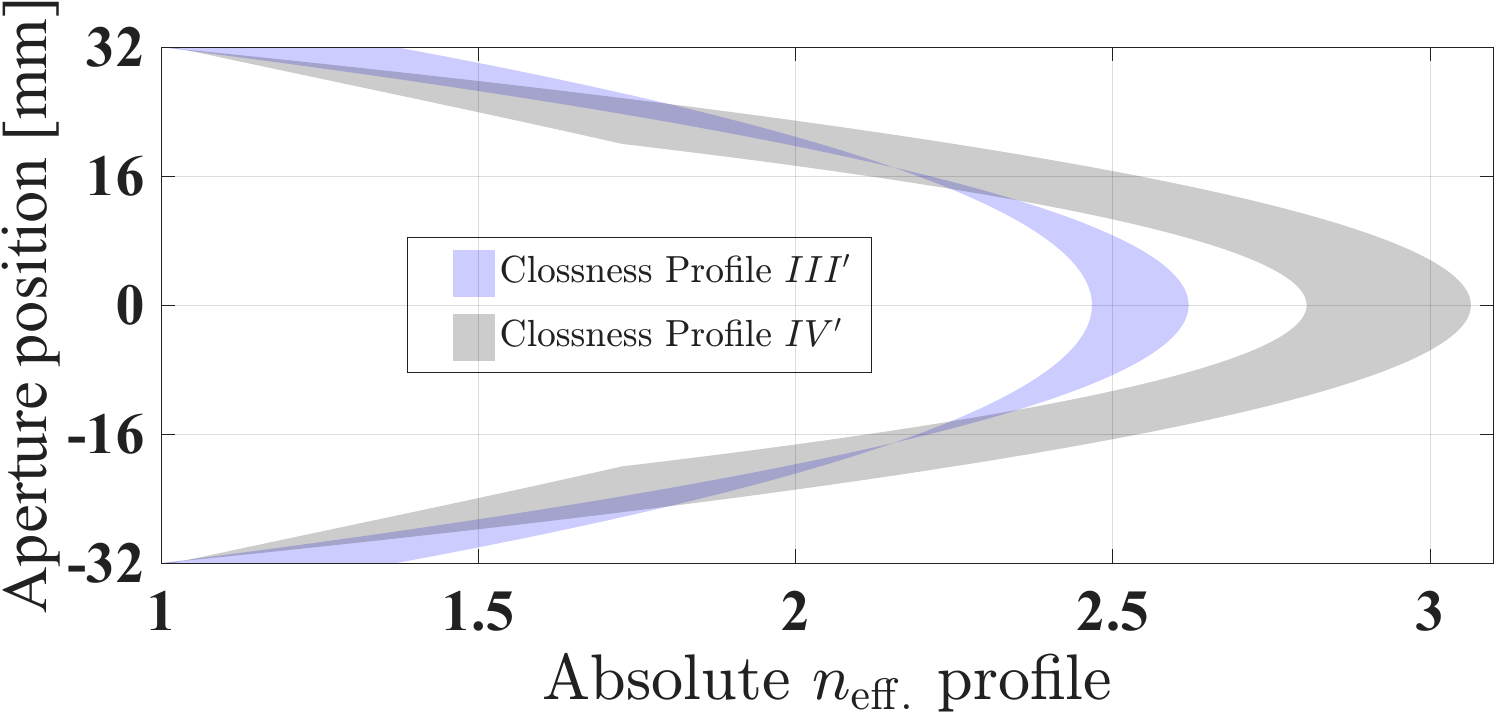}}
            \end{center}
            \vspace{-4.0mm}
            \caption{Comparison of the focusing performance between the ideal lens and the proposed GRIN lens in the mm-wave spectrum. Panels (a)–(d) illustrate the GRIN lens layers and their corresponding effective refractive index profiles for layer shifts of $[0,12,-12]$~mm, $[0,7.9,-7.9]$~mm, $[0,0,0]$~mm, and $[0,-12,12]$~mm (Profiles $I'-IV'$), respectively. Panel (e) presents a comparative analysis of the ideal and attained refractive index profiles, while Panel (f) depicts the aperture-dependent gradients of the effective refractive indices for both ideal and reconstructed cases. Panels (g) and (h) present the normalized reconstructed effective refractive indices and their closeness, defined as the total spatial difference between them.}
            \label{Fig: Fig07}
        \end{figure}

\begin{figure}[ht!]
            \begin{center}
                \subfigure[]{\label{Fig: Fig08.a}
                \includegraphics[width=0.38\linewidth,trim={0.0cm 0.0cm 0.0cm 0.0cm},clip=true]{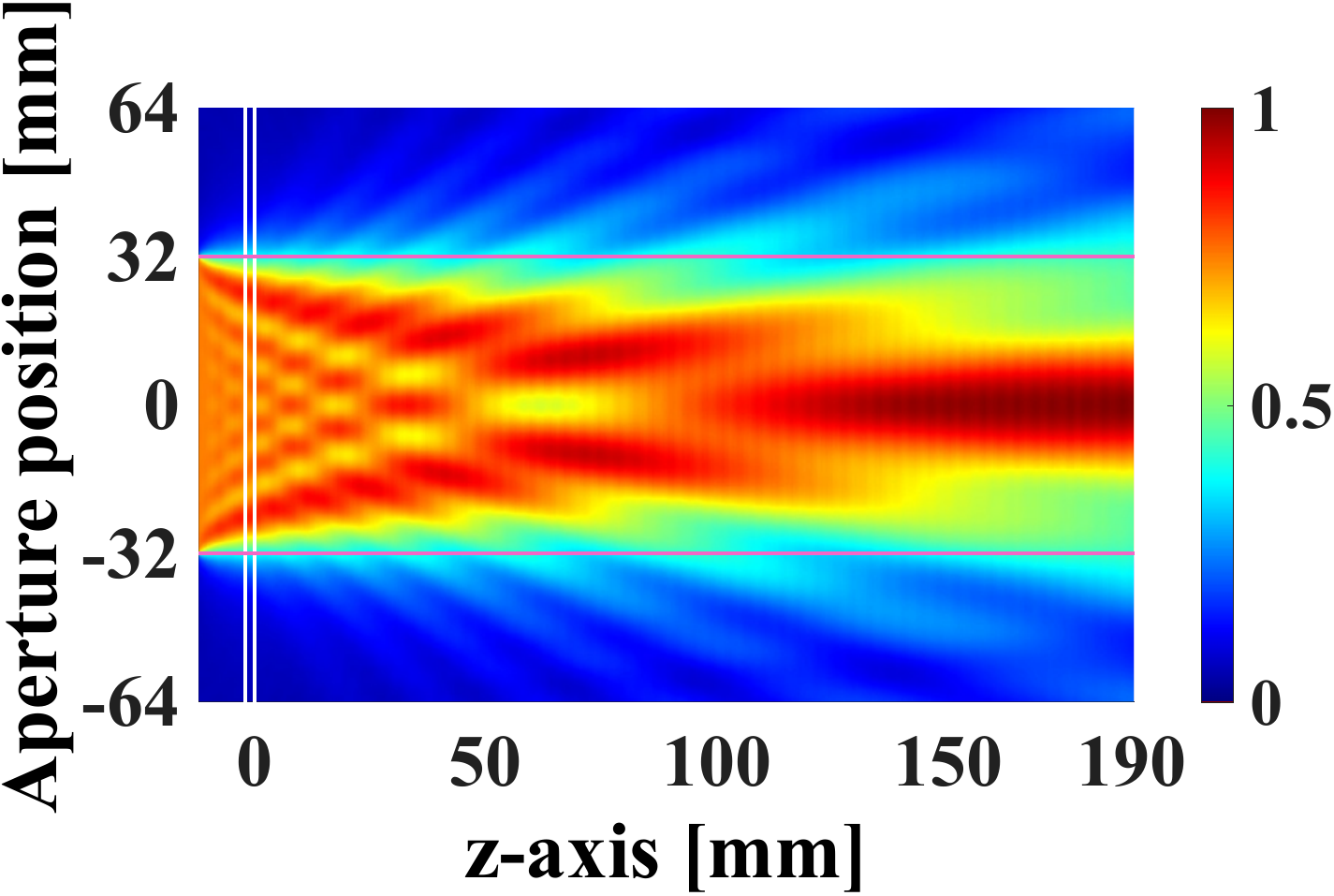}}
                \subfigure[]{\label{Fig: Fig08.b}
                \includegraphics[width=0.38\linewidth,trim={0.0cm 0.0cm 0.0cm 0.0cm},clip=true]{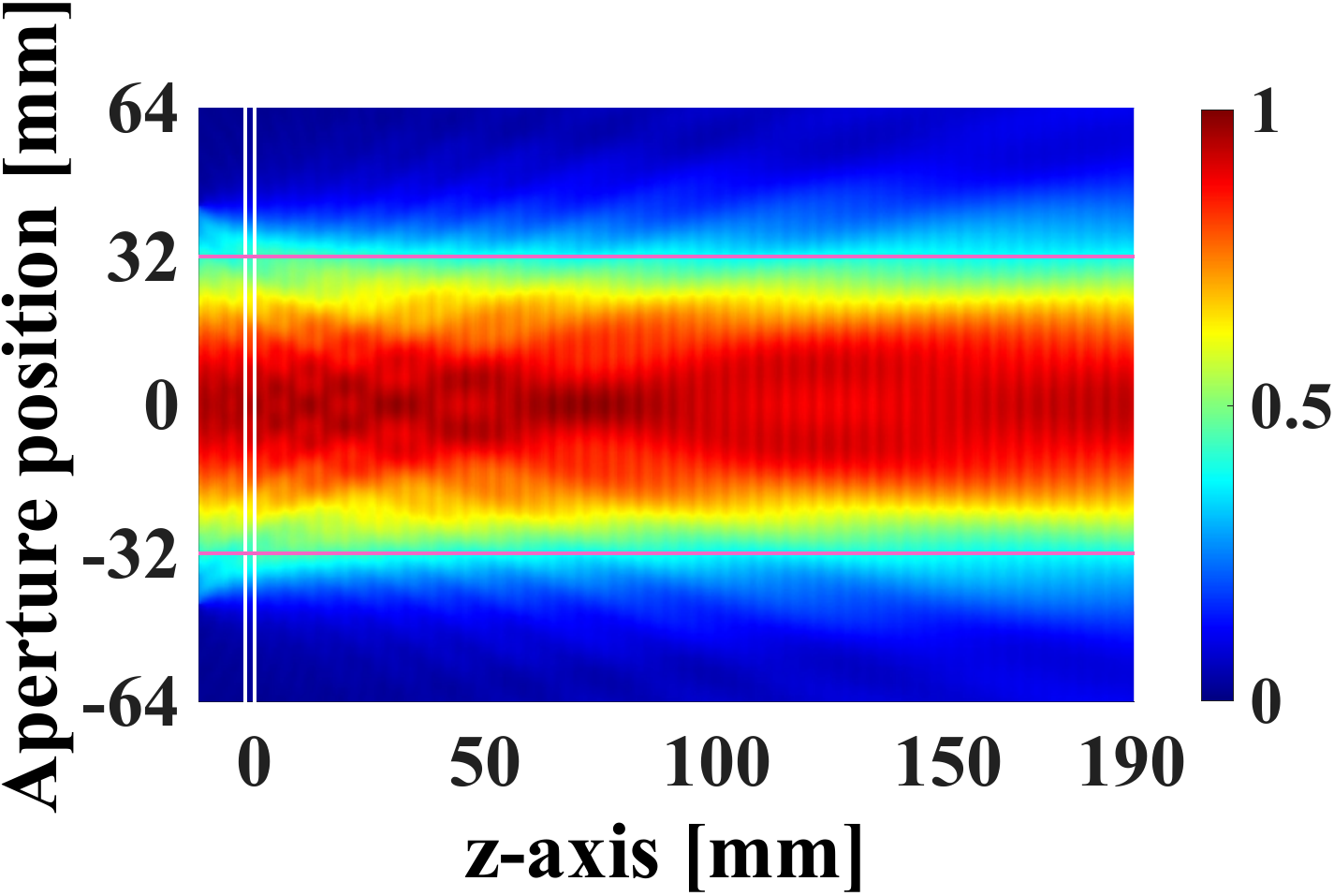}}
                
                \subfigure[]{\label{Fig: Fig08.c}
                \includegraphics[width=0.38\linewidth,trim={0.65cm 0.5cm 1.45cm 1.85cm},clip=true]{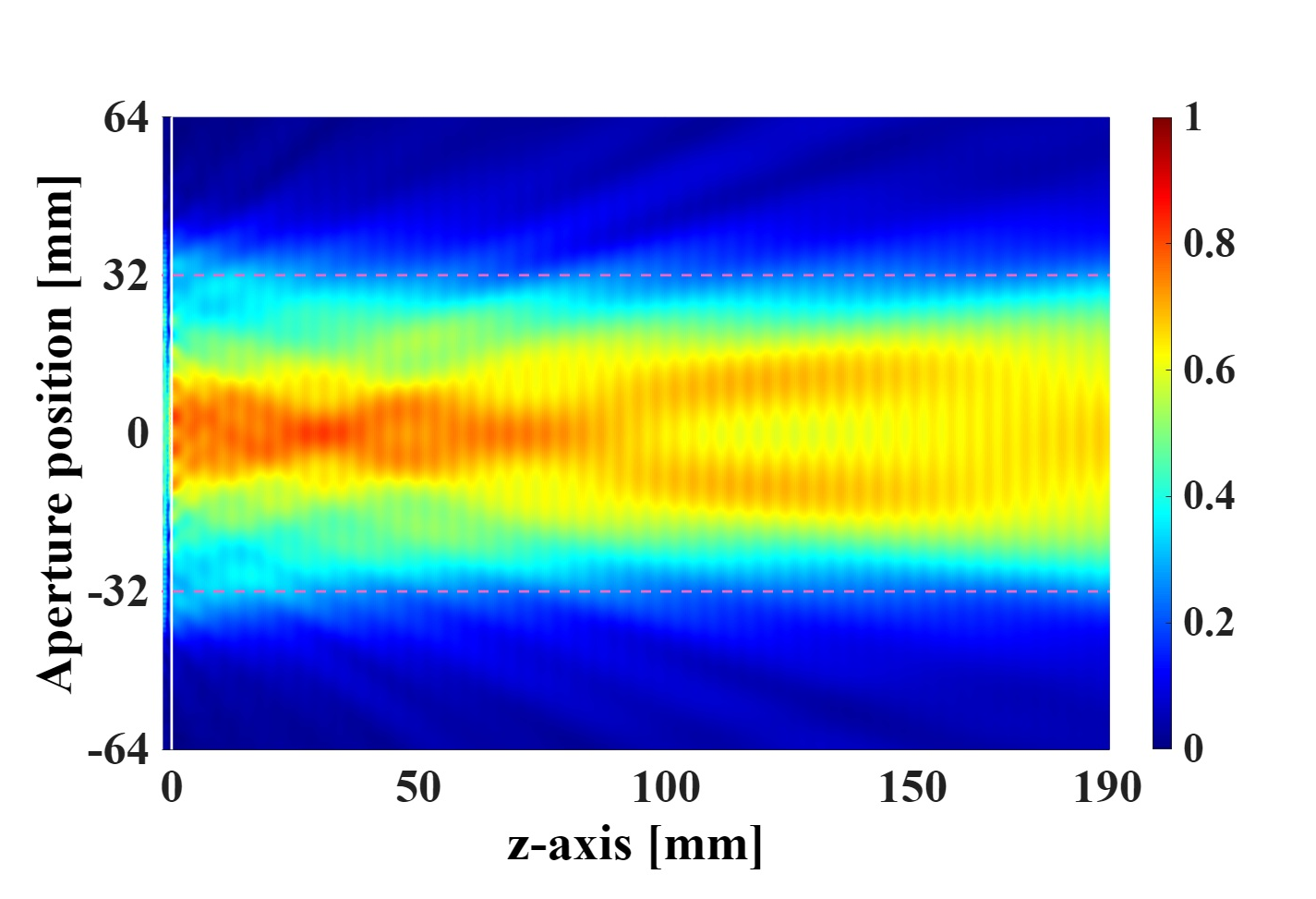}}
                \subfigure[]{\label{Fig: Fig08.d}
                \includegraphics[width=0.38\linewidth,trim={0.65cm 0.5cm 1.45cm 1.85cm},clip=true]{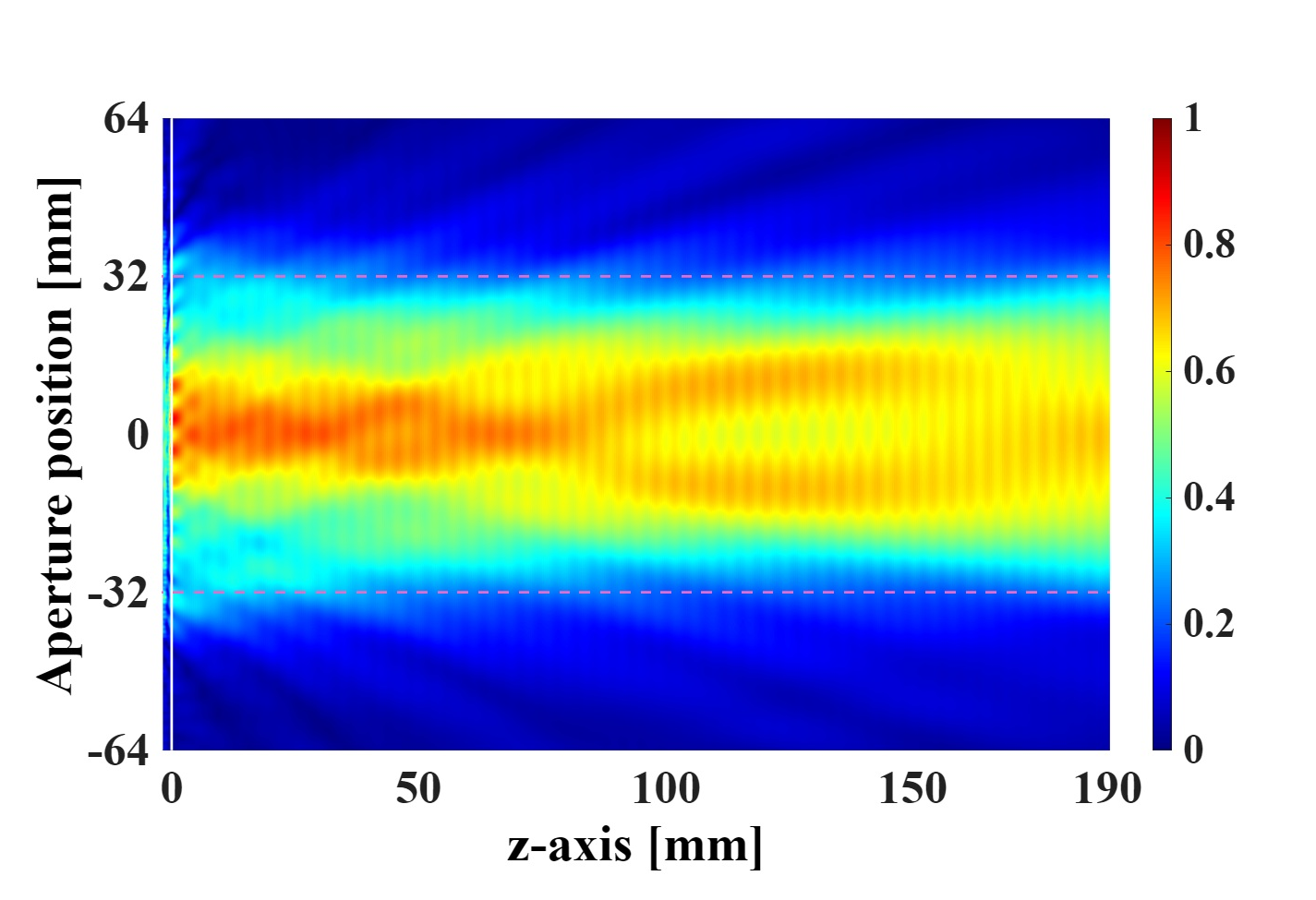}}
                
                \subfigure[]{\label{Fig: Fig08.e}
                \includegraphics[width=0.38\linewidth,trim={0.65cm 0.5cm 1.45cm 1.85cm},clip=true]{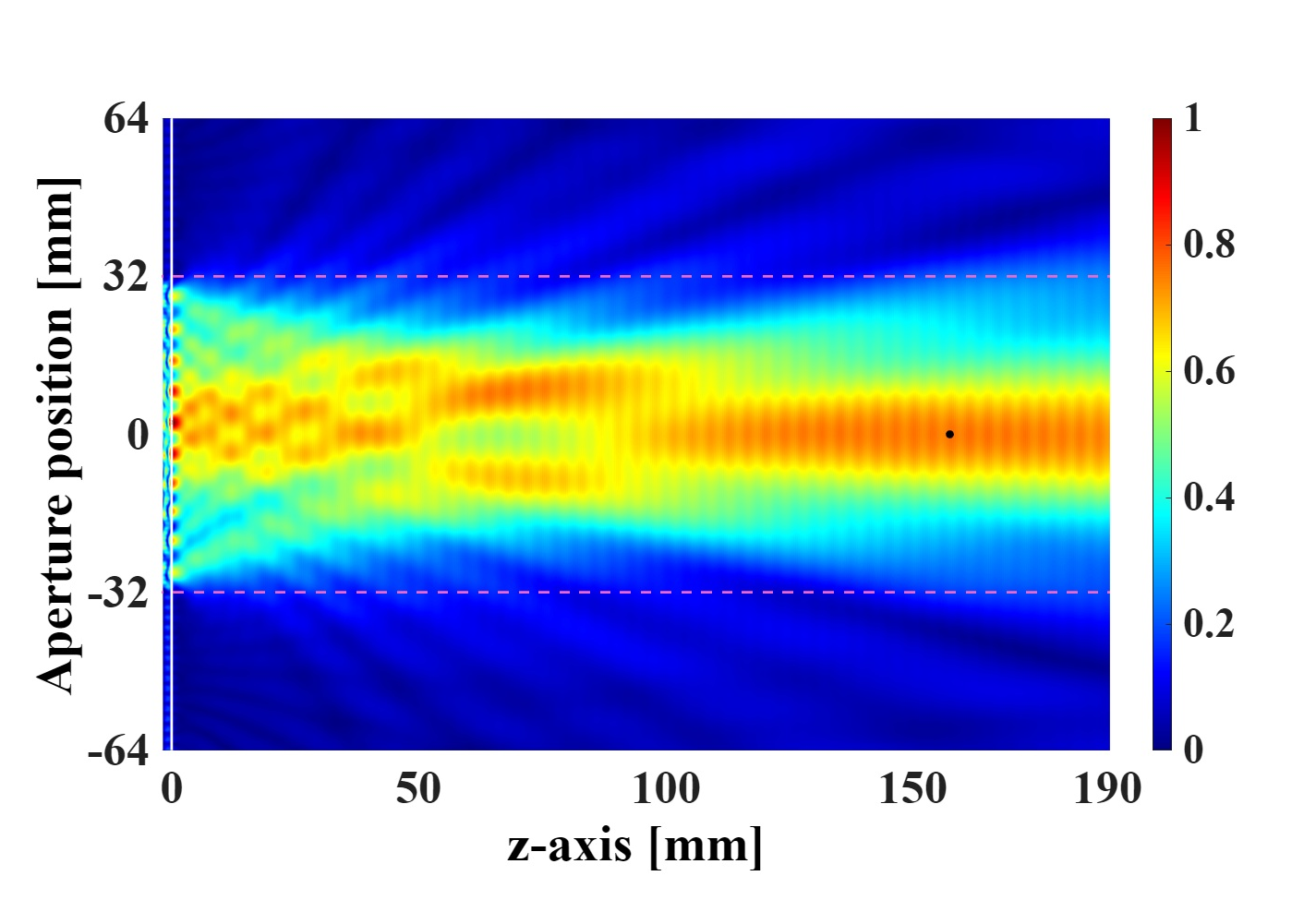}}
                \subfigure[]{\label{Fig: Fig08.f}
                \includegraphics[width=0.38\linewidth,trim={0.65cm 0.5cm 1.45cm 1.85cm},clip=true]{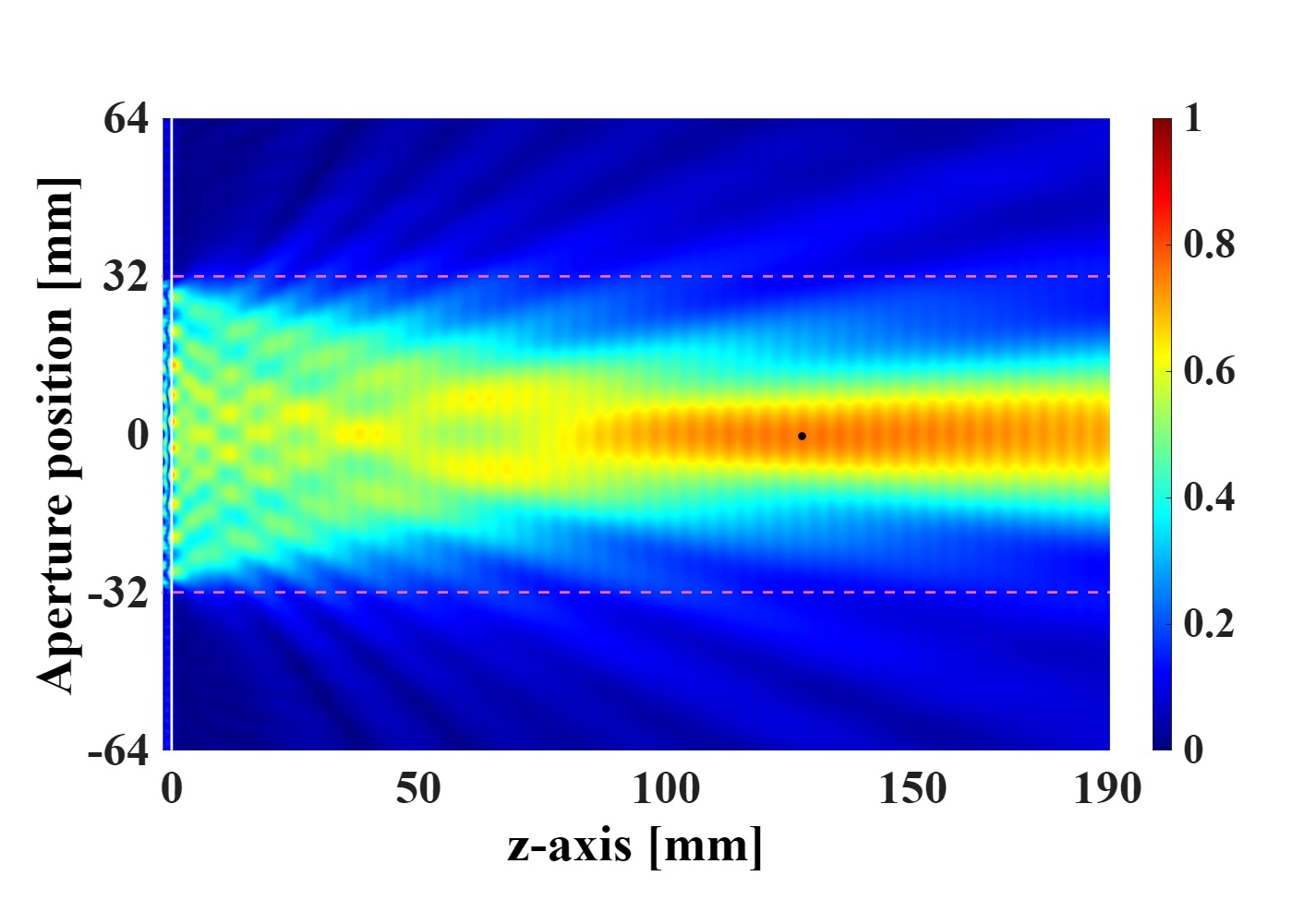}}
            \end{center}
            \vspace{-4.0mm}
            \caption{Normalized electric field ($E$-field) distributions generated by port excitation and the proposed GRIN lens operating in the mm-wave spectrum, 40~GHz, without/with a collimator: (a) Single-port excitation (without collimator), (b) 40-port excitation with a Gaussian magnitude profile (mean$=$0, standard deviation$=$25) (without collimator), (c) Profile $I'$ (without collimator), exhibiting divergent wavefront behavior, (d) Profile $II'$ (without collimator), demonstrating parallel wavefront transformation; (e) Profile $III'$ employing $L_{\rm x}$ collimator size yielding a focal length of 157.58~mm, and (f) Profile $IV'$ employing $L_{\rm x}$ collimator size, producing a focal length of 127.68~mm.}
            \label{Fig: Fig08}
        \end{figure}

        During the numerical investigations of the proposed GRIN lens, in addition to focusing, the authors have observed parasitic diffraction effects, which can significantly compromise the accuracy of beam characterization and misrepresent the lens’s true focusing performance, as shown in Figs.~\ref{Fig: Fig06.a} and~\ref{Fig: Fig08.a}. These effects appear when the lens is excited by a spatially limited beam with sharp shadow boundaries. To mitigate such undesired phenomenon, the authors have adopted an excitation scenario with a Gaussian amplitude profile along the $x$-axis. It has been found that such configuration effectively suppresses the mentioned parasitic effects.
        
        In simulations, the Gaussian excitation profile has been created by superposition of fields from several individual excitation ports weighted by a Gaussian envelope function, which yielded a smooth and well-behaved output field without significant artifacts, as illustrated in Figs.~\ref{Fig: Fig06.b} and~\ref{Fig: Fig08.b}. Consequently, this approach ensures that the analytical and numerical simulated beam patterns represent the true focusing capabilities of the GRIN lens more faithfully, enabling accurate evaluation of the proposed design.
        
        In the mid-IR simulations, a Gaussian excitation profile is realized using 20 ports with a Gaussian envelope characterized with the standard deviation $\sigma = 17$~$\mu\text{m}$. Within the GRIN lens aperture, 16 ports are positioned; however, two ports are placed on each side outside the lens aperture to maintain the uniformity of the excitation field across the aperture. In contrast, for the millimeter-wave simulations due to the larger aperture size in this spectral range, 40 ports with a Gaussian envelope with $\sigma = 25$~mm are employed. Among these, 30 ports are positioned within the aperture, while five ports are placed on each side outside the aperture.
        
        Figures~\ref{Fig: Fig06.c} and~\ref{Fig: Fig06.f} display the reconstructed normalized electric field distributions for the designed GRIN lens at the mid-IR spectrum, fabricated from conventional dielectric materials (specifically {\tt Si} and {\tt ZnSe}), corresponding to effective refractive index Profiles $I$-$IV$. Figures~\ref{Fig: Fig06.c} and~\ref{Fig: Fig06.d} present the electric field patterns demonstrating divergent and parallel wavefront propagation (afocal) behaviors, respectively. In contrast, Figures~\ref{Fig: Fig06.e} and~\ref{Fig: Fig06.f} depict cases in which focal spots are formed approximately at the anticipated focal lengths of 56.47~$\mu\text{m}$ and 24.37~$\mu\text{m}$, with relative errors of 4.37\% and 9.95\%, respectively. The relative deviation between the numerically calculated focal length $F_{\rm GRIN}$ and the ideal focal length $F_{\rm Ideal}$ is quantified through the percentage error $\Delta F=\frac{F_{\rm GRIN}-F_{\rm Ideal}}{F_{\rm Ideal}}\times 100$ [\%].
        
        Figures~\ref{Fig: Fig07.a} through~\ref{Fig: Fig07.d} present the refractive indices of each individual layer, along with their corresponding effective refractive indices for the mm-wave systems under Profiles $I'$ to $IV'$. Profiles $I'$-$IV'$ correspond to layer shift values of $X=[12,7.9,0,-12]$~mm, respectively. Regarding the reconstructed effective refractive indices of Profiles $I'$ and $II'$, the transmitted wavefronts are expected to exhibit divergent behavior and lens behavior with no net focusing effect, respectively. Figure~\ref{Fig: Fig07.e} illustrates the reconstructed effective refractive indices for Profiles $I'$-$IV'$, alongside the ideal refractive indices of conventional convex lenses with focal lengths of 154~mm and 120~mm, which serve as benchmarks. Figure~\ref{Fig: Fig07.f} shows the aperture-dependent gradients of the effective refractive indices for both the ideal and reconstructed cases. The deviation in the effective refractive index gradient for Profile $III'$ is negligible. In contrast, for Profile $IV'$, the gradient behavior suggests that the reconstructed effective refractive index cannot produce a sharp focal spot. Using cosine similarity as a metric to compare the shapes of the reconstructed effective refractive indices with ideal profiles at focal lengths of 154~mm and 120~mm yielded similarity values of 99.72\% and 99.84\% for Profiles $III'$ and $IV'$, respectively. Furthermore, when comparing the reconstructed profiles with the proposed ideal cases via normalized effective refractive indices, correlations of 100\% and 98.22\% were observed (Fig.~\ref{Fig: Fig07.g}). Finally, the closeness of the profiles was quantified as 8.46~mm and 20.94~mm, representing reasonably suitable results for confirming profile similarity (Fig.~\ref{Fig: Fig07.h}).
        
        Figures~\ref{Fig: Fig08.c} and~\ref{Fig: Fig08.f} present the reconstructed normalized electric field distributions for the GRIN lens designed for the mm-wave spectrum, using conventional dielectric materials (namely Rogers RO3010 and RO3006), corresponding to effective refractive index Profiles $I'$ through $IV'$. Figures~\ref{Fig: Fig08.c} and~\ref{Fig: Fig08.d} show the electric field patterns exhibiting divergent and parallel wavefront propagation behaviors, respectively. By contrast, Figures~\ref{Fig: Fig08.e} and~\ref{Fig: Fig08.f} illustrate scenarios where focal spots are generated approximately at the expected focal lengths of 157.58~mm and 127.68~mm, with associated relative errors of 2.27\% and 6.02\%, respectively.
        
        A metric for evaluating focusing ability (focusing performance) of the proposed lens is the Numerical Aperture (\textit{NA}), Eq.~\ref{Eq: Eq07}:
        \begin{equation}
            \textit{NA}
            =n\displaystyle\frac{\frac{L}{2}}{\sqrt{\left(\frac{L}{2}\right)^{\rm 2}+F^{\rm 2}}},
            \label{Eq: Eq07}
        \end{equation}
        where $n$ is the refractive index of the medium in which the metalens operates, with $n=1.00027304$ for air. Physically, \textit{NA} is a dimensionless number that quantifies the angular range over which a lens can accept or emit electromagnetic waves. Given the lens dimensions and its tunable focal length range, the computed \textit{NA} is within 28.42\% to 58.83\% and from 20.35\% to 25.77\% for the mid-IR and mm-wave spectra, respectively.
        
\begin{figure}[ht!]
            \begin{center}
                \subfigure[]{\label{Fig: Fig09.a}
                \includegraphics[width=0.485\linewidth,trim={0.0cm 0.0cm 0.0cm 0.0cm},clip=true]{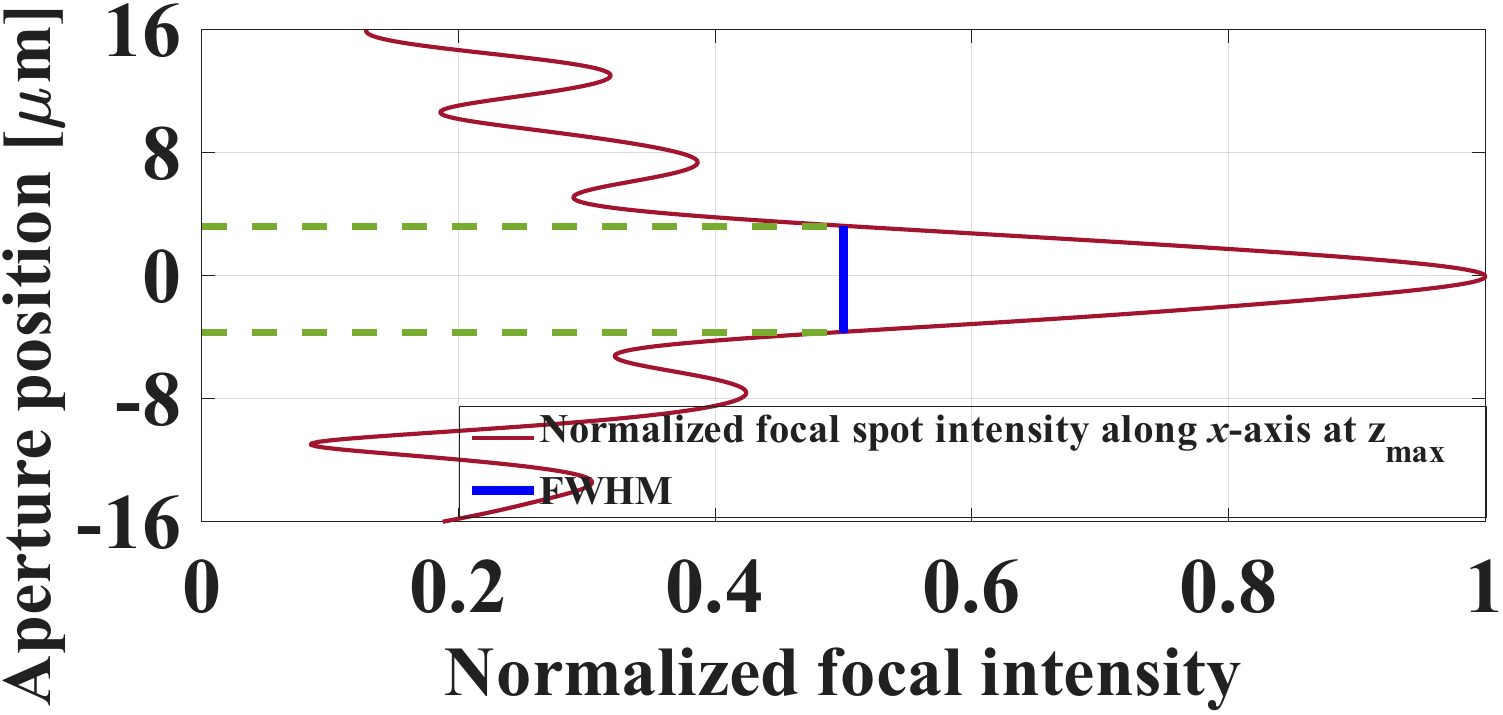}}
                \subfigure[]{\label{Fig: Fig09.b}
                \includegraphics[width=0.485\linewidth,trim={0.0cm 0.0cm 0.0cm 0.0cm},clip=true]{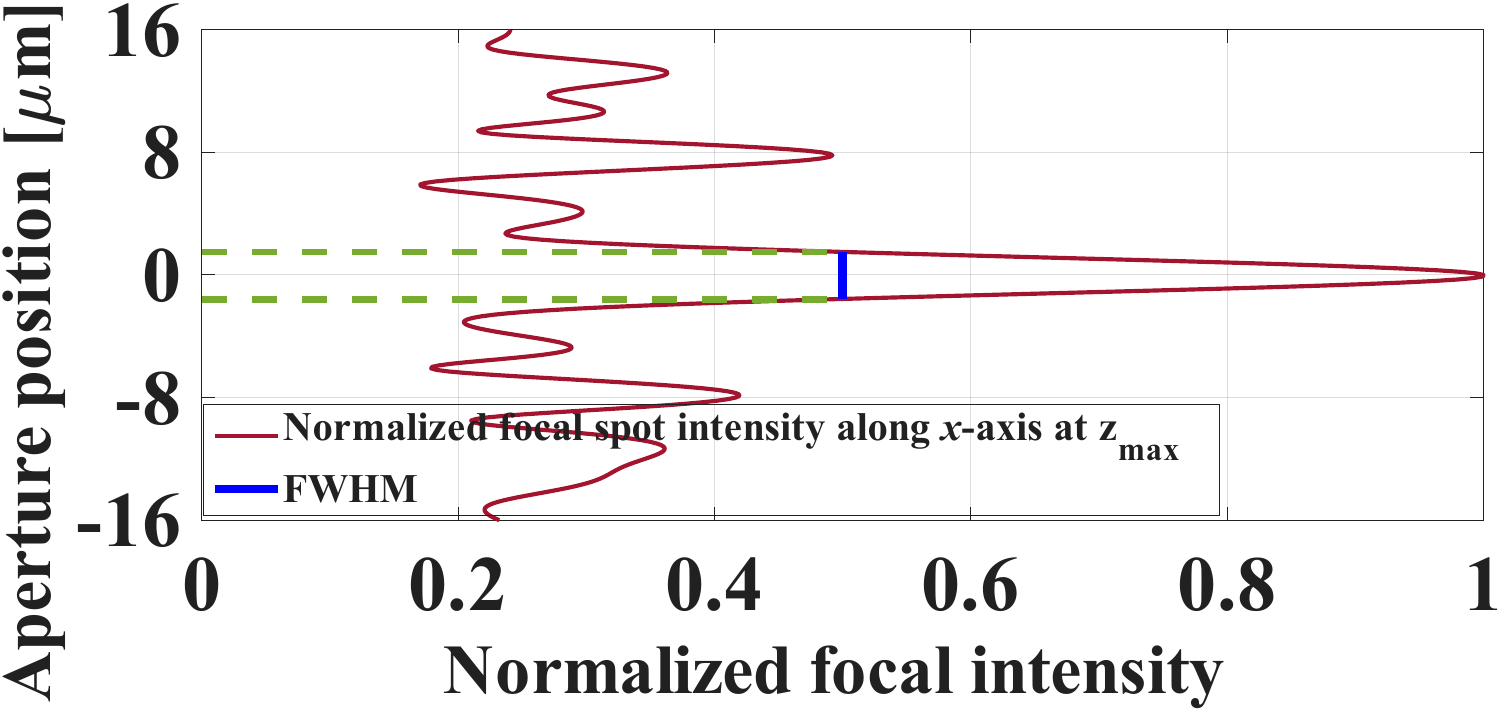}}
                
                \subfigure[]{\label{Fig: Fig09.c}
                \includegraphics[width=0.485\linewidth,trim={0.0cm 0.0cm 0.0cm 0.0cm},clip=true]{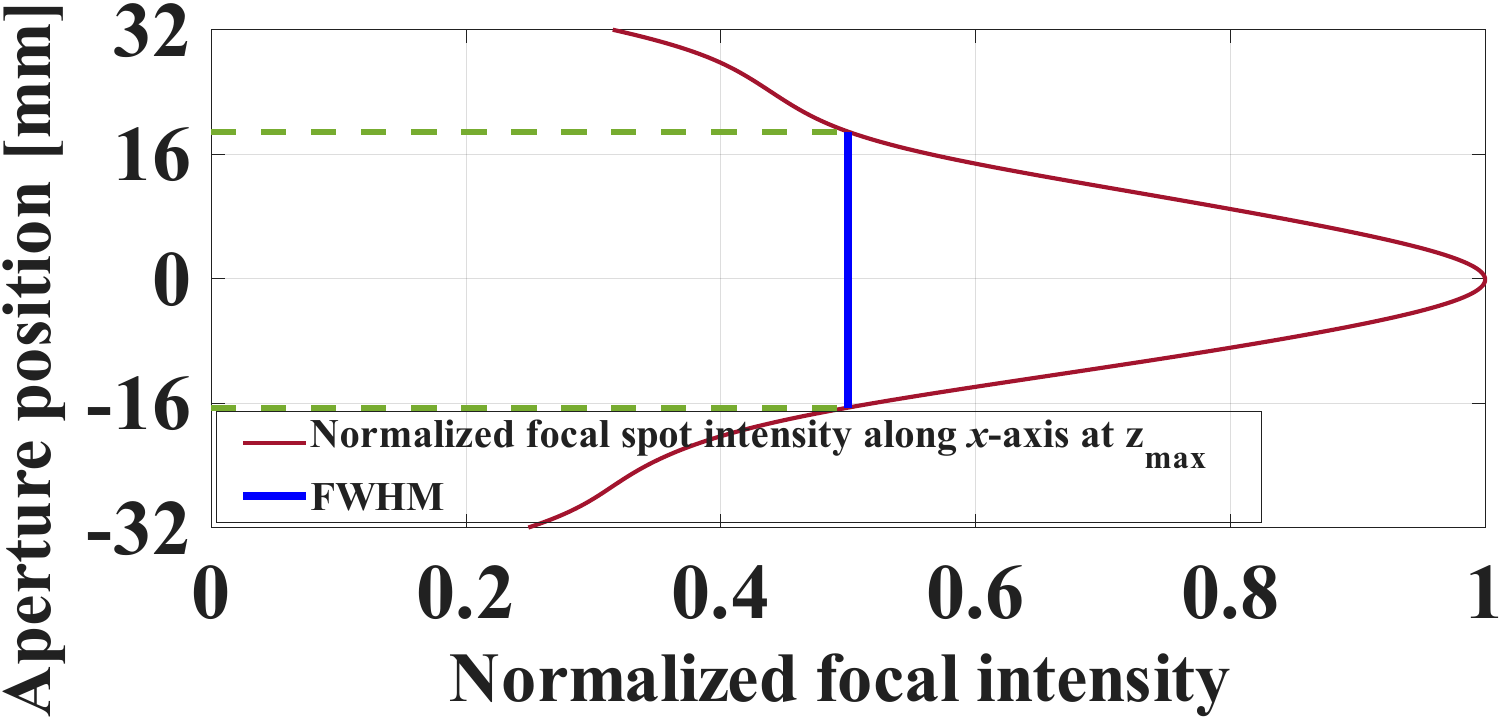}}
                \subfigure[]{\label{Fig: Fig09.d}
                \includegraphics[width=0.485\linewidth,trim={0.0cm 0.0cm 0.0cm 0.0cm},clip=true]{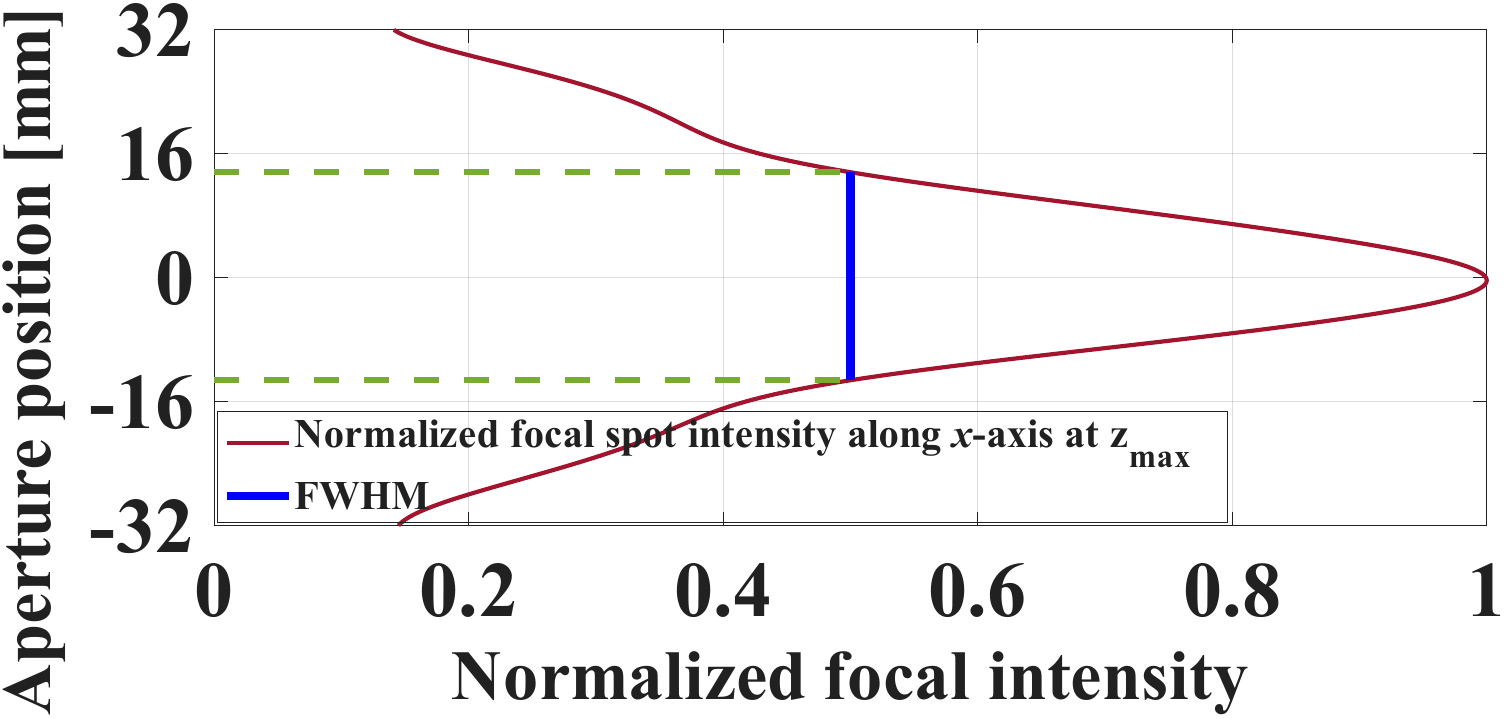}}
            \end{center}
            \vspace{-4.0mm}
            \caption{Normalized electric field ($E$-field) distributions along the lateral direction at focal spot of mid-IR and mm-wave spectra: (a) 6.9~$\mu\text{m}$ $\textit{FWHM}$ for Profile $III$, (b) 3.1~$\mu\text{m}$ $\textit{FWHM}$ for Profile $IV$, (c) 35.5~mm $\textit{FWHM}$ for Profile $III'$, and (d) 26.8~mm $\textit{FWHM}$ for Profile $IV'$.}
            \label{Fig: Fig09}
        \end{figure}

        Furthermore, the Full Width at Half Maximum ($\textit{FWHM}$) values extracted from Fig.~\ref{Fig: Fig09} are 6.9~$\mu\text{m}$ and 3.1~$\mu\text{m}$ for Profiles $III$ and $IV$ in the mid-infrared band, and 35.5~mm and 26.8~mm for Profiles $III'$ and $IV'$ in the millimeter-wave band, respectively. The broader $\textit{FWHM}$ in the mm-wave design results from ($i$) the lower spatial sampling of the gradient profile (128 unit cells versus 320 in the mid-IR design) and ($ii$) the use of a collimator and the larger lateral layer displacements required to shorten the focal length, which introduce gradient-slope errors and boundary truncation. These effects increase longitudinal aberration and broaden the $\textit{FWHM}$.
        
        The $\textit{FWHM}$ is related to the \textit{NA} and operating wavelength $\lambda$ by $\textit{FWHM}=\Upsilon\lambda/\textit{NA}$, where $\Upsilon$ characterizes the focusing quality and is obtained from $\Upsilon=\textit{FWHM}\cdot\textit{NA}/\lambda$. For the mid-IR design ($\lambda\approx2~\mu\text{m}$), the extracted $\Upsilon$ values are 0.98 and 0.91 for Profiles $III$ and $IV$, respectively, while for the mm-wave design ($\lambda\approx7.5$~mm) they are 0.96 and 0.92 for Profiles $III'$ and $IV'$, respectively. Lower $\Upsilon$ values correspond to tighter, more diffraction-limited focal spots, whereas larger values indicate broader $\textit{FWHM}$ due to aberrations and other nonideal effects.
        
        The Fresnel number ($F_{N}$) defined as:
        \begin{equation}
            F_{N}=\frac{D^{\rm 2}}{4\lambda R},
            \label{Eq: Eq08}
        \end{equation}
        where $D$ is the aperture diameter and $R$ is the distance to the observation point, distinguishes radiative near-field (Fresnel) and far-field (Fraunhofer) propagation, with $F_{N} \gtrsim 1$ indicating radiative near-field operation (Fresnel diffraction). Using Eq.~\ref{Eq: Eq08}, $F_{N}$ ranges from 2.37 to 5.82 for the mid-IR lens and from 1.14 to 2.43 for the mm-wave lens, indicating focusing in moderate Fresnel diffraction regime. Achieving a sharper focal spot at smaller layer displacements while preserving the unit-cell transmission efficiency requires increasing the aperture-to-wavelength ratio ($L/\lambda$).
    \subsection{Metallic Collimator}
        An ultrathin metallic collimator is introduced on the illumination side of the GRIN lens for mm-wave focusing. Full-wave simulations of the normalized $E$-field show that it serves three purposes: ($i$) suppressing irregular edge diffraction by confining the incident wave to the clear aperture ($L_{\rm x}$), particularly at the maximum lateral displacement ($\pm L/2$); ($ii$) masking the outer aperture region, where large displacements introduce boundary truncation and gradient-slope errors; and ($iii$) limiting wave propagation to the central region, where the Chebyshev-derived refractive-index profile best matches the target lens. The simulations further show that reducing the collimator aperture shifts the focal spot toward the lens and improves its spatial definition at the expense of peak intensity. Therefore, the collimator aperture must be chosen to balance focal position, spot size, and intensity. A collimator is not required for the divergent configurations.
        
        The influence of the collimator on mm-wave focusing was evaluated for Profile $IV'$ using four configurations: ($i$) without a collimator, ($ii$) with a 72~mm aperture (12.5\% larger than the lens aperture, $L+8$~mm), ($iii$) with an aperture equal to the lens aperture, and ($iv$) with a 56~mm aperture (12.5\% smaller than the lens aperture, $L-8$~mm), as summarized in Table~\ref{Tab: Table02}. The results confirm that decreasing the collimator aperture reduces the focal length, bringing the focal spot closer to the lens while reducing the peak intensity.
        \begin{table*}[!htbp]
            \centering
            \caption{Quantitative comparison of collimator impact on focusing performance.}
            \vspace*{2mm}
            \setlength{\tabcolsep}{3pt}
            \begin{tabular}{c|c|c|c}
                \toprule
                \toprule
                 \rowcolor{Gray}
                  & Aperture & Focal & Normalized \\
                 \rowcolor{Gray}
                  & length & length & focal intensity \\
                 \rowcolor{Gray}
                 \multirow{-3}{*}{Collimator} & [mm] & [mm] & [\%] \\
                \toprule
                 \xmark & - & 179.88 & 100 \\
                \hline
                 \rowcolor{Gray}
                 \checkmark & 72 & 153.58 & 93.51 \\
                \hline
                 \checkmark & 64 & 127.68 & 78.63 \\
                \hline
                 \rowcolor{Gray}
                 \checkmark & 56 & 108.63 & 65.24 \\
                \toprule
                \toprule
            \end{tabular}
            \begin{minipage}{7.5cm}
            \vspace{0.05cm}
            \end{minipage}
            \label{Tab: Table02}
        \end{table*}
    \subsection{Tunable Focal Length as Function of Layer Shift}
        In optical and diffractive systems, the Power Factor ($PF$) is defined as the refractive power of a lens, given by the reciprocal of its focal length ($F$, expressed in [$\mu\text{m}$] for the mid-IR range or in [mm] for mm-wave systems), i.e., $PF\equiv 1/F$. Through the reciprocal shifting of the second and third layers in convergent configurations, the lens's power factor varies proportionally to the rate of layer displacement. To accurately predict the $PF$ of the proposed GRIN lenses for other shift values, the focal lengths for shift values of $X=[-2,-4,-6]$~$\mu\text{m}$ (mid-IR) and $X=[-4,-6,-8]$~mm (mm-wave) applied to the second and third layers were numerically determined as [49.47,32.37,28.37]~$\mu\text{m}$ and [150.28,138.98,131.68]~mm, respectively. Table~\ref{Tab: Table03} summarizes the $PF$ variations and focal length estimation errors as functions of the lateral shifts applied to the second and third layers of the proposed GRIN lens, for both convergent and parallel configurations.
        
\begin{figure}[ht!]
            \begin{center}
                \subfigure[]{\label{Fig: Fig10.a}
                \includegraphics[width=0.48\linewidth,trim={0.0cm 0.0cm 0.0cm 0.0cm},clip=true]{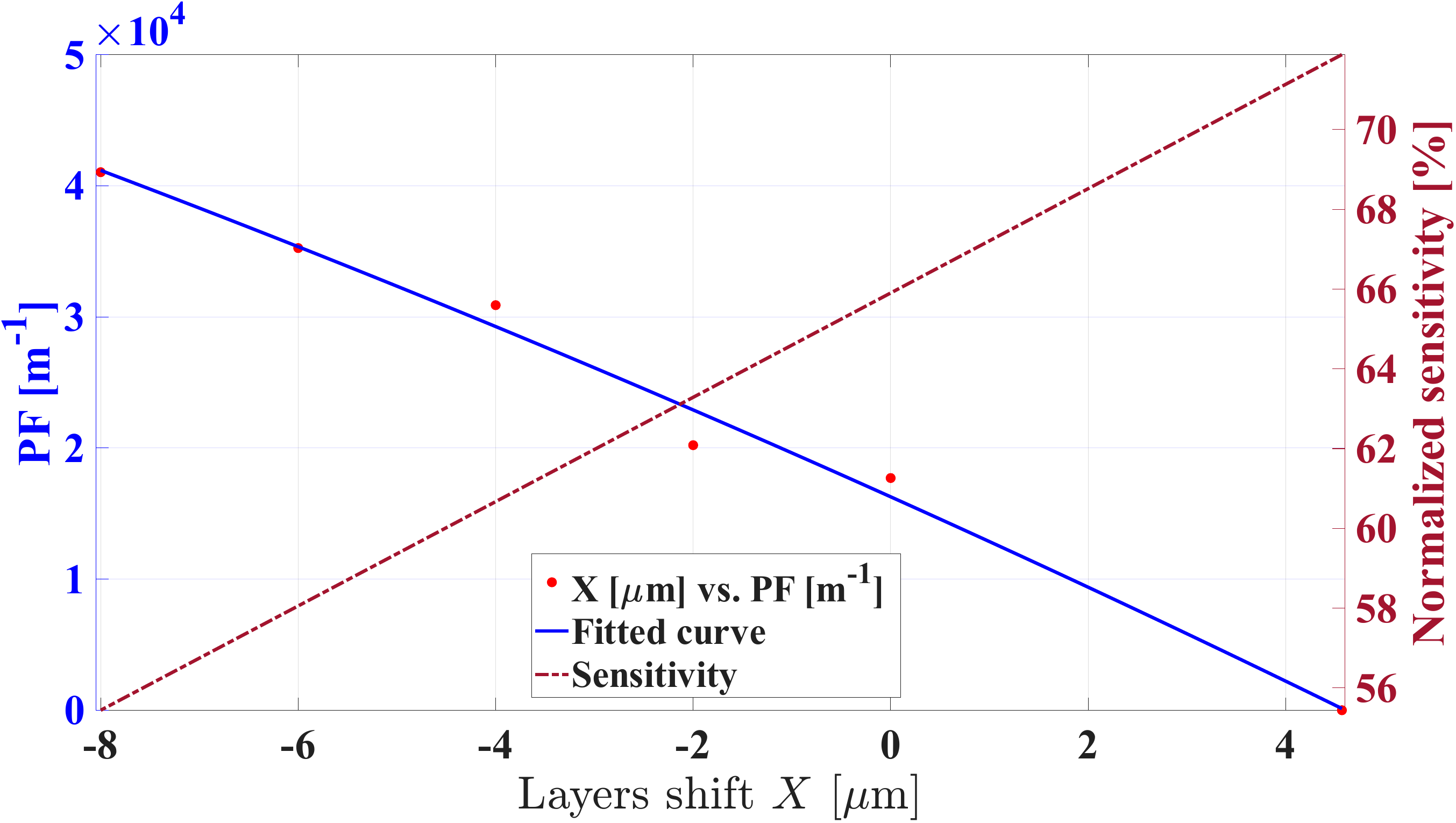}}
                \subfigure[]{\label{Fig: Fig10.b}
                \includegraphics[width=0.48\linewidth,trim={0.0cm 0.0cm 0.0cm 0.0cm},clip=true]{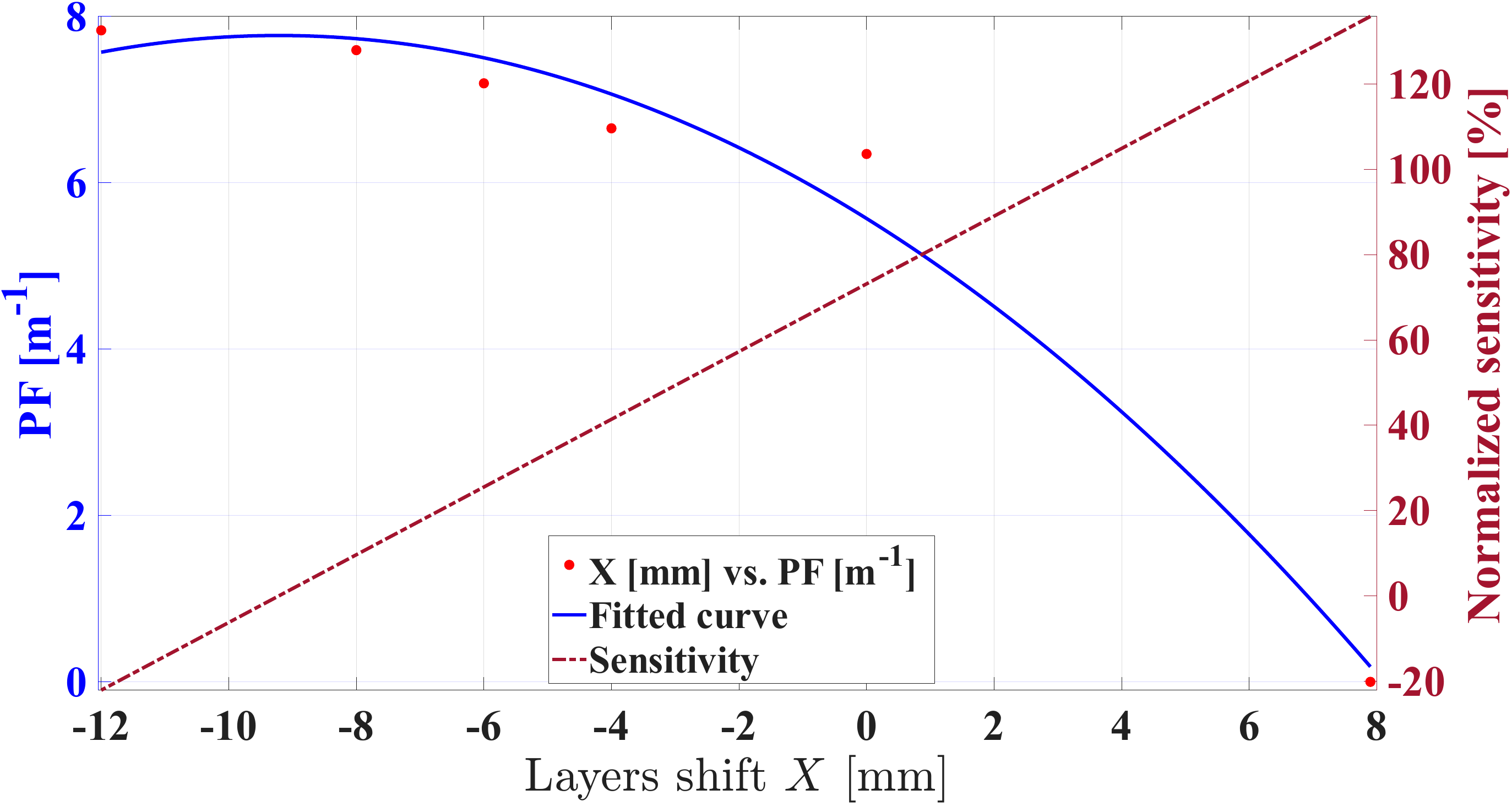}}
            \end{center}
            \vspace{-4.0mm}
            \caption{Power factor variation and sensitivity with respect to the second and third layers shift in parallel and convergent configurations: (a) Mid-IR at 150~THz and (b) mm-wave at 40~GHz spectra.}
            \label{Fig: Fig10}
        \end{figure}

        To estimate the focal length for different layer displacements, the displacement vs. focal length data were fitted with second-order polynomials. A second-order model was selected because lower-order polynomials did not accurately capture the nonlinear behavior near the maximum lateral displacement ($\pm L/2$), while maintaining accuracy over the entire tuning range. This was verified by comparing the Coefficient of Determination ($R^{\rm 2}$) and Root Mean Square Error (RMSE) of the second-order fit with the first-order model (Table~\ref{Tab: Table04}). The resulting models for the mid-IR and mm-wave systems are given in Eqs.~\ref{Eq: Eq09}--\ref{Eq: Eq10} and illustrated in Fig.~\ref{Fig: Fig10}. In both cases, the fitted curves show that small increases in the layer displacement magnitude ($|X|$) produce nearly constant changes in the $PF$ over the convergent tuning range, indicating approximately linear tuning.
        \begin{equation}
            PF_{\rm mid-IR}(X)=-33.5396X^{\rm 2}-3.3803\times10^{\rm 3}X+1.6278\times10^{\rm 4}
            \label{Eq: Eq09}
        \end{equation}
        \begin{equation}
            PF_{\rm mm-wave}(X)=-0.0259X^{\rm 2}-0.4774X+5.5703.
            \label{Eq: Eq10}
        \end{equation}
        \begin{table*}[!htbp]
            \centering
            \caption{Power factor variations and the error in focal length estimation as functions of the second and third layer shifts in parallel and convergent configurations.}
            \vspace*{2mm}
            \setlength{\tabcolsep}{3pt}
            \begin{tabular}{c|c|c|c|c|c|c}
                \toprule
                \toprule
                 \rowcolor{Gray}
                 \multicolumn{7}{c}{mid-IR spectrum} \\
                \toprule
                 $X$ shift values [$\mu\text{m}$] & 4.57 & 0 & -2 & -4 & -6 & -8 \\
                \hline
                \rowcolor{Gray}
                 $F_{\rm GRIN}$ [$\mu\text{m}$] & - & 56.47 & 49.47 & 32.37 & 28.37 & 24.37 \\
                \hline
                 $F_{\rm Ideal}$ [$\mu\text{m}$] & - & 54 & 46 & 30 & 26 & 22 \\
                \hline
                 \rowcolor{Gray}
                 $\Delta F$ [\%] & - & 4.37 & 7.01 & 7.32 & 8.35 & 9.95 \\
                \hline
                 $PF_{\rm GRIN}$~[m$^{\rm -1}$] & 0 & 17.71$\times10^{3}$ & 20.41$\times10^{3}$ & 30.89$\times10^{3}$ & 35.25$\times10^{3}$ & 41.03$\times10^{3}$ \\
                \toprule
                \toprule
                 \rowcolor{Gray}
                 \multicolumn{7}{c}{mm-wave spectrum} \\
                \toprule
                 $X$ shift values [mm] & 7.9 & 0 & -4 & -6 & -8 & -12 \\
                \hline
                \rowcolor{Gray}
                 $F_{\rm GRIN}$ [mm] & - & 157.58 & 150.28 & 138.98 & 131.68 & 127.68 \\
                \hline
                 $F_{\rm Ideal}$ [mm] & - & 154 & 145 & 133 & 125 & 120 \\
                \hline
                \rowcolor{Gray}
                 $\Delta F$ [\%] & - & 2.27 & 3.51 & 4.31 & 5.07 & 6.02 \\
                \hline
                 $PF_{\rm GRIN}$~[m$^{\rm -1}$] & 0 & 6.35 & 6.65 & 7.19 & 7.59 & 7.83 \\
                \toprule
                \toprule
            \end{tabular}
            \begin{minipage}{7.5cm}
            \vspace{0.05cm}
            \end{minipage}
            \label{Tab: Table03}
        \end{table*}
        
        \begin{table*}[ht!]
            \centering
            \caption{Fitting accuracy comparison across model orders.}
            \vspace*{2mm}
            \setlength{\tabcolsep}{3pt}
            \begin{tabular}{c|c|c|c|c}
                \toprule
                \toprule
                 \rowcolor{Gray}
                  Spectrum & Model order & $R^{\rm 2}$ & RMSE [m$^{\rm -1}$] & Adequacy assessment \\
                \toprule
                   & 1st-order (linear) & 0.9874 & 1.5181$\times$10$^{\rm 3}$ & Fairly adequate \\
                   & 2nd-order (quadratic) & 0.989040 & 1.413517$\times$10$^{\rm 3}$ & Fairly adequate \\
                \toprule
                   & 1st-order (linear) & 0.824 & 1.134 & Fairly adequate \\
                   & 2nd-order (quadratic) & 0.9775 & 0.4054 & Highly adequate \\
                \toprule
                \toprule
            \end{tabular}
            \begin{minipage}{7.5cm}
            \vspace{0.05cm}
            \end{minipage}
            \label{Tab: Table04}
        \end{table*}
        
        Normalized sensitivity is beneficial for analyzing GRIN lens focusing performance as it enables a direct and equitable comparison across different layer shifts. The normalized sensitivity metric is a dimensionless factor defined as the relative change in $PF$ per relative shift in the lens layer ($S_{\rm N}=\frac{\Delta PF/PF_{\rm b}}{\Delta X/X_{\rm b}}$), where $PF_{\rm b}$ and $X_{\rm b}$ represent the reference $PF$ and shift values for the mid-infrared ($PF=41.03\times10^{3}$~m$^{\rm -1}$ and $X_{\rm b}=-8~\mu\text{m}$) and millimeter-wave ($PF=7.83$~m$^{\rm -1}$ and $X_{\rm b}=-12~\text{mm}$) cases. To evaluate lens focusing performance, Figs.~\ref{Fig: Fig10.a} and~\ref{Fig: Fig10.b} depict the normalized sensitivity for the mid-IR and mm-wave structures across a range of shift values, which correspond to varying lens $PF$ quality. In convergent configurations, as the magnitude of the layer shifts ($|X|$) increases (i.e., $X$ decreases), the $PF$ of the lens increases.
        
        The key design parameters and simulated focusing performance of the proposed mechanically reconfigurable GRIN lens are summarized in Table~\ref{Tab: Table05}. Compared with prior works, the presented concept demonstrates competitive or improved focal accuracy and mechanical tunability without relying on complex feed networks or 3D printing. Furthermore, unlike optofluidic or graphene-based tunable GRIN lenses that require external fluid handling or voltage biasing, the proposed symmetric lateral displacement mechanism offers a robust mechanical alternative.
        
        Overall, the results in Table~\ref{Tab: Table05} indicate that the proposed GRIN lens achieves accurate focal control with percentage errors below 10\% across both spectral bands, positioning it favorably among state-of-the-art reconfigurable GRIN focusing systems. The tuning resolution of the focal point may be determined from the actuation mechanisms and the relationship linking layer displacements to focal length. Under ideal conditions, the estimated focal length tuning resolution may range from approximately 2-10~nm in the mid-IR and 1-5~$\mu\text{m}$ in the mm-wave. These figures represent the minimum focal shift attainable per mechanical step, assuming an ideal quasi-linear power-factor response. However, experimental validation remains necessary to confirm if these values are feasible in practice.
        \begin{landscape}
            \begin{table*}[!htbp]
                \scriptsize
                \centering
                \caption{Comparison with state-of-the-art GRIN lens.}
                \vspace*{2mm}
                \setlength{\tabcolsep}{3pt}
                \begin{tabular}{c|c|c|c|c|c|c|c|c|c}
                    \toprule
                    \toprule
                    \rowcolor{Gray}
                      & Approach &  &  &  &  &  &  &  &  \\
                    \rowcolor{Gray}
                      & or & \multirow{-2}{*}{Frequency} & Polarization & \multirow{-2}{*}{Reconfigurability} & \multirow{-2}{*}{Key} & Limitations & Remarks & \multirow{-2}{*}{Tuning} & \multirow{-2}{*}{Tuning} \\
                    \rowcolor{Gray}
                     \multirow{-3}{*}{Refs.} & Structure & \multirow{-2}{*}{range} &  & \multirow{-2}{*}{mechanism} & \multirow{-2}{*}{features} &  &  & \multirow{-2}{*}{range} & \multirow{-2}{*}{resolution} \\
                    \toprule
                      & Meta-optical & Visible &  &  & 33\% reduction in rigid & No & Inverse-designed & Depth of field & Not \\
                     \multirow{-2}{*}{\cite{Fröch_2023}} & fiber endoscope & wavelengths & \multirow{-2}{*}{--} & \multirow{-2}{*}{Static} & tip length; 22.5$^\circ$ FOV & tunability & meta-optics & > 30~mm & applicable \\
                    \hline
                    \rowcolor{Gray}
                      & GRIN lens &  &  &  & Antenna performance &  & High gain &  &  \\
                    \rowcolor{Gray}
                     \cite{Whiting_2023} & integrated with & Microwave & \multirow{-2}{*}{Dual} & Static & enhancement; Uniform  & \multirow{-2}{*}{No} & Improved aperture & \multirow{-2}{*}{Not} & \multirow{-2}{*}{Not} \\
                    \rowcolor{Gray}
                      & reflectarray feeds &  & \multirow{-2}{*}{polarization} &  & field distribution & \multirow{-2}{*}{tunability} & efficiency & \multirow{-2}{*}{applicable} & \multirow{-2}{*}{applicable} \\
                    \hline
                      & Inverse-designed &  &  &  & High focusing efficiency & No tunability & Computationally & Not & Not \\
                     \multirow{-2}{*}{\cite{Blackwell_2023}} & GRIN lens & \multirow{-2}{*}{THz} & \multirow{-2}{*}{Linear} & \multirow{-2}{*}{Static} & Quantum algorithm-based & Complex ML & intensive design & applicable & applicable \\
                    \hline
                    \rowcolor{Gray}
                      & Graphene-dielectric &  &  & Electrical tuning &  & Needs biasing & Active & Tunable &  \\
                    \rowcolor{Gray}
                     \cite{Moharrami_2020} & multilayer & THz & \multirow{-2}{*}{TM} & (chemical & \multirow{-2}{*}{Continuous tunability} & Material/Fabrication & material-based & photonic & \multirow{-2}{*}{Voltage} \\
                     \rowcolor{Gray}
                      & GRIN lens &  & \multirow{-2}{*}{polarized} & potential) & \multirow{-2}{*}{EMT-based design} & complexity & control & jets & \multirow{-2}{*}{dependent} \\
                    \hline
                      & Body-matched &  &  &  & Improves imaging & Requires pumps & Not practical &  & Fluid \\
                     \cite{Rezaeieh_2021} & GRIN lens & Microwave & Linear & \multirow{-2}{*}{Fluidic} & accuracy & Requires reservoirs & for & \multirow{-2}{*}{Not} & volume \\
                      & for imaging &  &  & \multirow{-2}{*}{tuning} & Plane-wave conversion & Leakage risk & compact systems & \multirow{-2}{*}{applicable} & dependent \\
                    \hline
                    \rowcolor{Gray}
                      & Tapered &  &  &  &  &  &  &  &  \\
                    \rowcolor{Gray}
                     \cite{Mousavi_2023} & GRIN lens & Microwave & \multirow{-2}{*}{Linear} & Static & \multirow{-2}{*}{Enhanced penetration} & \multirow{-2}{*}{Physically large} & \multirow{-2}{*}{125\% fractiona} & \multirow{-2}{*}{Not} & \multirow{-2}{*}{Not} \\
                    \rowcolor{Gray}
                      & antenna &  & \multirow{-2}{*}{(Slot-fed)} &  & \multirow{-2}{*}{Near-field imaging} & \multirow{-2}{*}{Complex fabrication} & \multirow{-2}{*}{bandwidth} & \multirow{-2}{*}{applicable} & \multirow{-2}{*}{applicable} \\
                    \hline
                      & Planar GRIN using  &  &  &  & High channel counts & Wavelength-dependent & Discusses & Number of channels &  \\
                      & selfoc semi-rod &  & Not &  & Simple geometry & performance & fundamental limits & scales inversely &  \\
                      & lens for on-axis and &  & specified &  & No integrated & Chromatic and off-axis & (SBP) for & with wavelength &  \\
                     \multirow{-4}{*}{\cite{Gomez_2008}} & off-axis multiple & \multirow{-4}{*}{Optical} &  & \multirow{-4}{*}{Mechanical} & surface components & aberrations & channel capacity &  & \multirow{-4}{*}{Discrete} \\
                    \hline
                    \rowcolor{Gray}
                      & Microfluidic &  &  &  & Simple fabrication & Optical aberrations & Fully tunable & 120-130~$\mu\text{m}$ & Syringe \\
                    \rowcolor{Gray}
                      & L-GRIN &  & \multirow{-2}{*}{Not} &  & Low fluid consumption & Temperature-sensitive & liquid GRIN & along the & pumps \\
                    \rowcolor{Gray}
                     \multirow{-3}{*}{\cite{Huang_2010}} & lens & \multirow{-3}{*}{Optical} & \multirow{-2}{*}{specified} & \multirow{-3}{*}{Fluidic} & Easy integration & Precise flow control & lens & $z$-axis & dependent \\
                    \hline
                      & GRIN Luneburg &  &  &  &  & Complex feed network &  & Tunability &  \\
                     \cite{Biswas_2023} & lens with & Ku-band & \multirow{-2}{*}{Dual} & \multirow{-2}{*}{Beam switching} & \multirow{-2}{*}{Multibeam operation} & Not & \multirow{-2}{*}{System-level} & is external & Discrete \\
                      & feed matrix &  & \multirow{-2}{*}{polarization} & \multirow{-2}{*}{via feed network} & \multirow{-2}{*}{Reduces mutual coupling} & lens-intrinsic tuning & \multirow{-2}{*}{beamforming} & (feed) &  \\
                    \hline
                    \rowcolor{Gray}
                      & Active GRIN &  &  &  & Low power consumption & Requires a complex & All 1024 elements & $\pm$40$^\circ$ &  \\
                     \rowcolor{Gray}
                      & using &  & \multirow{-2}{*}{Linear} & \multirow{-2}{*}{Electronic} & Simplified control network & multi-layer PCB & are controlled & beam steering & Discrete \\
                    \rowcolor{Gray}
                     \multirow{-3}{*}{\cite{Cho_2020}} & FSS & \multirow{-3}{*}{Microwave} & \multirow{-2}{*}{polarization} & \multirow{-2}{*}{tuning} & Wide 3-dB bandwidth & for FSS structure & simultaneously & range &  \\
                    \hline
                      & Multilayer planar &  &  & Mechanical lateral & Continuous focal tuning &  &  &  &  \\
                      & GRIN lens &  &  & displacement & No biasing & \multirow{-2}{*}{Requires precision} &  &  &  \\
                     \multirow{-3}{*}{This} & (EMT- and  & \multirow{-3}{*}{mid-IR} & \multirow{-3}{*}{Linear} & (opposite layers & No fluids & \multirow{-2}{*}{mechanics; residual} & \multirow{-3}{*}{Compact} & \multirow{-3}{*}{24.37--56.47~$\mu\text{m}$} & \multirow{-3}{*}{Actuator} \\
                     \multirow{-3}{*}{work} & Chebyshev-based) & \multirow{-3}{*}{mm-wave} & \multirow{-3}{*}{(y-pol.)} & shift) & Scalable & \multirow{-2}{*}{aberration at extremes} & \multirow{-3}{*}{<10\% error} & \multirow{-3}{*}{127.68--157.58~mm} & \multirow{-3}{*}{dependent} \\
                    \toprule
                    \toprule
                \end{tabular}
                \label{Tab: Table05}
            \end{table*}
        \end{landscape}
    \subsection{Analytical and Numerical Tolerance Analysis for Lateral Misalignment and Fabrication Errors}
        An analytical tolerance analysis based on partial differentiation was performed to evaluate the sensitivity of the effective refractive index and focal length to fabrication errors (see Supplementary Material). Gaussian perturbations of $\pm10\%$ were applied to the physical parameters, and 10,000 Monte Carlo trials yielded mean integrated global uncertainty margins of $\pm1.0328$ and $\pm1.0237$ for the mid-IR and mm-wave bands, respectively (Eq.~\ref{Eq: Eq11}).
        \begin{equation}
            \sigma_{n^{\rm eff.}}(x) = \sqrt{\sum_{k=1}^{N} \left( T_{k} \sigma_{n_{\rm eff.}^k}(x) \right)^{\rm 2} + \sum_{k=1}^{N} \left( n_{\rm eff.}^{k}(x) \sigma_{T_{k}} \right)^{\rm 2}}.
            \label{Eq: Eq11}
        \end{equation}
        
        The focal-length sensitivity was evaluated using the following equation (Eq.~\ref{Eq: Eq12}):
        \begin{equation}
            \sigma_{F}(x_{\rm i}) = \frac{|x_{i}|}{t |\frac{dn(x_{i})}{dx}| \sqrt{1 - \left(t \frac{dn(x_{i})}{dx}\right)^{\rm 2}}} \sqrt{ \left(\frac{\sigma_{t}}{t}\right)^{\rm 2} + \left(\frac{\sigma_{\frac{dn(x_{i})}{dx}}}{\frac{dn(x_{i})}{dx}}\right)^{\rm 2}}.
            \label{Eq: Eq12}
        \end{equation}
        In this equation, the accumulated focal standard deviation ($\sigma_{F}(x_{i})$) results from substrate thickness errors ($\sigma_{t}$) and local lithographic index-gradient errors ($\sigma_{\frac{dn(x_{i})}{dx}}$). The geometric scaling term shows that the sensitivity increases as $t\frac{dn(x_{i})}{dx}\rightarrow1$, corresponding to the onset of total internal reflection within the layers.
        
        The analytical model was validated by 10,000 Monte Carlo trials with Gaussian perturbations of up to $\pm10\%$ applied simultaneously to the layer thickness and index-gradient profiles. The simulations predict average systematic focal shifts of $\pm0.7570~\mu\text{m}$ and $\pm3.8247~\text{mm}$ for the mid-IR and mm-wave bands, respectively, with corresponding standard deviations of $\pm5.7426~\mu\text{m}$ and $\pm27.4~\text{mm}$. These results indicate that the Chebyshev-derived index profile is robust against the considered fabrication errors.
        
        Numerical studies were also performed for Profiles $IV$ and $IV'$ to evaluate the effects of lateral misalignment and fabrication errors on the focal length. Misalignment shifts of $\chi=\pm0.8~\mu\text{m}$ and $\chi=\pm1.2~\text{mm}$ (corresponding to 10\% of the nominal layer displacements) were introduced into the second and third layers for the mid-IR and mm-wave designs ($[0,X+\chi,-(X+\chi)]$), respectively. Fabrication errors were further evaluated by introducing Gaussian random variations of $\pm10\%$ in the groove height. The resulting focal-length variations confirm the tolerance of the proposed design to layer misalignment and fabrication errors (Table~\ref{Tab: Table06}).
        \begin{table*}[ht!]
            \centering
            \caption{Numerical tolerance analysis for lateral misalignment and manufacturing errors with Profiles $IV$ and $IV’$.}
            \vspace*{2mm}
            \setlength{\tabcolsep}{3pt}
            \begin{tabular}{c|cc|cc|cc}
                \toprule
                \toprule
                 \rowcolor{Gray}
                 - & \multicolumn{2}{c}{Fabrication tolerance} & \multicolumn{2}{c}{$+$ layer misalignment} & \multicolumn{2}{c}{$-$ layer misalignment} \\
                \toprule
                  & mid-IR & mm-wave & mid-IR & mm-wave & mid-IR & mm-wave \\
                 \multirow{-2}{*}{Spectrum} & [$\mu m$] & [mm] & [$\mu m$] & [mm] & [$\mu m$] & [mm] \\
                \hline
                 \rowcolor{Gray}
                 Main &  &  &  &  &  &  \\
                 \rowcolor{Gray}
                 focal length & \multirow{-2}{*}{24.37} & \multirow{-2}{*}{127.68} & \multirow{-2}{*}{24.37} & \multirow{-2}{*}{127.68} & \multirow{-2}{*}{24.37} & \multirow{-2}{*}{127.68} \\
                \hline
                 Noise  &  &  &  &  &  &  \\
                 contaminated & 25.37 & 126.68 & 26.37 & 129.68 & 22.37 & 125.68 \\
                 focal length &  &  &  &  &  &  \\
                \hline
                \rowcolor{Gray}
                 $\Delta F$~[\%] & $\sim$4.10 & $\sim$0.78 & $\sim$8.21 & $\sim$1.57 & $\sim$8.21 & $\sim$1.57 \\
                \toprule
                \toprule
            \end{tabular}
            \begin{minipage}{7.5cm}
            \vspace{0.05cm}
            \end{minipage}
            \label{Tab: Table06}
        \end{table*}
\section{Discussions and Conclusion}
    GRIN lenses have emerged as essential components in optical, infrared and microwave systems due to their ability to manipulate electromagnetic wavefronts through spatially varying refractive index profiles. In telecommunications, such lenses play an important role in reconfigurable focusing applications, enabling efficient conversion of spherical wavefronts into plane waves with minimal distortion. This capability supports low-profile and frequency-independent operation, key requirements for 5G and 6G transceiver systems.
    
    GRIN lenses further enhance phase-modulated lenses by suppressing sidelobe levels and improving spatial selectivity, attributed to their precise wavefront focusing. They also optimize coupling between free-space and guided-wave components in high-frequency data links, thereby improving signal integrity and reducing insertion losses.
    
    By providing compact, high-precision focusing, GRIN lenses can enhance the spatial resolution and field of view (FOV) of millimeter-wave biomedical imaging systems. Improved focusing increases image contrast and enables more accurate tumor detection and characterization in applications such as breast imaging, skin cancer detection, and security screening. In wearable systems, GRIN lenses can facilitate compact wireless body-area networks for continuous health monitoring, while their ability to concentrate microwave energy enables localized tissue heating for hyperthermia therapy. More broadly, their capability to control electromagnetic wave propagation makes GRIN lenses attractive for a wide range of mid-IR and millimeter-wave applications.
    
    In this article, we presented a methodology for designing mechanically reconfigurable GRIN lenses capable of dynamic focusing in the mid-IR and mm-wave bands. Using EMT, a multilayer architecture was designed in which the effective refractive index is spatially tailored through groove etching in individual lens layers. A comparative study of wavelets and polynomial basis functions identified first-kind Chebyshev polynomials as the most suitable choice for approximating the smooth refractive-index profiles required for reconfigurable focusing. We have shown that symmetric lateral displacement of mirrored third-order Chebyshev layers enables subwavelength focal tuning. Full-wave simulations confirmed continuous focal-length tuning, with relative errors as low as 4.37\% in the mid-IR band and 2.27\% in the mm-wave band.
    
    Essentially, the proposed planar GRIN lens implements the Alvarez tuning principle {\em in the refractive-index domain,} where lateral displacement of complementary layers enables continuous focal tuning. In a classical Alvarez lens, two cubic-phase plates are shifted laterally to generate a quadratic phase profile whose optical power varies linearly with displacement. In our lens, a stationary second-order Chebyshev layer provides the baseline focusing power, while two symmetrically displaced third-order Chebyshev layers modulate the quadratic coefficient of the effective refractive-index distribution. This separates the fixed focusing power from the focal tuning mechanism. Unlike the classical Alvarez lens, which relies on cancellation of cubic surface terms, the proposed design exploits mirrored Chebyshev polynomial refractive index profiles and eliminates the need for freeform optical surfaces. The planar architecture is compatible with lithographic fabrication and integrated photonic platforms, while effective medium theory (EMT) provides additional control over the local refractive index through groove geometry.
    
    The discussed GRIN lens concept may find applications in telecommunication and medical imaging applications. In a future work we will investigate means to reduce interface reflections. The refractive index distribution will be also synthesized using further-optimized polynomial profiles, which will minimize phase errors and aberrations. When combined with the minimax approximation property of Chebyshev polynomials, this may provide a novel framework for the design of reconfigurable planar GRIN lenses.
\section*{Acknowledgement}
    K. Kaboutari acknowledges co-financing by Funda\c{c}\~{a}o para a Ci\^{e}ncia e a Tecnologia (Portuguese Foundation for Science and Technology) through the Carnegie Mellon Portugal Program under the fellowship PRT/BD/154201/2022 and DOI identifier \href{https://doi.org/10.54499/PRT/BD/154201/2022}{10.54499/PRT/BD/154201/2022}. S. Maslovski and P. Pinho acknowledge funding by national funds through FCT – Fundação para a Ciência e a Tecnologia, I.P., and, when eligible, by EU funds under the project UID/50008/2025 – Instituto de Telecomunicações, with DOI identifier - \href{https://doi.org/10.54499/UID/50008/2025}{10.54499/UID/50008/2025}.

\end{document}